\documentclass[iop]{emulateapj}
\usepackage{epsfig}

\slugcomment{Submitted to the Astrophysical Journal Supplement}

\shorttitle{Tycho-2 Stars in the Galactic Bulge Survey}
\shortauthors{Hynes et al.}

\begin{document}

\title{Identification of Galactic Bulge Survey X-ray Sources with Tycho-2 Stars}

\author{
Robert I. Hynes\altaffilmark{1,2}, 
N.~J. Wright\altaffilmark{3}, 
T.~J. Maccarone\altaffilmark{4}, 
P.~G. Jonker\altaffilmark{5,6,3}, 
S. Greiss\altaffilmark{7}, 
D. Steeghs\altaffilmark{7,3}, 
M.~A.~P. Torres\altaffilmark{5,3}, 
C.~T. Britt\altaffilmark{2}, 
G. Nelemans\altaffilmark{6}}

\altaffiltext{1}{E-mail: rih@phys.lsu.edu} \altaffiltext{2}{Department
  of Physics and Astronomy, Louisiana State University, 202 Nicholson
  Hall, Tower Drive, Baton Rouge, LA 70803, USA}
\altaffiltext{3}{Harvard–Smithsonian Center for Astrophysics, 60
  Garden Street, Cambridge, MA 02138, U.S.A.}  
\altaffiltext{4}{School of Physics and Astronomy, University of
  Southampton, Southampton SO17 1BJ, United Kingdom}
\altaffiltext{5}{SRON, Netherlands Institute for Space Research,
  Sorbonnelaan 2, 3584 CA, Utrecht, The Netherlands}
\altaffiltext{6}{Department of Astrophysics, IMAPP, Radboud University
  Nijmegen, Heyendaalseweg 135, 6525 AJ, Nijmegen, The Netherlands}
\altaffiltext{7}{Astronomy and Astrophysics, Department of Physics,
  University of Warwick, Coventry, CV4 7AL, United Kingdom}

\begin{abstract} 
We identify 69 X-ray sources discovered by the Galactic Bulge Survey
(GBS) that are coincident with, or very close to bright stars in the
Tycho-2 catalog.  Additionally, two other GBS sources are resolved
binary companions to Tycho-2 stars where both components are
separately detected in X-rays.  Most of these are likely to be real
matches, but we identify nine objects with large and significant X-ray
to optical offsets as either detections of resolved binary companions
or chance alignments.  We collate known spectral types for these
objects, and also examine 2MASS colors, variability information from
the All-Sky Automated Survey (ASAS), and X-ray hardness ratios for the
brightest objects. Nearly a third of the stars are found to be
optically variable, divided roughly evenly between irregular
variations and periodic modulations. All fall among the softest
objects identified by the GBS.  The sample forms a very mixed
selection, ranging in spectral class from O9 to M3. In some cases the
X-ray emission appears consistent with normal coronal emission from
late-type stars, or wind emission from early-types, but the sample
also includes one known Algol, one W~UMa system, two Be stars, and
several X-ray bright objects likely to be coronally active stars or
binaries.  Surprisingly, a substantial fraction of the
spectroscopically classified, non-coincidental sample (12 out of 38
objects) have late B or A type counterparts.  Many of these exhibit
redder near-IR colors than expected for their spectral type and/or
variability, and it is likely that the X-rays originate from a
late-type companion star in most or all of these objects.
\end{abstract}


\section{Introduction}

The Galactic Bulge Survey (GBS; \citealt{Jonker:2011a}) is a wide but
shallow X-ray survey of two strips above and below the bulge.  The
design of the survey calls for two regions of Galactic longitude
$-3^{\circ} < l < 3^{\circ}$, and latitude $1^{\circ} < |b| <
2^{\circ}$.  Three quarters of the area was imaged by the {\it Chandra
  X-ray Observatory} ACIS-I camera with exposures of at least 2\,ks
and a detection flux limit of
$7.7\times10^{-14}$\,erg\,cm$^{-2}$\,s$^{-1}$ from 2007--2009, with
the remaining area being observed in Fall 2011 and Spring 2012.  This
has resulted in the detection of 1234 X-ray sources from the
2007--2009 observations \citep{Jonker:2011a} and an additional 424
objects from 2011--2012 (Jonker et al., in preparation).  By design,
the survey avoids the most crowded and absorbed regions of the
Galactic plane and so many of these sources have unique candidate
optical counterparts.  Follow-up optical spectroscopy and multi-epoch
photometry is finding a high fraction of sources with spectroscopic
peculiarities and/or photometric variability that appear to be true
counterparts to the X-ray sources.  Most sources are expected to be
Galactic in origin, although about some background active galactic
nuclei are present in the sample \citep{Maccarone:2012a}.

Among the optical counterparts, a fraction correspond to very bright,
often nearby stars, and these can be expected to be dominated by
detections of `normal' stars.  This paper reports on the properties of
those GBS objects coincident with stars in the Tycho-2 catalog
\citep{Hog:2000a}.  As such, it is limited to optical counterparts
brighter than 12th magnitude.  Many of these stars also appear in the
Henry Draper catalogs \citep{Cannon:1924a,Cannon:1936a,Nesterov:1995a}
and so have spectral classifications.  While our main interest is to
find any abnormal objects among the sample, by matching of our
predominantly faint X-ray sources with bright optical counterparts we
are naturally selecting for objects with rather low X-ray to optical
flux ratios, and will thus expect to mainly exclude exotica like X-ray
binaries.

While there were isolated early detections of stellar X-rays, the {\it
  Einstein} satellite really opened to door to systematic studies of
stellar X-ray emission, and revealed that across the
Hertzsprung-Russell diagram X-ray sources are the rule, rather than
the exception \citep{Vaiana:1981a}.  As the sensitivity of X-ray
telescopes has improved, more and more stars have been added to X-ray
catalogs, consequently filling in the region of parameter space
occupied by the lowest X-ray luminosity stars. In some cases this is
through deliberate surveys, and sometimes as serendipitous detections.
A comprehensive review of stellar X-ray astronomy is provided by
\citet{Gudel:2009a} and references therein.

Among early-type OB stars, the standard paradigm is that X-rays arise
from shocks within their strong winds at quite large radii
\citep[][and references therein]{Owocki:2001a}. Empirically, the
relation $L_{\rm X} / L_{\rm bol} \simeq 10^{-7}$ is found in O--B1
stars \citep{Pallavicini:1981a,Berghofer:1997a}, supporting a strong
coupling between the radiatively driven winds and the X-ray
generation.  Many details of the mechanism remain unclear, and the
high spectral resolution studies possible with {\it Chandra} and {\it
  XMM-Newton} have raised challenges to the standard paradigm
\citep[e.g.\ ][]{Pollock:2007a}. The X-ray spectra from these objects
are typically very soft, peaking below $1.5$\,keV. They can usually be
well fit by between one and three thermal plasma components
\citep[e.g.\ ][]{Sana:2006a,Antokhin:2008a}. In reasonably close OB
binaries, it is possible to also develop X-rays from the shock where
their winds interact, leading to the class of over-luminous colliding
wind binaries \citep{Stevens:1992a}.

The efficiency of X-ray production from winds drops substantially
among B stars, with $L_{\rm X} / L_{\rm bol}$ decreasing from
$10^{-6}$ for the most X-ray luminous B0 stars to
$10^{-10}$--$10^{-8}$ for spectral types later than B2
\citep{Cohen:1997a}.  \citet{Sana:2006a} found that in the young open
cluster NGC\,6231 less than 20\,\%\ of mid to late B stars were
detected in X-rays, and those that were had harder spectra than the O
stars, and were sometimes seen to flare.  They suggested that the
X-rays actually originate from pre-main-sequence companions, for which
the hard spectra and flaring are quite typical.

The Be stars form an important sub-class of B stars
\citep{Porter:2003a}. In these rapidly rotating stars an equatorial
excretion disk develops. These objects can be intrinsic X-ray emitters
in their own right, with typical Be stars in the sample of
\citet{Cohen:1997a} showing X-ray luminosities about three times
higher than comparable normal B stars.  More dramatically, when Be
star systems play host to neutron stars, more luminous X-rays from the
interaction between the compact object and the circumstellar material
can be seen, typically in periodic outbursts \citep{Coe:2000a}.
Recently a class of intermediate luminosity objects, the $\gamma$~Cas
analogs, have been identified \citep{Motch:2007a} including, it turns
out, one of our sample, HD\,161103 \citep{LopesDeOliveira:2006a}.
These objects show much harder X-ray spectra than typical OB or Be
stars, and short timescale X-ray variability.  Their nature remains
unclear; they could be unusual single Be stars, or counterparts to the
Be X-ray binaries containing white dwarfs instead of neutron stars.

X-ray emission from solar- and late-type main sequence stars is
ubiquitous and originates from a magnetically confined plasma at a
temperature of several million Kelvin known as a corona
\citep{Vaiana:1981a}. The corona is powered by a dynamo, which itself
is generated by a combination of rotation and convection within the
convective envelope. A strong correlation between rotation and stellar
X-ray activity was first noted by \citet{Pallavicini:1981a} and has
since been quantified in detail (e.g.\ \citealt{Wright:2011a}),
providing strong support for the dynamo-induced nature of stellar
activity. As stars spin down as they age
(e.g. \citealt{Skumanich:1972a}), the dynamo weakens and X-ray
activity reduces from the high levels seen in young stars ($L_{\rm X}
/ L_{\rm bol} \sim 10^{-3}$, e.g.\ \citealt{Feigelson:2005a}) to the
levels typically seen in the Sun and old field stars ($L_{\rm X} /
L_{\rm bol} \sim 10^{-5} - 10^{-7}$, e.g.\ \citealt{Wright:2010a}). 
Generally late-type evolved stars are weak X-ray emitters. Exceptions
can be expected to be unusually rapidly rotating single stars, of
which the FK~Com stars present the most extreme example, or forced
into rapid rotation in close binaries such as RS~CVn systems. A
comprehensive review of X-ray emission from stellar coronae is
provided by \citet{Gudel:2004a}.

In fully radiative stars (earlier than mid F-type) the dynamo
mechanism breaks down and X-ray emission becomes very weak.  Vega, for
example, remains undetected as an X-ray source, with a {\em Chandra}
upper limit to its luminosity of $2\times10^{25}$\,erg\,s$^{-1}$
\citep{Pease:2006a}.  X-ray detections of late B and A stars are then
quite unexpected, and in most cases the origin of the X-rays will be a
later type companion star.

More generally, the overarching scientific interest of the GBS is to
extend our knowledge of binaries, and binary evolution, and as
discussed above, we would expect many of the X-ray brightest objects
in our sample to be binaries in some form.  RS~CVn systems have
already been mentioned.  BY~Dra systems are similar, containing
main-sequence stars.  Algols, too are typically seen as X-ray
emitters, with the X-rays originating from the cooler star in the
binary.  The most extreme close binaries, contact W~UMa systems, are
also expected to have strong coronal X-ray emission.  The Algols
illustrate an important point.  Here, it is commonly the hotter
component that is seen in the optical spectrum, but the cool component
that gives rise to coronal X-rays.  Generally, we cannot always assume
that the optically brightest star is the X-ray source.

The goals of this paper are to summarize the matches between the GBS
and Tycho-2 stars, to discuss their properties and attempt to classify
each object, to examine if any of these stars show anomalous X-ray
emission for their spectral type, and ultimately to either identify
objects containing compact companions or eliminate these objects from
our sample as candidate interacting binaries.  We begin by discussing
the catalogs and matching criteria used in Section~\ref{MatchSection}.
To classify the sources we then examine X-ray to optical flux ratios
(Section~\ref{RatioSection}), optical/infrared color-color diagrams
(Section~\ref{ColorColorSection}), and X-ray hardness ratios
(Section~\ref{HardnessSection}).  We draw this, and other available
information together for individual sources in
Section~\ref{SourceSection}, and finally discuss the classes of
objects found in Section~\ref{DiscussionSection} and summarize our
findings in Section~\ref{Conclusion}.

\section{Optical and infrared counterparts to GBS X-ray sources}
\label{MatchSection}

\begin{figure}
\includegraphics[width=3.6in]{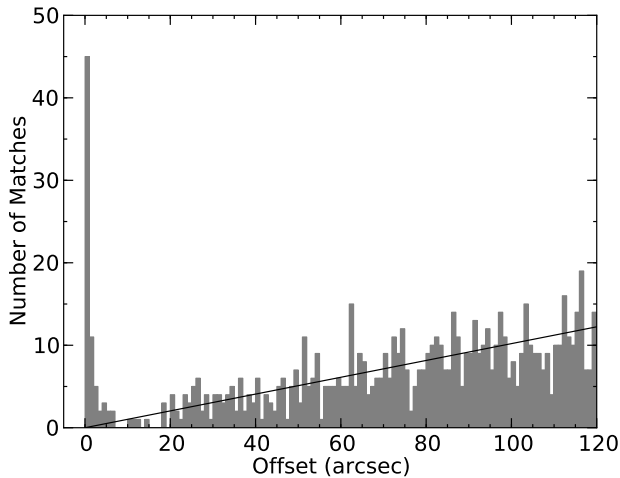}
\caption{Number of matches found between GBS X-ray sources and stars
  in the Tycho-2 catalog, as a function of offset.  Bins are 1\,arcsec
  wide.  The solid line is the expected number of chance coincidences
  based on the number of matches found with offsets of 1--15\,arcmin.}
\label{OffsetHistFig}
\end{figure}

\subsection{Duplicate X-ray detections}

Our starting point is the X-ray source lists from the GBS. We merge
the published list, which we will refer to as the northern sample
\citep{Jonker:2011a} with additional sources detected in the final
southern strip of the survey, the southern sample (Jonker et al.\ in
preparation). For ease of recognition we will use abbreviated source
names of CXnnnn for sources from the northern sample, and CXBnnnn for
sources in the southern sample. Besides being more compact, this
notation has the advantage that each list is sorted in order of X-ray
brightness, so the CX number conveys information that is absent from
the longer positional names such as CXOGBS~J175024.4--290216.

In performing our optical matches, we identified two cases where two
of the X-ray sources appear to coincide with the same Tycho star. In
both cases, the two X-ray sources were detected in different images,
and one has a much larger off-axis angle than the other, and
consequently less precise coordinates. In these cases we believe that
they are duplicate detections of the same source, with position
differences just too large for \citet{Jonker:2011a} to have identified
them as duplicates. The redundant sources are CX191, which appears to
have been an off-axis detection of CX360, and CX230, which appears to
be an off-axis detection of CX33. Another similar case has been found
with the optically fainter catalysmic variable candiate CX93, for
which CX153 appears to be an off-axis detection (Ratti et al.\ in
preparation). On examination of the GBS source catalog, we identified
a further fifteen likely duplications which share the characteristics
of being from different Chandra observation IDs and having one or both
detection at sizeable off-axis angle. These are: CX16 = CX515, CX38 =
CX225, CX61 = CX136, CX175 = CX464, CX177 = CX384, CX345 = CX233,
CX270 = CX571, CX597 = CX273, CX280 = CX382, CX394 = CX285, CX292 =
CX323, CX304 = CX303, CX1155 = CX483, CX636 = CX635, and CX646 =
CX851. In each case, we have listed the detection with smallest
off-axis angle first, so the second source of each pair can be
considered the duplicate.  In most cases, the detection closer to the
axis is also the stronger one.  A few more cases were found where
close objects appear to be detected in the same observation ID, and we
believe these are distinct X-ray sources. CX916 and CX917 are one such
pair that match an optical double, HD316666A and B. We removed all
eighteen of the apparent duplicate objects from the catalog leaving
1216 unique sources in the northern sample.

Following this experience, we were careful to check the southern
sample for repeat detections of objects from the northern sample, and
for duplicates.  As a result, the southern sample includes 424 new and
unique X-ray sources with no credible duplications.  After duplicate
removal, our combined catalogs total 1640 unique sources.  All
statistics reported here on numbers of matches expected and found are
based on this combined unique subset of the full source lists.

 \subsection{Matches with the Tycho-2 and Henry Draper Catalogs}

We primarily consider matches with the Tycho-2 Astrometric Catalog
\citep{Hog:2000a}.  This provides a reasonably uniform catalog of the
2.5 million brightest stars in the sky, with excellent positions and
lower quality, although still useful, two color photometry.  We also
check for matches with the original Henry Draper Catalog
\citep{Cannon:1924a}, the Henry Draper Catalog Extension
\citep{Cannon:1936a}, and the Henry Draper Extension Charts
\citep[see][and references therein]{Nesterov:1995a} as these provide
spectroscopic classifications for many of the matches found in the
Tycho-2 Catalog.

We consider a candidate match to be an X-ray source within 10\,arcsec
of a Tycho-2 star to allow for the large uncertainties in {\it
  Chandra} positions for large off-axis angles
\citep[see][]{Hong:2005a}.  Proper motions are applied to the Tycho-2
stars pair-by-pair to obtain the correct optical coordinates
corresponding to the epoch of each Chandra observation.  The proper
motions between the J2000 reference epoch of the Tycho-2 catalog, and
the 2006--2012 Chandra observations are mostly small compared to the
Chandra positional uncertainty, but not always, with the largest
accumulated proper motion (for CXB93, a nearby M dwarf) being
3.3\,arcsec.

\begin{deluxetable*}{lllcllcll}
\tabletypesize{\scriptsize}
\tablecaption{\bf Bright Optical Counterparts to GBS Northern Sources
\label{MatchTableNorth}}
\tablewidth{0pt}
\tablehead{\colhead{}  & \colhead{}  & \colhead{} & \colhead{Offset} & \colhead{} & \colhead{} & \colhead{} & \colhead{} &\colhead{}\\
\colhead{GBS}  & \colhead{Tycho-2}  & \colhead{HD} & \colhead{(arcsec)} & \colhead{$V$} & \colhead{$(B-V)$} & \colhead{$\log (F_{\rm x} / F_{\rm opt})$} &
\colhead{Spectral Type} & \colhead{Refs.}}
\startdata
4 & 6839-00084-1 & 316072\phm{A} & 0.46 & \phn$ 9.93\pm0.03$ & \phs$1.21\pm0.07$\phm{$^a$} & -2.54 & G9\,{\sc iii} & 1\phm{, 9, 10} \\
6 & 6836-00576-1 & 161103\phm{A} & 0.21 & \phn$ 8.40\pm0.01$ & \phs$0.33\pm0.02$\phm{$^a$} & -3.34 & B0.5\,{\sc iii-v}e & 2, 3\phm{, 10} \\
7 & 6839-00513-1 & \nodata & 0.30 & $11.20\pm0.10$ & \phs$0.88\pm0.19$\phm{$^a$} & -2.23 & K0\,{\sc v} & 1\phm{, 9, 10} \\
9 & 6839-00257-1 & 315997\phm{A} & 0.30 & $11.36\pm0.11$ & \phs$0.32\pm0.14$\phm{$^a$} & -2.20 & A5 & 4\phm{, 9, 10} \\
10 & 6839-00191-1 & 315992\phm{A} & 0.46 & $10.13\pm0.04$ & \phs$1.10\pm0.08$\phm{$^a$} & -2.75 & G7 & 4, 5\phm{, 10} \\
12 & 7377-01129-1 & 318207\phm{A} & 0.11 & \phn$ 9.46\pm0.02$ & \phs$1.09\pm0.04$\phm{$^a$} & -3.10 & G5 & 4\phm{, 9, 10} \\
25 & 7381-00792-1 & \nodata & 0.56 & $11.76\pm0.17$ & \phs$0.85\pm0.29$\phm{$^a$} & -2.50 & \nodata & \nodata \\
26 & 7377-00936-1 & 161212\phm{A} & 0.14 & \phn$ 8.87\pm0.02$ & \phs$0.64\pm0.03$\phm{$^a$} & -3.65 & G3\,{\sc v} & 6\phm{, 9, 10} \\
27 & 6839-00636-1 & \nodata & 0.80 & $12.19\pm0.23$ & \phs$0.63\pm0.36$\phm{$^a$} & -2.33 & \nodata & \nodata \\
31 & 6839-00348-1 & \nodata & 0.10 & \phn$ 9.97\pm0.03$ & \phs$0.47\pm0.05$\phm{$^a$} & -3.25 & O9\,{\sc v} & 7\phm{, 9, 10} \\
32 & 6839-00218-1 & \nodata & 0.34 & $10.97\pm0.08$ & \phs$0.66\pm0.11$\phm{$^a$} & -2.83 & \nodata & \nodata \\
33 & 6840-01414-1 & 316341\phm{A} & 0.08 & \phn$ 9.10\pm0.02$ & \phs$0.50\pm0.03$\phm{$^a$} & -3.62 & B0\,{\sc v}e & 8\phm{, 9, 10} \\
53 & 6836-01113-1 & 161117\phm{A} & 3.33 & \phn$ 8.85\pm0.02$ & \phs$0.39\pm0.02$\phm{$^a$} & -3.88 & F5\,{\sc v} & 6\phm{, 9, 10} \\
59 & 6832-00663-1 & \nodata & 3.53 & $10.24\pm0.05$ & \phs$0.35\pm0.06$\phm{$^a$} & -3.35 & \nodata & \nodata \\
72 & 7377-00222-1 & 316356\phm{A} & 0.18 & $10.40\pm0.06$ & \phs$0.37\pm0.08$\phm{$^a$} & -3.36 & F8 & 4\phm{, 9, 10} \\
77 & 6839-00487-1 & 159571\phm{A} & 6.32 & \phn$ 9.05\pm0.02$ & \phs$0.11\pm0.02$\phm{$^a$} & -3.90 & B8\,{\sc v} & 6\phm{, 9, 10} \\
82 & 6849-01294-1 & \nodata & 0.97 & $12.32\pm0.19$ & \phs$0.49\pm0.28$\phm{$^a$} & -2.58 & \nodata & \nodata \\
91 & 6849-01082-1 & 314883\phm{A} & 1.11 & $10.58\pm0.05$ & \phs$0.47\pm0.06$\phm{$^a$} & -3.35 & F8 & 4\phm{, 9, 10} \\
115 & 6839-00265-1 & 316070\phm{A} & 0.60 & \phn$ 9.98\pm0.03$ & \phs$0.36\pm0.04$\phm{$^a$} & -3.66 & A2 & 4\phm{, 9, 10} \\
156 & 6840-01006-1 & 161907\phm{A} & 6.24 & \phn$ 8.05\pm0.01$ & \phs$0.28\pm0.02$\phm{$^a$} & -4.48 & F0\,{\sc v} & 6\phm{, 9, 10} \\
183 & 6840-01069-1 & 162120\phm{A} & 1.04 & \phn$ 8.34\pm0.01$ & \phs$0.18\pm0.02$\phm{$^a$} & -4.43 & A2V & 6\phm{, 9, 10} \\
205 & 7377-00199-1 & 161852\phm{A} & 1.63 & \phn$ 6.65\pm0.01$ & \phs$0.32\pm0.02$\phm{$^a$} & -5.18 & F2\,{\sc iv/v} & 6\phm{, 9, 10} \\
256 & 6853-00214-1 & 162761\phm{A} & 0.26 & \phn$ 7.89\pm0.01$ & \phs$1.08\pm0.02$\phm{$^a$} & -4.77 & K0\,{\sc iii} & 6\phm{, 9, 10} \\
272 & 6836-00635-1 & \nodata & 0.82 & $11.53\pm0.14$ & \phs$0.81\pm0.26$\phm{$^a$} & -3.32 & \nodata & \nodata \\
275 & 6835-00321-1 & 160627\phm{A} & 1.96 & \phn$ 8.11\pm0.01$ & \phs$0.36\pm0.02$\phm{$^a$} & -4.69 & F0\,{\sc v} & 6\phm{, 9, 10} \\
296 & 6840-01244-1 & 316432\phm{A} & 0.38 & $11.99\pm0.26$ & \phs$0.11\pm0.30$\phm{$^a$} & -3.18 & F0 & 4\phm{, 9, 10} \\
333 & 6839-00340-1 & 159509\phm{A} & 1.55 & \phn$ 9.69\pm0.03$ & \phs$0.42\pm0.04$\phm{$^a$} & -4.10 & A4\,{\sc iii} & 6\phm{, 9, 10} \\
337 & 6839-00205-1 & 315995\phm{A} & 1.08 & $10.88\pm0.08$ & \phs$0.24\pm0.09$\phm{$^a$} & -3.63 & A0 & 4\phm{, 9, 10} \\
352 & 6853-00288-1 & 316675\phm{A} & 0.34 & $12.03\pm0.33$ & $-0.22\pm0.33$\tablenotemark{a} & -3.23 & F8 & 4\phm{, 9, 10} \\
360 & 6840-00525-1 & \nodata & 4.07 & $11.11\pm0.12$ & \phs$0.65\pm0.18$\phm{$^a$} & -3.60 & \nodata & \nodata \\
388 & 6839-00615-1 & \nodata & 4.79 & $11.51\pm0.15$ & \phs$0.39\pm0.19$\phm{$^a$} & -3.44 & \nodata & \nodata \\
402 & 7376-00402-1 & 316033\phm{A} & 0.68 & \phn$ 9.99\pm0.04$ & \phs$0.91\pm0.06$\phm{$^a$} & -4.04 & G5 & 4\phm{, 9, 10} \\
452 & 7377-00626-1 & \nodata & 0.24 & $11.04\pm0.12$ & \phs$1.63\pm0.40$\phm{$^a$} & -3.69 & \nodata & \nodata \\
467 & 6835-00391-1 & \nodata & 0.84 & $10.07\pm0.04$ & \phs$0.58\pm0.07$\phm{$^a$} & -4.08 & \nodata & \nodata \\
485 & 6839-00212-1 & 316059\phm{A} & 0.03 & $10.48\pm0.05$ & \phs$0.60\pm0.07$\phm{$^a$} & -3.92 & F8 & 4\phm{, 9, 10} \\
506 & 6835-00186-1 & 160390\phm{A} & 0.33 & \phn$ 9.32\pm0.02$ & \phs$0.41\pm0.03$\phm{$^a$} & -4.34 & A2/A3\,{\sc iii} & 6\phm{, 9, 10} \\
514 & 6849-01034-1 & 314884\phm{A} & 1.47 & $10.04\pm0.03$ & \phs$0.01\pm0.03$\phm{$^a$} & -4.17 & B9 & 4\phm{, 9, 10} \\
524 & 6853-00080-1 & \nodata & 0.78 & $10.46\pm0.08$ & \phs$1.30\pm0.21$\phm{$^a$} & -4.00 & \nodata & \nodata \\
622 & 6839-00019-1 & \nodata & 2.37 & $11.66\pm0.18$ & \phs$0.73\pm0.30$\phm{$^a$} & -3.52 & \nodata & \nodata \\
632 & 6838-01055-1 & 315998\phm{A} & 1.28 & \phn$ 9.81\pm0.03$ & \phs$1.43\pm0.08$\phm{$^a$} & -4.26 & K1\,{\sc iii} & 5\phm{, 9, 10} \\
680 & 6853-01070-1 & \nodata & 0.44 & $11.35\pm0.18$ & \phs$0.54\pm0.26$\phm{$^a$} & -3.74 & \nodata & \nodata \\
698 & 6853-01354-1 & \nodata & 0.49 & $11.52\pm0.18$ & \phs$0.25\pm0.21$\phm{$^a$} & -3.67 & \nodata & \nodata \\
728 & 7377-00394-1 & \nodata & 2.68 & $11.29\pm0.14$ & \phs$1.34\pm0.39$\phm{$^a$} & -3.76 & \nodata & \nodata \\
768 & 7377-00310-1 & \nodata & 0.47 & $11.41\pm0.15$ & \phs$1.91\pm0.50$\phm{$^a$} & -3.72 & \nodata & \nodata \\
785 & 6835-00311-1 & 160769\phm{A} & 0.54 & \phn$ 8.86\pm0.02$ & \phs$0.42\pm0.03$\phm{$^a$} & -4.74 & F3\,{\sc v} & 6\phm{, 9, 10} \\
863 & 6838-00781-1 & 158902\phm{A} & 0.39 & \phn$ 7.23\pm0.01$ & \phs$0.29\pm0.02$\phm{$^a$} & -5.39 & B3--5\,{\sc i}a/b{\sc--ii} & 6, 9, 10 \\
904 & 6849-01144-1 & \nodata & 0.67 & $11.25\pm0.09$ & \phs$0.73\pm0.14$\phm{$^a$} & -3.91 & \nodata & \nodata \\
916 & 6849-00227-1 & 316666A & 0.73 & $10.06\pm0.05$ & \phs$0.34\pm0.06$\phm{$^a$} & -4.38 & F0 & 4\phm{, 9, 10} \\
925 & 6849-00157-1 & \nodata & 2.84 & $10.68\pm0.09$ & \phs$1.04\pm0.19$\phm{$^a$} & -4.13 & \nodata & \nodata \\
1087 & 6835-00082-1 & \nodata & 0.80 & $10.89\pm0.09$ & \phs$0.71\pm0.13$\phm{$^a$} & -4.05 & \nodata & \nodata \\
1092 & 6835-00596-1 & \nodata & 0.20 & $10.88\pm0.09$ & \phs$0.79\pm0.14$\phm{$^a$} & -4.06 & \nodata & \nodata \\
1113 & 6835-00312-1 & \nodata & 0.53 & $11.73\pm0.16$ & \phs$0.53\pm0.22$\phm{$^a$} & -3.68 & \nodata & \nodata \\
1219 & 7376-00879-1 & \nodata & 0.14 & $11.18\pm0.13$ & \phs$0.92\pm0.24$\phm{$^a$} & -3.94 & \nodata & \nodata \\
\enddata
\tablenotetext{a}{This color is almost certainly spurious due to strong variability in this object.}
\tablerefs{
(1)~\citet{Torres:2006a}
(2)~\citet{Steele:1999a},
(3)~\citet{LopesDeOliveira:2006a},
(4)~\citet{Nesterov:1995a},
(5)~\citet{Siebert:2011a},
(6)~\citet{Houk:1982a},
(7)~\citet{Vijapurkar:1993a},
(8)~\citet{Levenhagen:2006a},
(9)~\citet{MacConnell:1976a},
(10)~\citet{Garrison:1977a},
}
\end{deluxetable*}

\begin{deluxetable*}{lllcllcll}
\tabletypesize{\scriptsize}
\tablecaption{\bf Bright Optical Counterparts to GBS Southern Strip Sources
\label{MatchTableSouth}}
\tablewidth{0pt}
\tablehead{\colhead{}  & \colhead{}  & \colhead{} & \colhead{Offset} & \colhead{} & \colhead{} & \colhead{} & \colhead{} &\colhead{}\\
\colhead{GBS}  & \colhead{Tycho-2}  & \colhead{HD} & \colhead{(arcsec)} & \colhead{$V$} & \colhead{$(B-V)$} & \colhead{$\log (F_{\rm x} / F_{\rm opt})$} &
\colhead{Spectral Type} & \colhead{Refs.}}
\startdata
B5 & 7375-00399-1 & 315961 & 0.14 & $10.17\pm0.05$ & \phs$1.12\pm0.10$ & -2.95 & K5 & 1 \\
B9 & 7377-01107-1 & 161853 & 0.67 & \phn$ 7.95\pm0.01$ & \phs$0.13\pm0.02$ & -3.85 & O8\,{\sc iii} & 2 \\
B17 & 6853-01752-1 & 316565 & 4.98 & $10.62\pm0.09$ & \phs$0.53\pm0.12$ & -3.10 & F8 & 1 \\
B93 & 7381-00585-1 & 318327A & 0.49 & $10.66\pm0.07$ & \phs$1.04\pm0.13$ & -3.62 & M3\,{\sc v} & 3 \\
B116 & 6849-01175-1 & 314886 & 0.27 & $10.19\pm0.03$ & \phs$0.36\pm0.04$ & -3.92 & A5 & 1 \\
B128 & 6832-00621-1 & \nodata & 1.42 & \phn$ 9.89\pm0.04$ & \phs$0.51\pm0.05$ & -4.06 & \nodata & \nodata \\
B181 & 6853-00523-1 & 162962 & 0.35 & \phn$ 9.95\pm0.05$ & \phs$0.84\pm0.08$ & -4.16 & A0 & 4 \\
B200 & 7376-00433-1 & \nodata & 0.42 & $12.07\pm0.22$ & \phs$0.24\pm0.29$ & -3.31 & \nodata & \nodata \\
B211 & 7381-00288-1 & 160826 & 2.61 & \phn$ 8.51\pm0.01$ & \phs$0.07\pm0.02$ & -4.73 & B9\,{\sc v} & 5 \\
B225 & 6853-03032-1 & \nodata & 2.51 & $11.01\pm0.09$ & \phs$0.44\pm0.12$ & -3.83 & \nodata & \nodata \\
B233 & 6853-00479-1 & 316692 & 1.13 & $10.21\pm0.06$ & $-0.01\pm0.06$ & -4.15 & A0 & 1 \\
B287 & 7376-00194-1 & 158982 & 1.39 & \phn$ 9.40\pm0.02$ & \phs$0.27\pm0.03$ & -4.48 & A2\,{\sc iv/v} & 5 \\
B296 & 7377-00827-1 & 161839 & 5.95 & \phn$ 9.63\pm0.03$ & \phs$0.11\pm0.04$ & -4.51 & B5/7\,{\sc ii/iii} & 5 \\
B302 & 6849-01627-1 & \nodata & 0.42 & $11.87\pm0.17$ & \phs$0.39\pm0.22$ & -3.61 & \nodata & \nodata \\
B306 & 6853-00059-1 & 163613 & 0.39 & \phn$ 8.56\pm0.01$ & \phs$0.25\pm0.02$ & -4.94 & B1\,{\sc i--ii} & 6 \\
B422 & 7375-00782-1 & 315956 & 0.60 & \phn$ 9.66\pm0.03$ & \phs$0.43\pm0.04$ & -4.52 & F2 & 1 \\
\enddata
\tablerefs{
(1)~\citet{Nesterov:1995a},
(2)~\citet{Parthasarathy:2000a}
(3)~\citet{Raharto:1984a},
(4)~\citet{Cannon:1949a},
(5)~\citet{Houk:1982a},
(6)~\citet{Hoffleit:1956a},
}
\end{deluxetable*}

We identify 70 Chandra sources that lie within 10\,arcsec of 69
Tycho-2 stars (both CX916 and CX917 lie close to the same star; see
below).  Information about these matches is summarized in
Table~\ref{MatchTableNorth} for the objects from \citet{Jonker:2011a}
and Table~\ref{MatchTableSouth} for the southern objects from Jonker
et al.\ (in preparation).  Most of these actually match to within
1\,arcsec, with a few at substantially larger deviations: the median
X-ray/optical offset is 0.60\,arcsec, while the mean offset is
1.19\,arcsec.  After removing the ten objects identified as either
chance coincidences or alignments with resolved companions (see
Section~\ref{CoincidenceSection}), the median X-ray/optical offset
drops to 0.49\,arcsec, and the mean to 0.74\,arcsec.  These numbers
are consistent with the absolute astrometric calibration of {\it
  Chandra}, for which the 90\,\%\ uncertainty is 0.6\,arcsec
\footnote{http://cxc.harvard.edu/cal/ASPECT/celmon/}.

Of these sources, 28 also clearly coincide with a source in one of the
Henry Draper catalogs.  In addition in one case, CX77, the positional
offset between the Tycho-2 position and the nearest HD star, HD159571
is 45\,arcsec, but the photometry of the two stars agree.  The
discrepancy cannot be due to proper motion as the Tycho-2 star only
has a proper motion of 2.9\,mas\,yr$^{-1}$.  Examining Digital Sky
Survey images of the region, there is no bright star at the quoted
position of HD159571, so we believe the coordinates quoted are
inaccurate and count this as a match, in agreement with
\citet{Fabricius:2002b}.

Finally, we identify two additional GBS sources with resolved
companions to X-ray detected Tycho-2 stars.  CX917 appears to coincide
with HD\,316666B, the companion to HD\,316666A (CX916).  CXB36
coincides with HD\.318327B, the companion to HD\,318327A (CXB93).

\subsection{Chance coincidences}
\label{CoincidenceSection}

With the precise localization possible with {\it Chandra}, we expect
few of our matches to be coincidences with bright stars.  We can
assess the probability of chance coincidences by comparing the number
of matches between GBS X-ray sources and Tycho-2 stars within a
specified matching radius to the number of matches found within an
annulus around the X-ray source.  This preserves information about the
spatial distributions of X-ray and optical sources to provide a robust
estimate, for this region of the sky, of the probability of a chance
coincidence.

We plot the number of matches found as a function of offset in
Figure~\ref{OffsetHistFig}. The central peak contains 45 candidate
matches with offsets below 1\,arcsec, 11 with offsets of 1--2\,arcsec,
and 14 with offsets of 2--7\,arcsec. There are no candidate matches
between 7--10\,arcsec, so we have a total of 70 candidate matches with
Tycho-2 stars. Comparing this central peak in the distribution with
the number found at larger radii it is clear that few can be expected
to be coincidences. Statistically we would expect $\sim5$ chance
coincidences within 10\,arcsec, a 20\,\% chance of finding one match
within 2\,arcsec, and a 5\,\%\ chance of finding one within 1\,arcsec.
Given these statistics it seems likely that a few of the matches with
offsets of 2--7\,arcsec are due to chance coincidence, but most are
probably not. In some cases the match may be with a resolved binary
companion rather than to the Tycho-2 star itself.

Those sources observed by Chandra at small off-axis angles with good
detections are most suspect. We can assess the likelihood of a chance
match on a case-by-case basis using the results of \citet{Hong:2005a}
who present 95\,\%\ confidence radii as a function of off-axis angle
and number of counts.  Of the offsets greater than 2\,arcsec, CX917
can immediately be excluded as this corresponds to the resolved
companion HD316666B, and the large offset comes from matching it with
HD316666A. We find that the offsets for CX728, CX925, CXB211, and
CXB225 are not significant at the 95\,\%\ confidence level so these
may be true counterparts to the X-ray source observed at large
off-axis angles.  For the remainder of the objects with offsets
greater than 2\,arcsec, in every case the offset is at least twice the
95\,\%\ confidence ranges so these nine objects (CX53, 59, 77, 156,
360, 388, 622, and CXB17, and 296) are either matches with resolved
binary companions or true coincidences.

We also checked the ten objects with offsets between 1 and 2\,arcsec
by the same criteria.  All of these objects were observed at greater
than 5\,arcmin off-axis angle, and none of the offsets seen are
significant at the 95\,\%\ confidence level, so we believe these are
all true matches, as was expected statistically.

\subsection{Spectral types}

The first step in obtaining a credible classification for the objects
found is a spectral type for the optical counterpart. The Henry Draper
Catalog, Extension, and Charts provide a baseline classification but
this is often crude and lacks luminosity classes. For objects in the
original Henry Draper Catalog, improved classifications with
luminosity classes are available in the Michigan Catalogue of
two-dimensional spectral types for the HD stars \citep{Houk:1982a}.
We also checked each object against the online Catalog of Stellar
Classifications \citep{Skiff:2011a} and for a number of objects,
especially the handful of early-type stars, one or more modern
classifications were available.  Where we list no spectral types in
Table~\ref{MatchTableNorth} or \ref{MatchTableSouth}, we could find no
published spectral classification in the literature.

\subsection{Infrared counterparts from 2MASS}

We also collate infrared photometry of our objects from the Two Micron
All Sky Survey \citep[2MASS]{Skrutskie:2006a}.  While in general the
GBS is working with the superior IR data from the Vista Variables in
the Via Lactea project \citep[VVV]{Saito:2011a}, for the objects
considered here, VVV photometry is expected to be saturated and so
2MASS is generally more reliable.

\subsection{Variability from ASAS}

Finally we checked each matched Tycho-2 star against the All Sky
Automated Survey (ASAS) Catalog of variable stars
\citep{Pojmanski:2002a}.  Our findings are summarized in
Table~\ref{VarTable}.  We found seven matches with identified ASAS
variables: CX4, CX6, CX9, CX33, CX352, CXB287, and CXB422.  Of these,
CX6 and CX352 are known variables (V3892~Sgr and V779~Sgr
respectively), and CX33 was a previously suspected variable
(NSV~23882).  CX9 is attributed a 5.72\,day period in the ASAS
catalog, but we could not reproduce this.  \citet{Otero:2006a} instead
find a period of 2.87\,days (almost, but not quite, half the ASAS
period) which we can reproduce.  We were not able to confirm the
putative variability in either CXB287, or CXB422 using the ASAS data.

\begin{figure}
\includegraphics[width=3.6in]{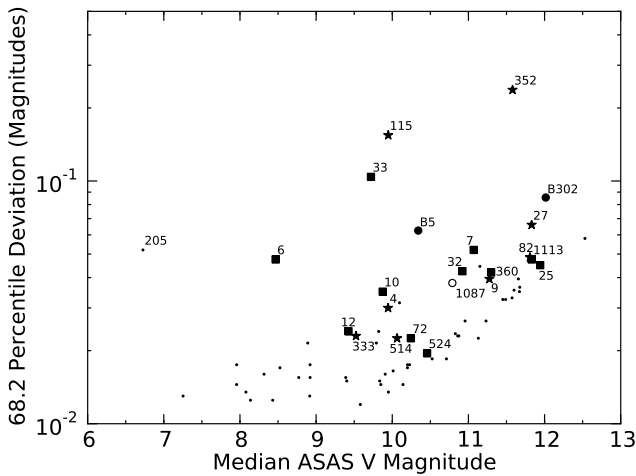}
\caption{Strength of variability in ASAS lightcurves as a function
  of magnitude.  68.2\,\%\ of points lie within 1 deviation of the
  median.  Solid stars are objects showing periodic variability, solid
  squares show apparently real aperiodic variability, and open circles
  show excess variance for their magnitude, but not compelling pattern
  of variability in their lightcurve.  Note that CX205 is not variable
  but is beyond the saturation limit of the ASAS survey.}
\label{VarianceFig}
\end{figure}

To place limits on the variability present in each of the other
objects, we also examined the ASAS photometry of each one.  We show in
Fig.~\ref{VarianceFig} a measure of the scatter of the lightcurves as
a function of magnitude.  All of our Tycho-2 stars were included in
the ASAS photometry database, although the brightest are saturated and
unusable.  CX205 is by far the worst case at magnitude 6.7.  CX156,
CX183, CX256, CX275, and CX863 all suffer from mild saturation but we
can still exclude large amplitude variability in these objects.  We
initially used the standard deviation of the lightcurves to quantify
the variability (considering only A grade photometry), but found that
there were several objects showing large standard deviations caused by
a small number of points with large excursions several magnitudes
below the typical values.  Many stars show these excursions, even well
understood objects such as a the W~UMa system V3892~Sgr (CX352), so we
believe they are spurious.  We opt to instead quantify the variance by
measuring the deviation from the median enclosing 68.2\,\%\ of the
points.  For a Gaussian distribution this is equal to the standard
deviation, but it is more robust to the presence of small numbers of
outliers with extremely large deviations.  Using this statistic, the
plot became much cleaner than using the standard deviation.

\begin{deluxetable*}{llcll}
\tabletypesize{\scriptsize}
\tablecaption{\bf ASAS Counterparts to Optically Bright GBS Sources
\label{VarTable}}
\tablewidth{0pt}
\tablehead{\colhead{}  & \colhead{}  & \colhead{68.2\,\%} & \colhead{Period} & \colhead{}\\
\colhead{GBS}  & \colhead{ASAS}  & \colhead{Variance} & \colhead{(Days)} & \colhead{Comments}}
\startdata
4 & 173931-2909.9 & 0.030 & $16.0834(55)$ & Previously identified ASAS variable\\
6 & 174446-2713.7 & 0.047 & \nodata & Known Be variable, V3892~Sgr\\
7 & 173826-2901.8 & 0.052 & \nodata & New ASAS variable\\
9 & 173508-2923.5 & 0.039 & $2.87233(14)$ & Previously identified ASAS variable\\
10 & 173629-2910.5 & 0.035 & \nodata & New ASAS variable\\
12 & 174347-3140.4 & 0.024 & \nodata & New ASAS variable\\
25 & 174503-3159.6 & 0.045 & \nodata & New ASAS variable\\
26 & 174533-3058.9 & 0.016 & \nodata & \\
27 & 173653-2848.7 & 0.066 & $31.132(30)$ & New ASAS variable \\
31 & 173804-2907.1 & 0.013 & \nodata & \\
32 & 174105-2815.1 & 0.043 & \nodata & New ASAS variable\\
33 & 174836-2957.5 & 0.104 & \nodata & Previously identified ASAS variable\\
53 & 174449-2635.3 & 0.015 & \nodata & \\
59 & 174500-2612.5 & 0.017 & \nodata & \\
72 & 174820-3028.6 & 0.022 & \nodata & New ASAS variable\\
77 & 173638-2859.9 & 0.018 & \nodata & \\
82 & 175710-2725.4 & 0.048 & $6.299(33)$ & New ASAS variable \\
91 & 175611-2714.5 & 0.016 & \nodata & \\
115 & 173941-2851.2 & 0.154 & $3.12678(14)$ & Known eclipsing Algol, V846~Oph\\
156 & 174928-2918.9 & 0.013 & \nodata & \\
183 & 175041-2916.7 & 0.016 & \nodata & \\
205 & 174917-3035.8 & 0.052 & \nodata & \\
256 & 175348-2841.3 & 0.018 & \nodata & \\
272 & 174354-2701.7 & 0.035 & \nodata & \\
275 & 174205-2650.7 & 0.013 & \nodata & \\
296 & 174951-2956.2 & 0.035 & \nodata & \\
333 & 173618-2834.3 & 0.023 & $14.6565(48)$ & New ASAS variable \\
337 & 173527-2930.8 & 0.023 & \nodata & \\
352 & 175537-2818.1 & 0.237 & $0.44503030(36)$ & Known W~UMa variable, V779~Sgr\\
360 & 175114-2919.2 & 0.042 & \nodata & New ASAS variable\\
388 & 173700-2906.1 & 0.023 & \nodata & \\
402 & 173533-3023.6 & 0.013 & \nodata & \\
452 & 174546-3108.5 & 0.032 & \nodata & \\
467 & 174124-2657.9 & 0.014 & \nodata & \\
485 & 173719-2829.4 & 0.018 & \nodata & \\
506 & 174046-2801.8 & 0.016 & \nodata & \\
514 & 175637-2711.8 & 0.022 & $0.889524(27)$ & New ASAS variable \\
524 & 175419-2836.9 & 0.019 & \nodata & New ASAS variable\\
622 & 173654-2952.7 & 0.037 & \nodata & \\
632 & 173354-2923.9 & 0.015 & \nodata & \\
680 & 175412-2839.3 & 0.032 & \nodata & \\
698 & 175221-2904.5 & 0.039 & \nodata & \\
728 & 174812-3024.5 & 0.058 & \nodata & \\
768 & 174438-3114.2 & 0.033 & \nodata & \\
785 & 174250-2750.6 & 0.022 & \nodata & \\
863 & 173303-2939.1 & 0.013 & \nodata & \\
904 & 175624-2710.4 & 0.026 & \nodata & \\
916 & 175542-2804.4 & 0.021 & \nodata & \\
925 & 175450-2753.6 & 0.019 & \nodata & \\
1087 & 174141-2736.8 & 0.038 & \nodata & \\
1092 & 174110-2647.1 & 0.023 & \nodata & \\
1113 & 174012-2802.2 & 0.048 & \nodata & New ASAS variable\\
1219 & 173317-3020.7 & 0.022 & \nodata & \\
\noalign{\smallskip}
B5 & 173209-3028.5 & 0.062 & \nodata & New ASAS variable\\
B9 & 174917-3115.3 & 0.015 & \nodata & \\
B17 & 175430-2923.9 & 0.032 & \nodata & \\
B93 & 174613-3206.1 & 0.018 & \nodata & \\
B116 & 175708-2708.9 & 0.024 & \nodata & \\
B128 & 174651-2546.8 & 0.014 & \nodata & \\
B181 & 175455-2912.2 & 0.016 & \nodata & \\
B200 & 173351-3050.5 & 0.045 & \nodata & \\
B211 & 174329-3213.9 & 0.013 & \nodata & \\
B225 & 175619-2828.2 & 0.026 & \nodata & \\
B233 & 175521-2834.4 & 0.018 & \nodata & \\
B287 & 173334-3032.0 & 0.015 & \nodata & \\
B296 & 174909-3117.3 & 0.024 & \nodata & \\
B302 & 175841-2754.1 & 0.086 & \nodata & New ASAS variable\\
B306 & 175810-2808.5 & 0.017 & \nodata & \\
B422 & 173117-3019.3 & 0.012 & \nodata & \\
\enddata
\end{deluxetable*}

Several variables showed up in this way that were not included in the
ASAS variability catalog, most notably V846~Oph (CX115), a known
eclipsing Algol.  We also performed a visual examination of all
lightcurves as a function of time, and a period search.  For each
object, we filter the data by the grade of photometry, typically
keeping just grades A and B.  Period searches were carried out
separately for 1--100\,days, and 2--24\,hours using Lomb-Scargle
periodograms, and verified by inspection of both unbinned and binned
folded lightcurves.  In this process, we picked up another four
periodic variables that had not passed the ASAS acceptance criteria:
CX27, CX82, CX333, and CX514, are all low-amplitude, periodic systems.
CX7, CX10, CX12, CX25, CX32, CX72, CX360, CX524, CX1113, and CXB5 all
show irregular variability of some form, either slow trends, irregular
flickering-like behavior, or in the case of CX72 a single discrete
outburst.  Our final test was to examine Lomb-Scargle periodograms of
each year of data separately.  This is useful where a period may not
be coherent on long timescales, for example if the starspot geometry
on a star is changing.  Only one periodic variable was added to the
list, CX7, which showed quite complex behavior.  We will discuss this
in Section~\ref{CXSevenSection}.

There remain several objects for which the variance is larger than
typical for stars of their magnitude, but for which we could not
confirm real variability in the lightcurve.  The most pronounced cases
are CX1087, CXB17, CXB200, and CXB296.  These objects may show
unresolved short timescale variability.  We note that both CXB17 and
CXB296 appear to be chance coincidences with Tycho-2 stars.

In all, we identify 9 periodic variables, 5 definite irregular ones,
and a further 7 suspected irregular variables.  Two X-ray bright
objects, CX4 and CX7, show both periodic and aperiodic behavior.  In
total, a third ($20/60$) of the non-coincidental matches appear to be
variable in ASAS data.  This is probably an underestimate given the
presence of other objects which appear to show an excess variance
compared to other stars at a similar magnitude, but neither a
detectable period, nor clear, resolved aperiodic variability.

\section{Flux ratios}
\label{RatioSection}

\begin{figure}
\includegraphics[width=3.6in]{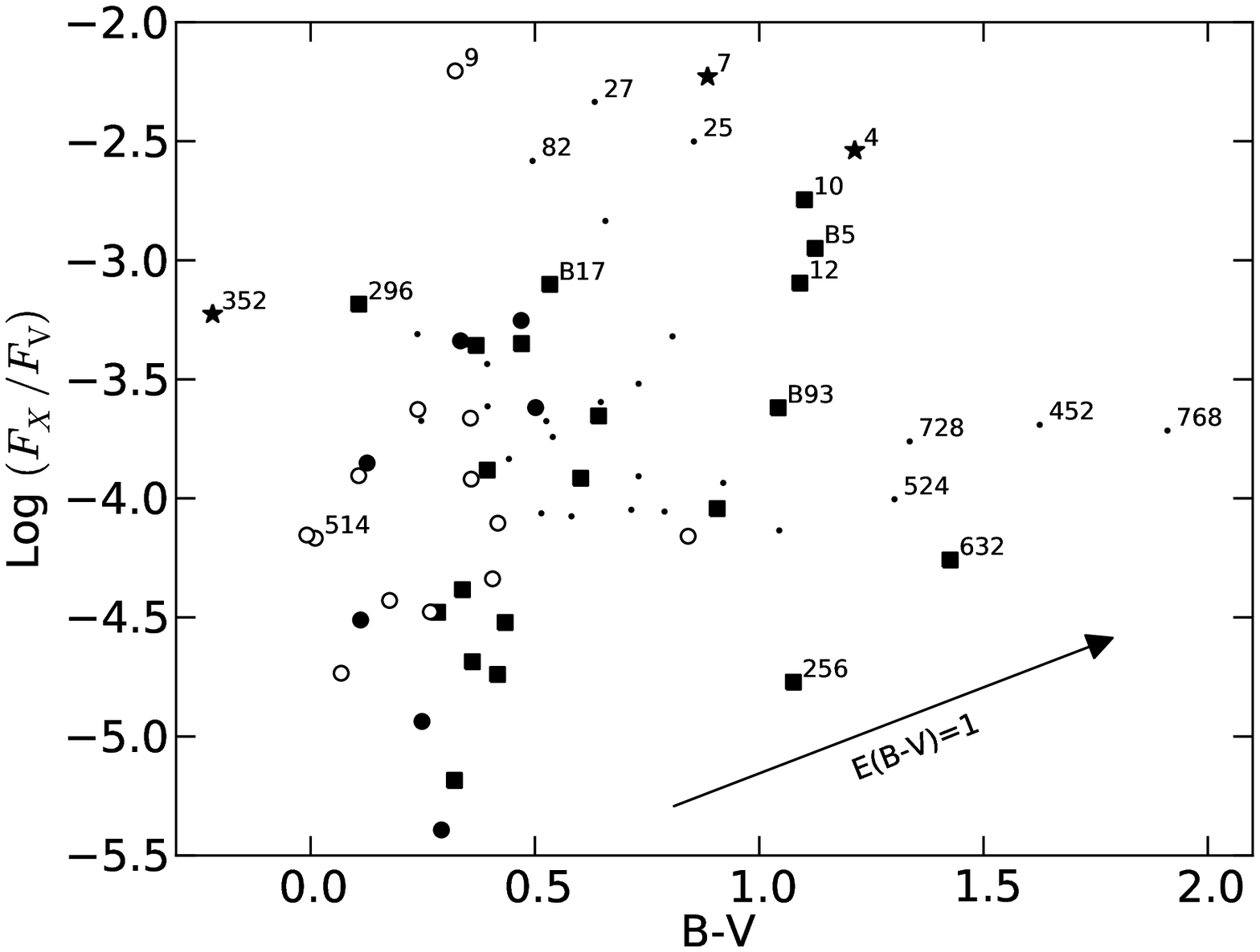}
\includegraphics[width=3.6in]{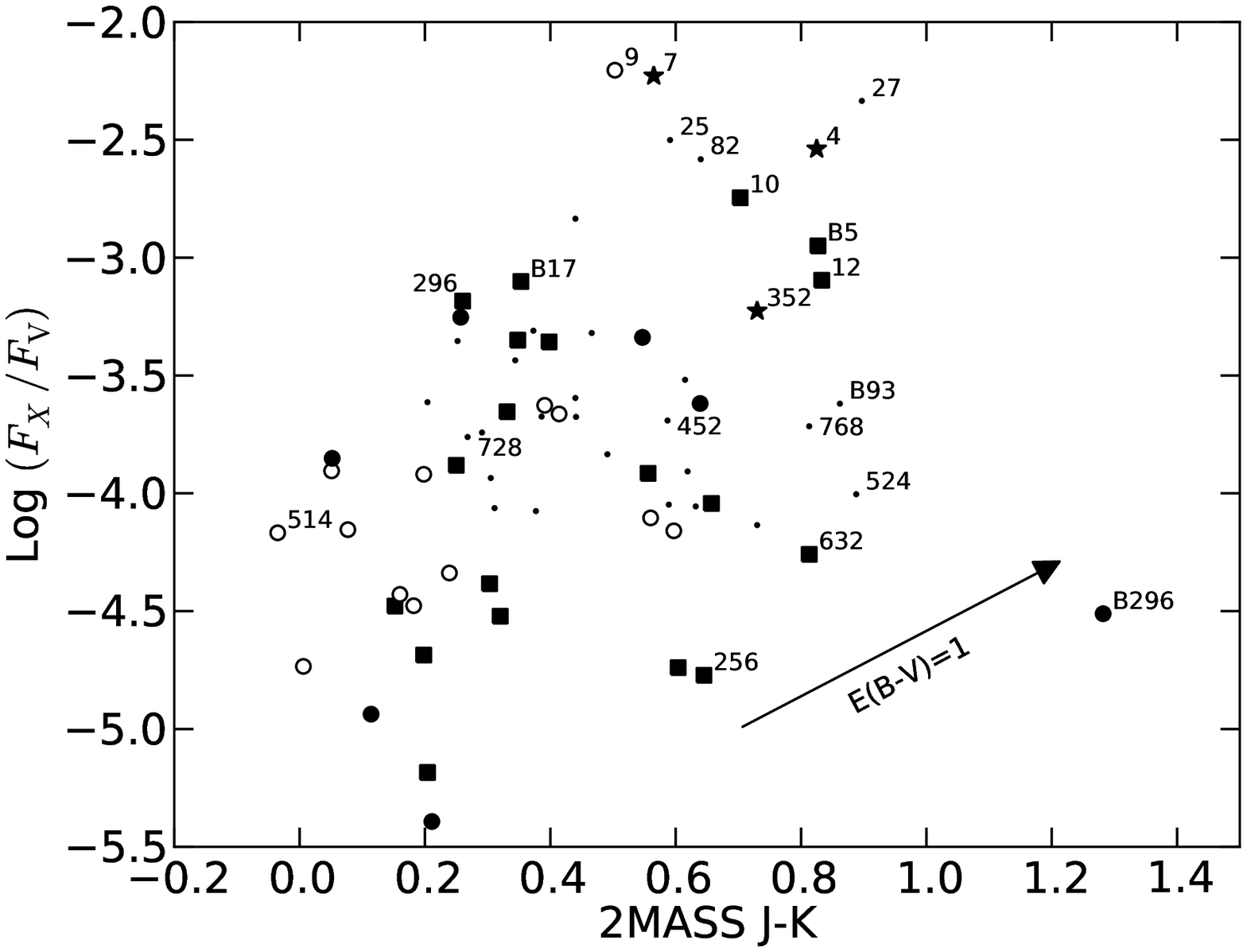}
\caption{Relation between X-ray to optical flux ratio and optical
  color (upper) and IR color (lower).  Stars with no spectral
  classification are shown with points.  Of the remainder, filled
  circles indicate O and early B stars, open circles late B and A
  stars, stars are unusual late type stars, and squares are late type
  stars with no variability or other noted peculiarities.  We label
  selected objects only for clarity.  These are mainly outliers from
  the distribution.  The arrows indicate the effect of a reddening of
  $E(B-V)=1$ on the color and flux ratio for an assumed 10\,MK APEC
  spectrum.}
\label{ColorVsRatioFig}
\end{figure}

Ratios of X-ray to optical flux provide a very useful diagnostic of
whether X-ray emission can be attributed to normal stellar activity
(in winds from early type stars and coronae of late type stars), or
require a more exotic explanation.  Such diagnostics are very crude in
the absence of spectral information as there are then uncertainties in
the nature of the optical star, its bolometric flux, and the amount of
optical extinction and X-ray absorption.  Nonetheless, as a crude cut
we show the optical to X-ray flux ratio as a function of $B-V$ and
$J-K$ colors in Fig.~\ref{ColorVsRatioFig}.

A group of stars stands out in both diagrams as having high $F_{\rm X}
/ F_{\rm opt}$ relative to the remainder. These are CX4, 7, 9, 10, 12,
25, 27, 82, and CXB5.  In defining these, we have simply picked the
outliers in Figure~\ref{ColorVsRatioFig}, and have not tried to set an
arbitrary cut-off in $\log(F_{\rm opt} / F_{V})$.

CX9 is an A star in an eclipsing binary. As we will discuss, we
suspect this is an Algol system. CX7 is a K0 star that was identified
as a pre-main-sequence star by \citet{Torres:2006a}. CX4 appears to be
a rapidly rotating G9 giant. The X-rays are likely coronal with strong
activity driven by rapid rotation.  All three of these objects for
which we have classifications appear to be unusual objects in some
respect. CX25, CX27, and CX82 have similar $F_{\rm X} / F_{\rm opt}$
ratios, so these too are probably unusual objects warranting further
attention.  CX10, CX12, and CXB5 are also outliers.  As will discuss
in Section~\ref{HardnessSection}, this group of sources besides being
among the brighter objects in our sample, also have the hardest
spectra, excluding the massive stars, and as we have seen several of
these are also variable.

A few stars appear to have anomalous colors.  CX352 is continuously
variable, so all of its colors are suspect unless taken at the same
orbital phase, or phase-averaged.  CX452, 728, 768, and 1219 all have
redder optical colors than expected based on their IR color (see
Section~\ref{ColorColorSection}) and so their Tycho-2 $B-V$ colors are
suspect.  They are not strong outliers when plotted in $J-K$.  CXB296
is a strong outlier when plotted against $J-K$ only due to its very
strong IR excess.  This is suspected to be a Be star, but in any case
appears to be a chance alignment with CXB296.

The bottom left of this diagram is populated by massive OB stars and A
stars.  Most of the objects falling in this region already have
spectral classifications, and it is likely that we have identified all
of the OB stars in the sample.  This is not surprising, as OB stars
are a rare population for which considerable effort has been expended
in optical observations to identify members.  Some of the unclassified
stars may be A stars, but it is likely that the majority are mid- to
late-type stars, with unremarkable $F_{\rm X} / F_{\rm opt}$ ratios
suggesting normal coronal emission.

For the stars for which we have spectral types, and even better,
luminosity classes, these diagnostics become more powerful.  Knowing
the spectral type of the star, we can deduce its bolometric flux using
standard relations from \citet{Cox:2000a}.  We can also compare
observed and intrinsic colors to deduce the reddening, $E(B-V)$, and
hence estimate the interstellar X-ray absorption using the scaling of
\citet{Bohlin:1978a}: $N_{\rm H} = 5.8\times 10^{21}$\,cm$^{-2}\times
E(B-V)$.  We convert {\it Chandra} count rates to absorption-corrected
fluxes using PIMMS.  We assume a 10\,MK APEC
spectrum\footnote{http://www.atomdb.org} and evaluate the unabsorbed
flux as a function of absorption column.  10\,MK was chosen not only
as representative of stellar coronal temperatures, but as roughly the
temperature at which ACIS-I is most sensitive to coronal spectra.  As
a result, the flux calculated assuming a 10\,MK spectrum should
provide a robust lower limit to the flux (and hence the flux ratio)
expected for hotter or cooler coronae.

\begin{figure}
\includegraphics[width=3.6in]{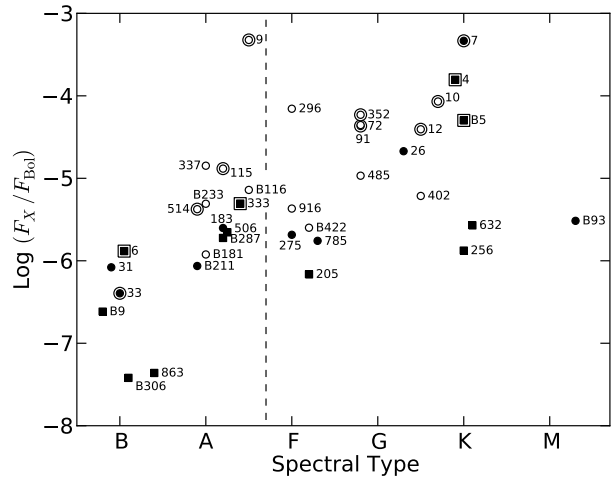}
\caption{Relation between X-ray to bolometric flux ratio as a function
  of spectral type for those stars with full (spectral and luminosity)
  or partial (spectral only) classifications.  Filled squares are
  evolved stars, filled circles are main-sequence stars, and open
  circles are stars with no luminosity classification.  Objects with
  an outer circle or square are identified variables.  Objects with
  high X-ray/optical offsets that are most likely to be chance
  alignments have been omitted.  The vertical dashed line is drawn for
  a spectral type of A7, to indicate the cut-off of expected coronal
  X-ray emission from the visible star. }
\label{SpTRatioFig}
\end{figure}

Using this information we then plot in Fig.~\ref{SpTRatioFig} the
absorption-corrected ratio of X-ray to bolometric luminosity as a
function of spectral-type. We indicate on this figure a vertical line
at spectral type A7, marking the effective cut-off of coronal
activity.  It is notable that there is a cluster of sources with
late-B and A spectral types in the gap that should exist between the
OB and Be stars to the left, and the coronal stars to the right.  Of
these, CX9 stands out by more than an order of magnitude compared to
the others.  We also note the very pronounced tendency for the stars
with high $F_{\rm X} / F_{\rm Bol}$ ratios to be variable, with CX296
as a notable exception.  Since the optical counterpart to CX296 is
among the faintest in our sample, it is quite possible its variability
is undetectable in the ASAS data.

\section{Color-color diagrams}
\label{ColorColorSection}

\begin{figure}
\includegraphics[width=3.6in]{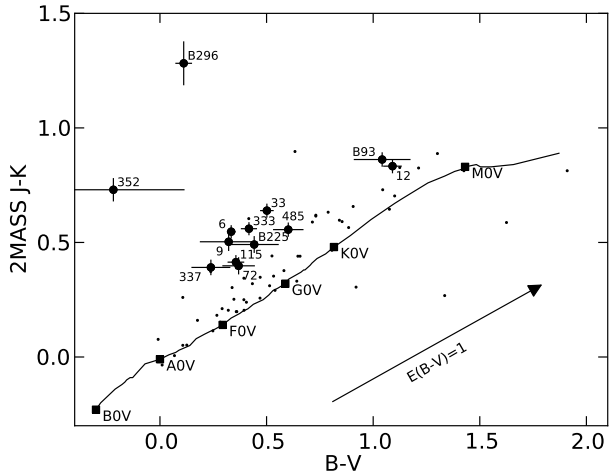}
\caption{Optical/IR color-color diagram.  Tycho-2 photometry has been
  converted to the standard UBV system, 2MASS photometry is left in
  its own system.  We overlay main-sequence colors for comparison.
  All points are shown, but for clarity we only show error bars and
  labels on those points lying more than 3\,$\sigma$ from the line.
  The arrow shows the effect of reddening by $E(B-V)=1$ for the
  extinction curve of \citet{Cardelli:1989a}.}
\label{ColorColorFig}
\end{figure}

We combine Tycho-2 and 2MASS colors in Fig.~\ref{ColorColorFig} to
examine possible deviations from the colors of a single star.  Tycho-2
photometry has been converted into the $UBV$ system using the
transformations of \citet{Hog:2000b}, while 2MASS photometry is left
in its own $JHK_{\rm S}$ system.  We also overlay main-sequence colors
taken from the online compilation of Eric
Mamajek\footnote{http://www.pas.rochester.edu/\textasciitilde
  emamajek/\\EEM\_dwarf\_UBVIJHK\_colors\_Teff.dat}.

The sources generally scatter around the stellar lines, although with
many more objects above the lines than below them, corresponding to IR
colors that are redder than expected based on optical colors.  This
cannot be explained by interstellar reddening as dereddening the
colors will take them further from the stellar lines.  It most
probably indicates a cooler IR excess, either from a cooler companion
star (in most cases) or from a circumstellar disk in the case of Be
stars.  In a few cases the colors may be compromised by variability.
CX352 is the worst case as it has a continuously variable
large-amplitude lightcurve, and its Tycho-2 colors are bluer than
expected from its spectral type.  The other large amplitude variable
is the eclipsing Algol CX115.  In this case, it is possible colors are
valid if they were obtained out of eclipse, and we certainly expect a
cool companion to be present since it is an Algol system.

Unfortunately for many of our objects the Tycho-2 colors are too
uncertain to draw strong conclusions about individual objects
(although the statement that more sources lie above the line than
below it stands in a statistical sense).  We select those individual
objects which lie more than 3\,$\sigma$ from either line for further
consideration, and highlight them in Fig.~\ref{ColorColorFig}.  All of
the objects below the line fail this test, since they all have large
color uncertainties.  The remaining outliers are all above the line,
corresponding to unusually red infrared colors.  The outliers are CX6,
CX9, CX12, CX33, CX72, CX115, CX333, CX337, CX352, CX485, CXB93,
CXB225, and CXB296.

We immediately discount CX352 as compromised by variability.  CX6 and
CX33 are both known Be stars so are expected to have an infrared
excess from a circumstellar disk.  HD\,161839 appears to only be a
coincidental match with CXB296, but also has a very large infrared
excess (seen beyond the 2MASS bands).  CX115 is an Algol system, so
should have a cooler companion star.  Of the others, CX9, CX333, and
CX337 are in the A star clump in Fig.~\ref{SpTRatioFig}.  All are
objects for which negligible intrinsic X-ray emission is expected, and
so the presence of a cooler companion is naturally to be expected.
CX12, CX72, and CX485 are late F to G stars with rather high $F_{\rm
  X} / F_{\rm Bol}$ ratios and so it is also quite credible that these
are active binaries with cooler companions.

\begin{figure}
\includegraphics[width=3.6in]{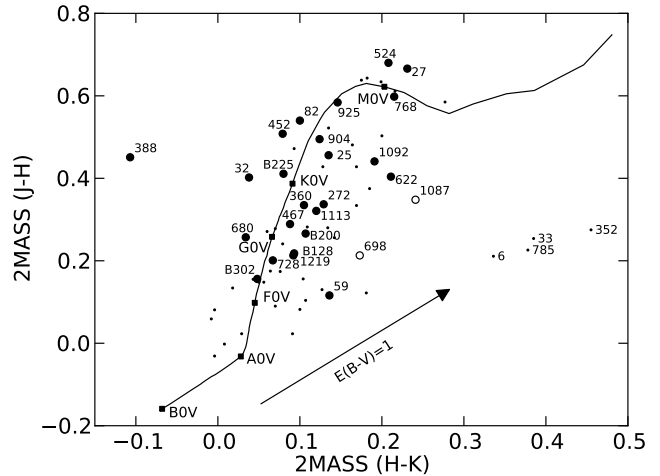}
\caption{Infrared color-color diagram.  We overlay main-sequence
  colors for comparison.  All points are shown, but for clarity we
  only show labels on objects that do not have a
  spectral classification, or that are strong outliers.  Open circles
  are objects for which the 2MASS photometry is flagged as suspect.
  The arrow shows the effect of reddening by $E(B-V)=1$ for the
  extinction curve of \citet{Cardelli:1989a}.  Main sequence colors
  are indicated following the compilation of Eric Mamajek (see text).}
\label{IRColorColorFig}
\end{figure}

An additional diagnostic that is independent of uncertainties in the
Tycho-2 photometry is a $J-H$ versus $H-K$ IR color-color diagram
(Figure~\ref{IRColorColorFig}).  This exploits changes in the shape of
the $JHK$ portion of the spectrum as a function of spectral class
\citep{Straizys:2009a}.  This also is more sensitive to the nature of
a cool component.  We show the main-sequence line of Eric Mamajek (see
above), and note that giants tend to be redder in $J-H$ than
main-sequence stars, so will fall higher in the diagram
\citep{Straizys:2009a}.  Most objects lie on the stellar line or to
the right of it (due to reddening) as expected.  The group of outliers
to the far right is a mixed group.  CX6 and CX33 are both Be stars
with dust emission.  CX352 is a large amplitude variable.  The other
outliers to the right have suspect 2MASS photometry.  There is no
obvious explanation for the location of CX388 to the left of the
diagram, although we note this star is likely a chance coincidence in
any case.  Among the objects with no spectral classification and good
2MASS photometry, all plausible matches with GBS objects show IR
colors consistent with lightly reddened FGKM stars, suggesting that in
all or most of these cases the X-rays originate from a coronally
active late type star.  This supports our earlier suggestion that we
have likely identified all of the OB stars in this sample.

\section{Hardness Ratios}
\label{HardnessSection}

\begin{figure}
\includegraphics[width=3.6in]{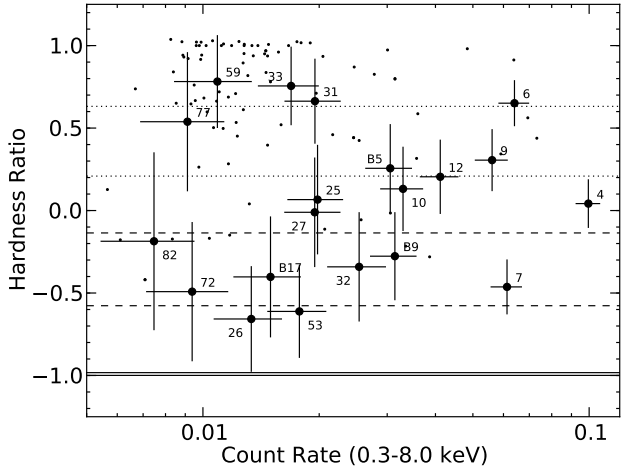}
\caption{Hardness ratios of the brightest sources. Our hardness ratios
  are defined as the 1.25--8.0\,keV rate minus 0.3--1.25\,keV rate
  divided by the 0.3--8.0\,keV rate. Note that this is a softer
  definition than \citet{Jonker:2011a} use. We show only those sources
  for which more than 20 counts were recorded. Sources considered in
  this work have uncertainties plotted and are annotated. Other
  sources are shown by points only to reduce the clutter of the plot.
  The solid lines are 1\,MK APEC spectra, the dashed lines are 10\,MK
  APEC spectra, and the dotted lines are 40\,MK APEC spectra. For each
  pair, the lower line has zero absorption, the upper line has to
  $N_{\rm H}=5.8\times 10^{21}$\,cm$^{-2}$ corresponding to
  $E(B-V)=1$.}
\label{HRFig}
\end{figure}

For the majority of our sources we do not have useful X-ray spectra.
{\it Chandra} is not optimized for spectroscopy with the short
exposures we are using, so those sources with large enough count rates
to yield a useful spectrum in just 2\,ks tend to be quite piled up,
compromising the spectral information.  We can, however, extract
useful hardness ratios for the brightest sources in the GBS, and this
was done by \citet{Jonker:2011a} for the northern sample.  The objects
considered here with Tycho-2 counterparts were among the softest
detected, with all having negative hardness ratios as defined by
\citet{Jonker:2011a}, so for this work we recalculated hardness ratios
for both samples to provide better discrimination amongst soft
sources.  The definition we use is the difference between the
1.25--8.0\,keV count rate and the 0.3--1.25\,keV rate divided by the
whole 0.3--8.0\,keV rate.  We calculate these ratios for all sources
for which more than 20 photons are detected, and show these data in
Fig.~\ref{HRFig} with the Tycho-2 sources highlighted.  We also show
hardness ratios for a number of combinations of coronal temperature
and absorption for comparison.

We can see immediately that, as noted above, the Tycho-2 sources are
amongst the softest in the GBS. This is as expected if they are
dominated by stellar coronal emission.  Conversely, most of the
bright, soft sources detected by the GBS have Tycho-2
counterparts. The hardest sources in our sample include CX6 and CX33,
both of which are Be stars; CX31 which is also a massive star; and
CX59 and CX77 both of which were flagged as possible chance alignments
with a larger than expected difference between X-ray and optical
positions.

The next group in hardness are CX4, CX9, CX10, CX12, CX25, CX27, and
CXB5.  Strikingly all of these objects were highlighted in
Section~\ref{RatioSection} as objects with the highest $F_{\rm X} /
F_{\rm opt}$ ratios.  All have late spectral types, or colors
consistent with late types (except CX9 for which a late-type companion
is indicated by eclipses), and so these appear to be active objects
with luminous, hot coronae.

\section{Individual objects}
\label{SourceSection}

\subsection{Source Distances and Locations}

\begin{deluxetable*}{llccccl}
\tabletypesize{\scriptsize}
\tablecaption{\bf Distances and luminosities based on Hipparcos parallaxes
\label{HipparcosDistanceTable}}
\tablewidth{0pt}
\tablehead{\colhead{}    & \colhead{}              & \colhead{}         & \colhead{Parallax} & \colhead{Distance} & \colhead{$L_{\rm X}$  } & \colhead{} \\
           \colhead{GBS} & \colhead{Spectral Type} & \colhead{$E(B-V)$} & \colhead{(mas)}    & \colhead{(pc)}     & \colhead{(erg\,s$^{-1}$)} & \colhead{Comments}}
\startdata
 26  & \phm{(}G3\,V                   &     $-0.01\pm0.03$ & \phn$9.35\pm1.21$     & $107\pm14$         & $(2.5\pm0.7)\times10^{29}$      & \nodata \\
 53  & (F5\,V)                 &     $-0.06\pm0.04$ & \phn$9.74\pm1.15$     & $103\pm12$         & $(1.4\pm0.3)\times10^{29}$      & Companion \\
156  & (F0\,V)                 &     $-0.04\pm0.03$ & $10.42\pm1.10$        & \phn$96\pm10$          & $(3.8\pm0.8)\times10^{29}$      & Coincidence or companion \\
205  & \phm{(}F2\,V                   &     $-0.03\pm0.04$ & $20.18\pm0.59$        & $49.6\pm1.5$       & $(1.2\pm0.1)\times10^{28}$      & \nodata \\
275  & \phm{(}F0\,V                   & \phs$0.04\pm0.03$  & \phn$6.15\pm0.78$     & $163\pm21$         & $(1.1\pm0.3)\times10^{29}$      & \nodata \\
785  & \phm{(}F3\,V                   & \phs$0.01\pm0.05$  & \phn$7.24\pm1.45$     & $138\pm28$         & $(3.5\pm1.4)\times10^{28}$      & \nodata \\
\noalign{\smallskip}
B93  & \phm{(}M3\,V                   & \phs$0.00\pm0.05$  & \phn\phn$68\pm45$\phn & $15^{+28}_{-6}$     & $(8^{+60}_{-5})\times10^{26}$    & Color from \citet{Koen:2010a} \\
B422 & \phm{(}F2                      & \phs$0.08\pm0.05$  & \phn$3.87\pm1.56$     & $260^{+170}_{-70}$  & $(1.0^{1.6}_{-0.5})\times10^{29}$      & \nodata \\
\end{deluxetable*}

In a few cases {\it Hipparcos} parallaxes exist for our optical
counterparts.  These provide good estimates of distance, and hence
X-ray luminosity (Table~\ref{HipparcosDistanceTable}).  In the absence
of a parallax distance, if we have a spectral and luminosity
classification then we can estimate a spectroscopic parallax distance
(Table~\ref{SpectroscopicDistanceTable}).  To do this we use intrinsic
$(B-V)$ colors from \citet{Fitzgerald:1970a} to estimate $E(B-V)$, and
then absolute magnitudes from \citet{Straizys:1981a} to derive a
distance.  In the absence of a luminosity class, we can obtain a lower
limit on the distance by assuming a main-sequence star.

For the non-Hipparcos objects for which we estimate $E(B-V)$ based on
spectral classifications, we also list the Bulge reddening estimated
by \citet{Gonzalez:2011a} and \citet{Gonzalez:2012a} based on red
clump stars in VVV data.  In every case, our inferred reddening is
much less than the Bulge reddening, consistent with the relatively
short distances deduced for these objects.  Note that this is also
true for the early-type stars at inferred distances of several
kiloparsecs.

Where appropriate, we will discuss the location of the more distant
(non-local) sources.  We assume the distance of the Sun from the
Galactic center is $8.33\pm0.35$\,kpc \citep{Gillessen:2009a}.  We
adopt a distance of the Sun above the plane of $26\pm3$\,pc
\citep{Majaess:2009a}.  These numbers are of some importance as we are
conducting a Bulge survey, not a plane survey, and so our lines of
sight diverge from the plane at larger distances.  Consequently, we
would only expect to see relatively nearby early-type stars in our
sample, at least among the higher latitude sources (e.g.\ CX863, CXB9,
CXB306).  The spiral structure of the Milky Way along our sight-lines
can be seen in the maps of \citet{Churchwell:2009a}.  Beyond the local
stars of the Orion Spur, we encounter the Sagittarius Arm at around
1.2\,kpc, the Scutum-Centaurus Arm around 2.8\,kpc, the Norma Arm
around 4.4\,kpc, and the Near 3\,kpc Arm around 5.8\,kpc.  Among the
luminous stars considered here, the Be stars CX6 and CX33 (and the
coincidental counterpart to CXB296), together with CX863 appear
consident with the Sagittarius Arm, while CX31, CXB9, and CXB306
appear to lie in the Scutum-Centaurus Arm.

\begin{deluxetable*}{llccccl}
\tabletypesize{\scriptsize}
\tablecaption{\bf Distances and luminosities based on spectroscopic parallaxes
\label{SpectroscopicDistanceTable}}
\tablewidth{0pt}
\tablehead{\colhead{}    & \colhead{}              & \colhead{Inferred} & \colhead{Line-of-Sight} & \colhead{Distance\tablenotemark{a}} & \colhead{$L_{\rm X}$\tablenotemark{a}  } & \colhead{} \\
           \colhead{GBS} & \colhead{Spectral Type} & \colhead{$E(B-V)$} & \colhead{$E(B-V)$}      & \colhead{(pc)}     & \colhead{(erg\,s$^{-1}$)} & \colhead{Comments}}
\startdata
  4  & \phm{(}G9\,III          & \phs$0.23\pm0.08$\phm{$^{b}$}      & $2.59\pm0.28$       &    \phn470  & \phn\phd$3\times10^{31}$ & \nodata \\
  6  & \phm{(}B0.5\,III--Ve    & \phs$0.61\pm0.03$\phm{$^{b}$}      & $2.16\pm0.28$       &       1100  & \phn\phd$2\times10^{32}$ & B0.5\,Ve assumed \\
  7  & \phm{(}K0\,V            & \phs$0.07\pm0.20$\phm{$^{b}$}      & $1.95\pm0.28$       &    \phn100  & \phn\phd$7\times10^{29}$ & \nodata \\
  9  & \phm{(}A5               & \phs$0.17\pm0.14$\phm{$^{b}$}      & $1.60\pm0.22$       &    \phn600  & \phn\phd$3\times10^{31}$ & A5\,V assumed\\
 10  & \phm{(}G7\,V            & \phs$0.37\pm0.08$\phm{$^{b}$}      & $1.71\pm0.25$       & \phn\phn51  & \phn\phd$2\times10^{29}$ & RAVE spectral type \\
 12  & \phm{(}G5               & \phs$0.41\pm0.05$\tablenotemark{b} & $3.63\pm0.33$       & \phn\phn41  &       $1.3\times10^{29}$ & G5\,V assumed \\
 31  & \phm{(}O9\,V            & \phs$0.78\pm0.05$\phm{$^{b}$}      & $1.89\pm0.24$       &       2600  & \phn\phd$4\times10^{32}$ & \nodata \\
 33  & \phm{(}B0\,Ve           & \phs$0.80\pm0.04$\phm{$^{b}$}      & $2.65\pm0.28$       &       1300  & \phn\phd$8\times10^{31}$ & \nodata \\
 72  & \phm{(}F8               &    $-0.16\pm0.09$\phm{$^{b}$}      & $2.86\pm0.41$       &    \phn180  & \phn\phd$3\times10^{29}$ & \nodata \\
 77  & (B8\,V)                 & \phs$0.22\pm0.04$\phm{$^{b}$}      & $1.54\pm0.24$       &    \phn450  & \phn\phd$3\times10^{30}$ & Coincidence \\
 91  & \phm{(}F8               &    $-0.06\pm0.07$\phm{$^{b}$}      & $2.78\pm0.42$       &    \phn200  & \phn\phd$3\times10^{29}$ & Resolved binary \\
115  & \phm{(}A2 + G7\,IV      & \phs$0.31\pm0.05$\phm{$^{b}$}      & $2.39\pm0.26$       &    \phn350  &       $1.4\times10^{30}$ & Algol \\
183  & \phm{(}A2\,V            & \phs$0.13\pm0.04$\phm{$^{b}$}      & $2.35\pm0.31$       &    \phn210  &       $1.9\times10^{30}$ & \nodata \\
256  & \phm{(}K0\,III          & \phs$0.07\pm0.06$\phm{$^{b}$}      & $1.19\pm0.22$       &    \phn240  & \phn\phd$2\times10^{29}$ & \nodata \\
296  & \phm{(}F0               &    $-0.21\pm0.30$\phm{$^{b}$}      & $1.86\pm0.27$       &    \phn700  &       $1.7\times10^{30}$ & F0\,V assumed \\
333  & \phm{(}A4\,III          & \phs$0.30\pm0.05$\phm{$^{b}$}      & $1.67\pm0.21$       &    \phn420  & \phn\phd$9\times10^{29}$ & \nodata \\
337  & \phm{(}A0               & \phs$0.25\pm0.09$\phm{$^{b}$}      & $1.57\pm0.24$       &    \phn730  & \phn\phd$3\times10^{30}$ & A0\,V assumed \\
402  & \phm{(}G5               & \phs$0.23\pm0.07$\tablenotemark{b} & $1.79\pm0.29$       & \phn\phn70  & \phn\phd$2\times10^{28}$ & G5\,V assumed \\
485  & \phm{(}F8               & \phs$0.07\pm0.08$\phm{$^{b}$}      & $1.84\pm0.23$       &    \phn170  & \phn\phd$8\times10^{28}$ & F8\,V assumed \\
506  & \phm{(}A2/3\,III        & \phs$0.34\pm0.06$\phm{$^{b}$}      & $1.98\pm0.25$       &    \phn370  & \phn\phd$4\times10^{30}$ & \nodata \\
514  & \phm{(}B9               & \phs$0.08\pm0.05$\phm{$^{b}$}      & $2.27\pm0.31$       &    \phn720  &       $1.3\times10^{30}$ & B9\,V assumed \\
632  & \phm{(}K1\,III          & \phs$0.34\pm0.11$\phm{$^{b}$}      & $1.37\pm0.23$       &    \phn390  & \phn\phd$5\times10^{29}$ & RAVE spectral type \\
863  & \phm{(}B3--5\,Ia/b--II  & \phs$0.36\pm0.04$\phm{$^{b}$}      & $1.39\pm0.23$       &       1300  & \phn\phd$5\times10^{30}$ & B3\,II assumed \\
916  & \phm{(}F0               & \phs$0.02\pm0.06$\phm{$^{b}$}      & $1.78\pm0.29$       &    \phn270  &       $1.0\times10^{29}$ & F0\,V assumed \\
\noalign{\smallskip}
B5   & \phm{(}K5               &    $-0.03\pm0.16$\tablenotemark{b} & $1.70\pm0.36$    &    \phn500  &       $1.0\times10^{31}$ & K5\,III assumed \\
B9   & \phm{(}O8\,III          & \phs$0.44\pm0.02$\phm{$^{b}$}      & $2.43\pm0.29$    &       2700  & \phn\phd$3\times10^{32}$ & \nodata \\
B17  & (F8)                    & \phs$0.00\pm0.12$\phm{$^{b}$}      & $0.90\pm0.21$    &    \phn200  & \phn\phd$5\times10^{29}$ & F8\,V assumed; Coincidence \\
B116 & \phm{(}A5               & \phs$0.21\pm0.05$\phm{$^{b}$}      & $2.06\pm0.34$    &    \phn340  & \phn\phd$4\times10^{29}$ & A5\,V assumed \\
B181 & \phm{(}A0               & \phs$0.85\pm0.09$\phm{$^{b}$}      & $1.18\pm0.23$    &    \phn200  & \phn\phd$2\times10^{29}$ & A0\,V assumed \\
B211 & \phm{(}B9\,III          & \phs$0.15\pm0.04$\phm{$^{b}$}      & $3.13\pm0.37$    &    \phn330  & \phn\phd$3\times10^{29}$ & \nodata \\
B233 & \phm{(}A0               & \phs$0.00\pm0.07$\phm{$^{b}$}      & $1.31\pm0.25$    &    \phn760  & \phn\phd$9\times10^{29}$ & A0\,V assumed\\
B287 & \phm{(}A2\,IV/V         & \phs$0.22\pm0.04$\phm{$^{b}$}      & $1.64\pm0.28$    &    \phn340  & \phn\phd$2\times10^{29}$ & \nodata \\
B296 & (B5/7\,II/III)          & \phs$0.26\pm0.06$\phm{$^{b}$}      & $2.67\pm0.35$    &       1100  & \phn\phd$2\times10^{30}$ & B6\,III assumed; Coincidence \\
B306 & \phm{(}B1\,I--II        & \phs$0.49\pm0.04$\phm{$^{b}$}      & $1.17\pm0.22$    &       2800  & \phn\phd$2\times10^{31}$ & B1\,II assumed \\
\tablenotetext{a}{We have not assessed uncertainties in distance or X-ray luminosity for these sources as we cannot reliably quantify uncertainties in the absolute magnitude.}
\tablenotetext{b}{For G and later spectral types, the intrinsic color is quite sensitive to the assumed luminosity class, so $E(B-V)$ is unreliable where this is not known.}
\end{deluxetable*}

\begin{figure*}
\includegraphics[width=2.4in]{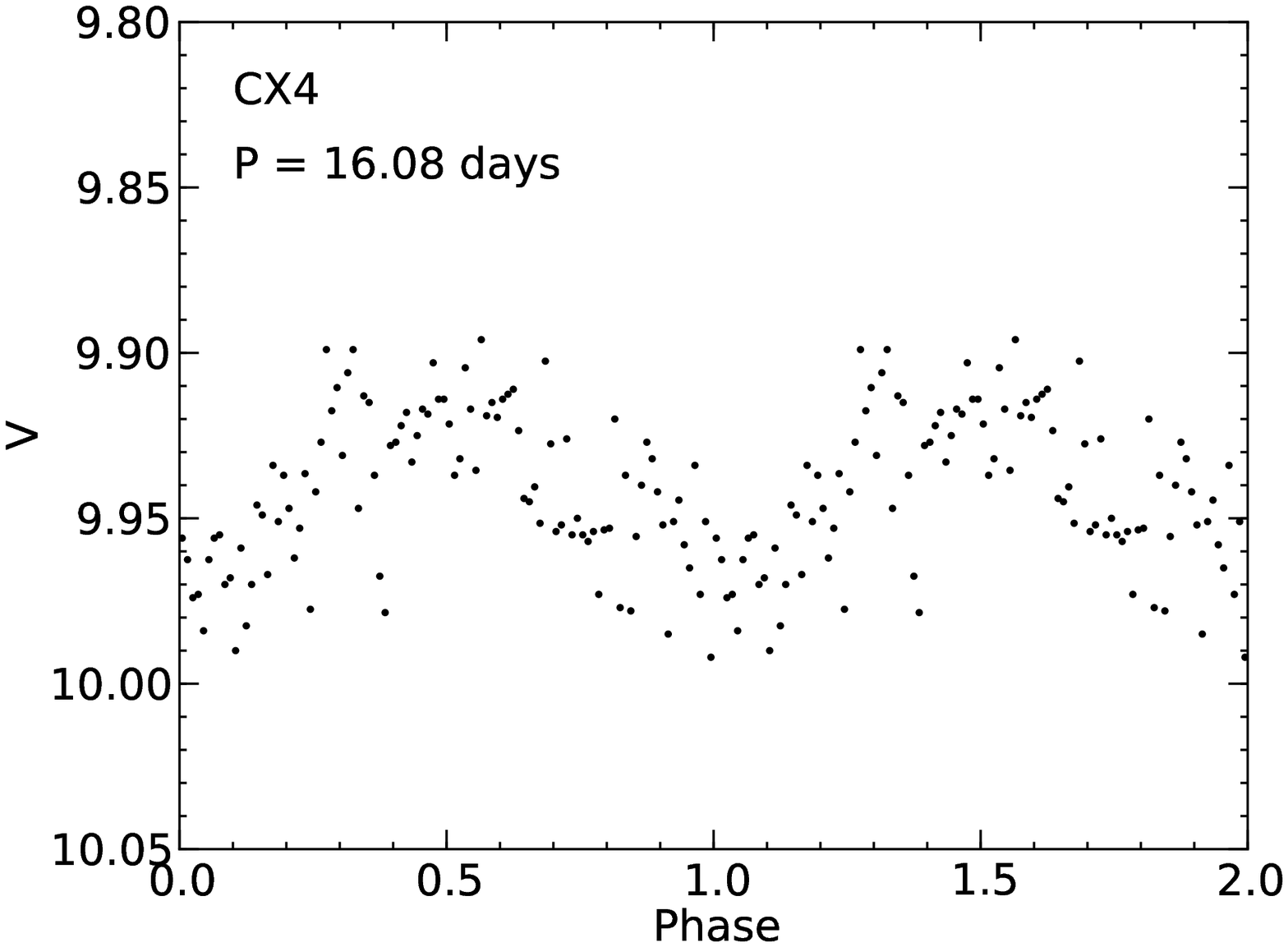}
\includegraphics[width=2.4in]{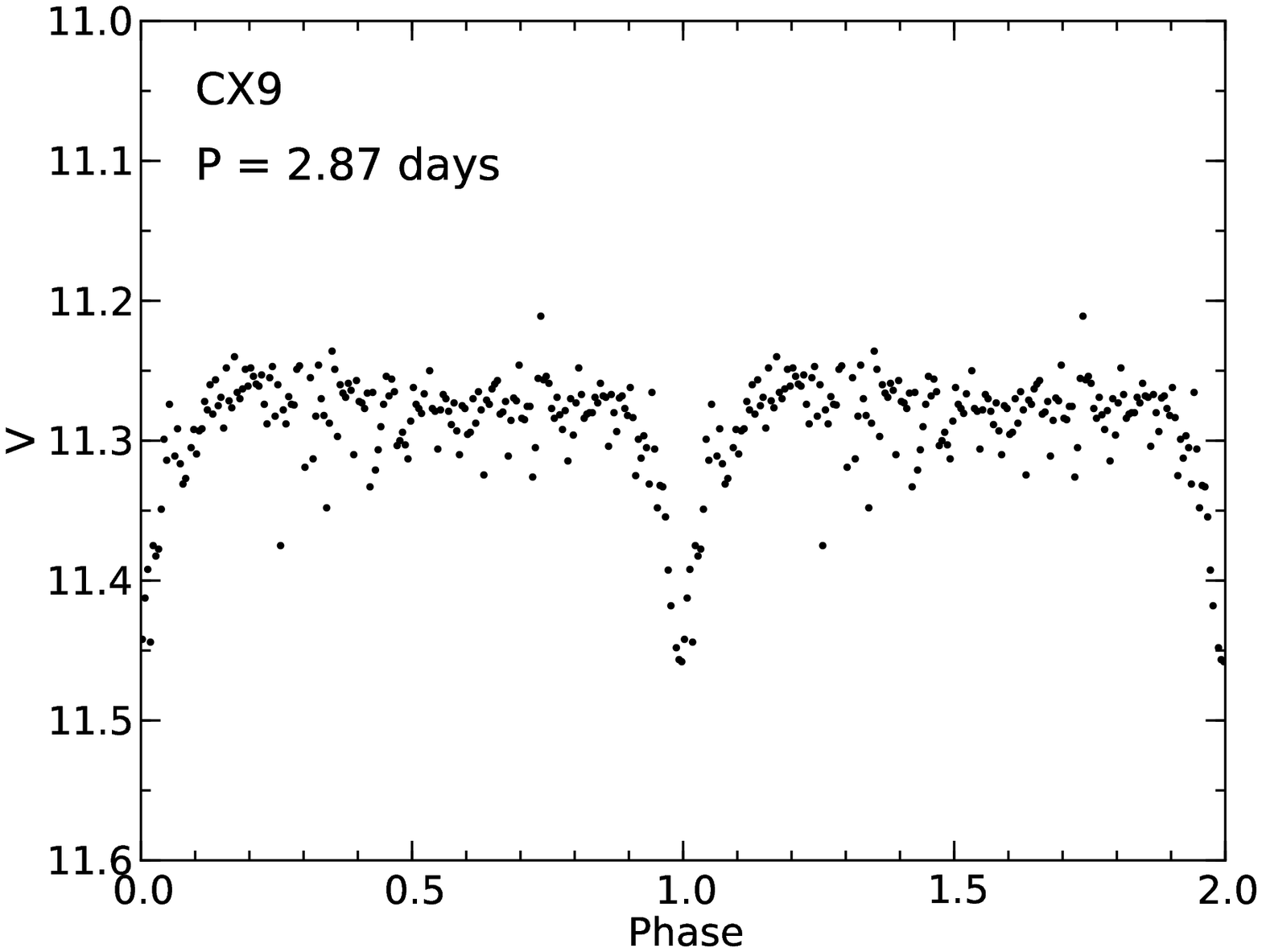}
\includegraphics[width=2.4in]{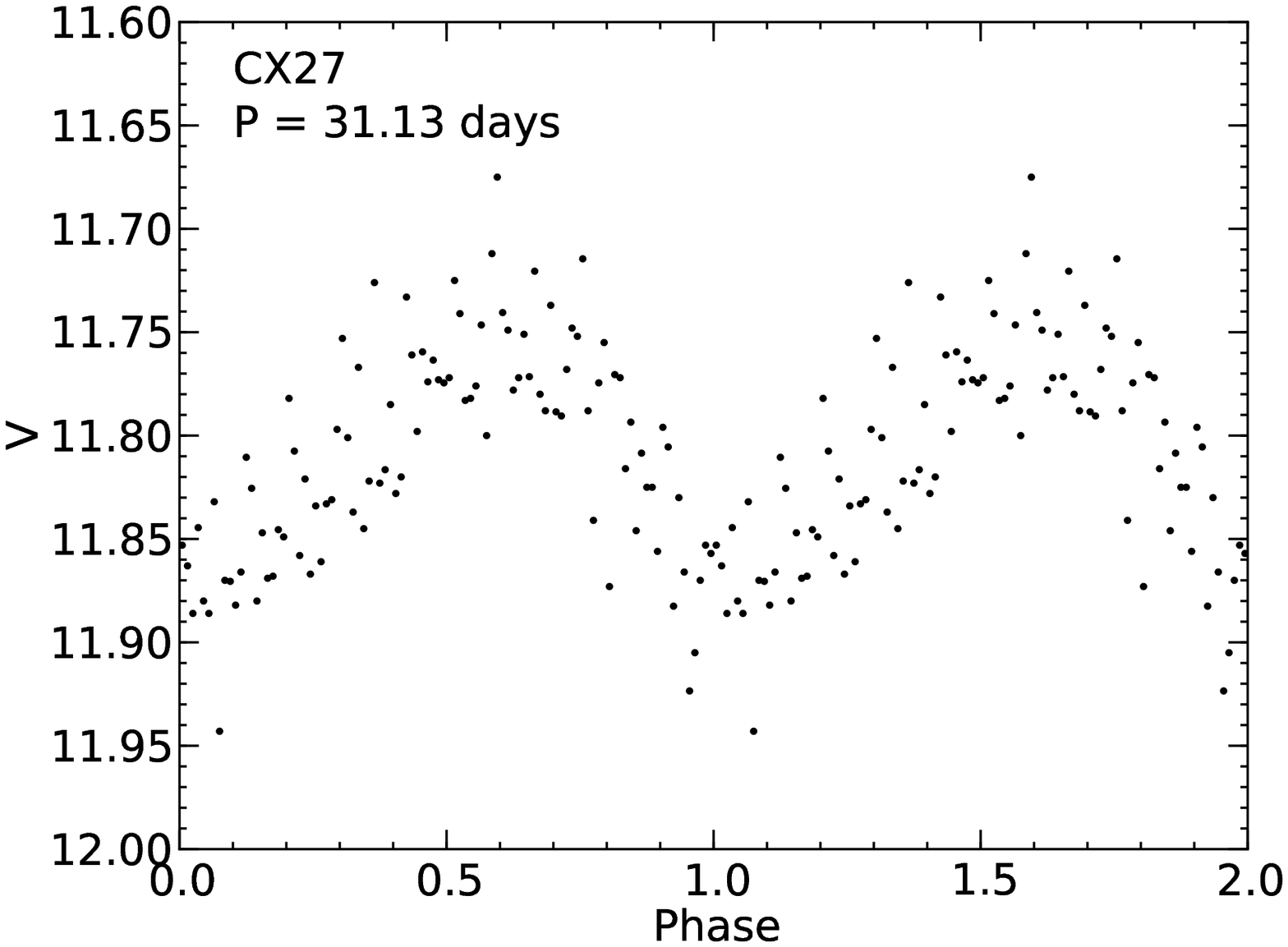}
\includegraphics[width=2.4in]{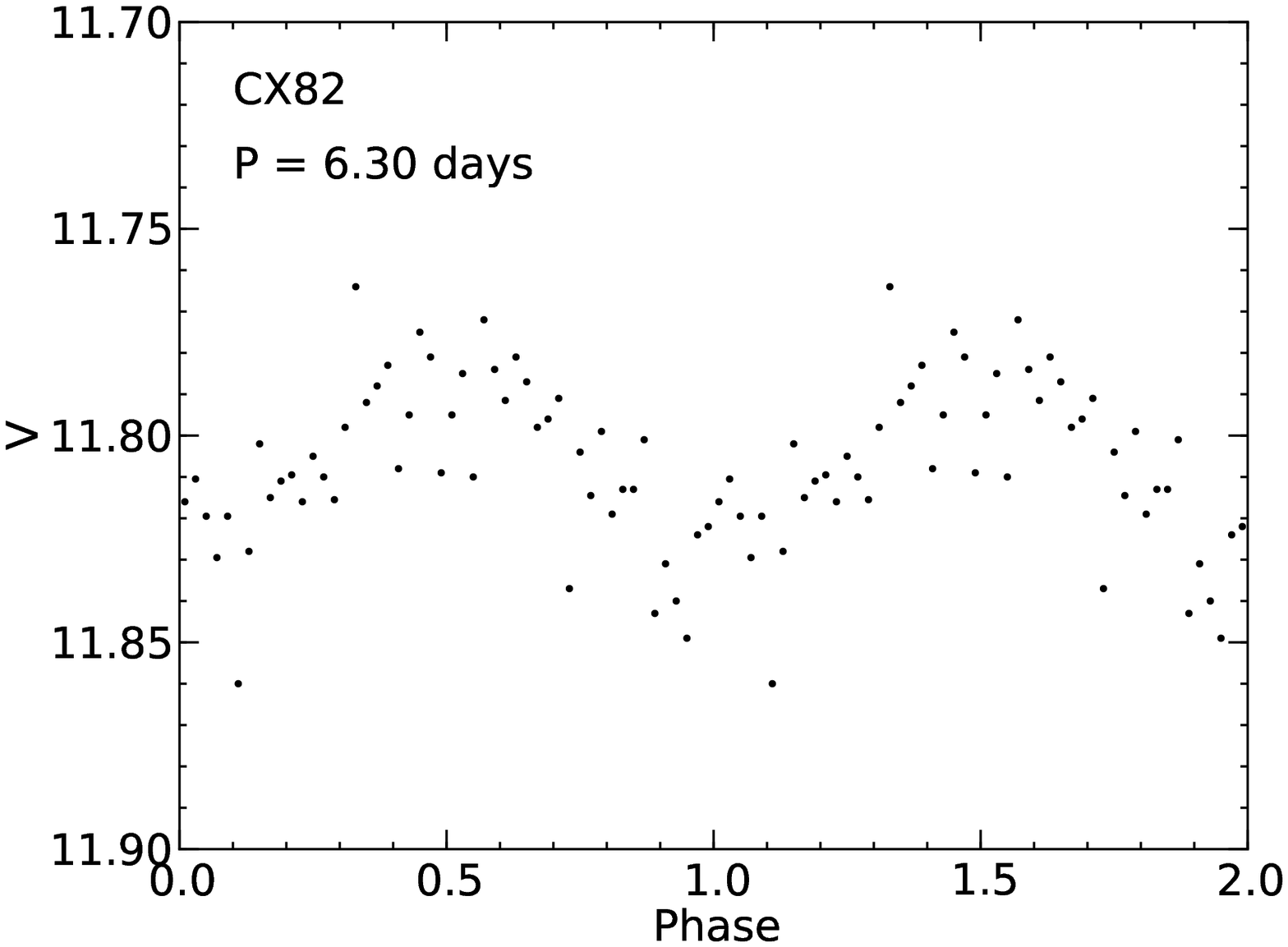}
\includegraphics[width=2.4in]{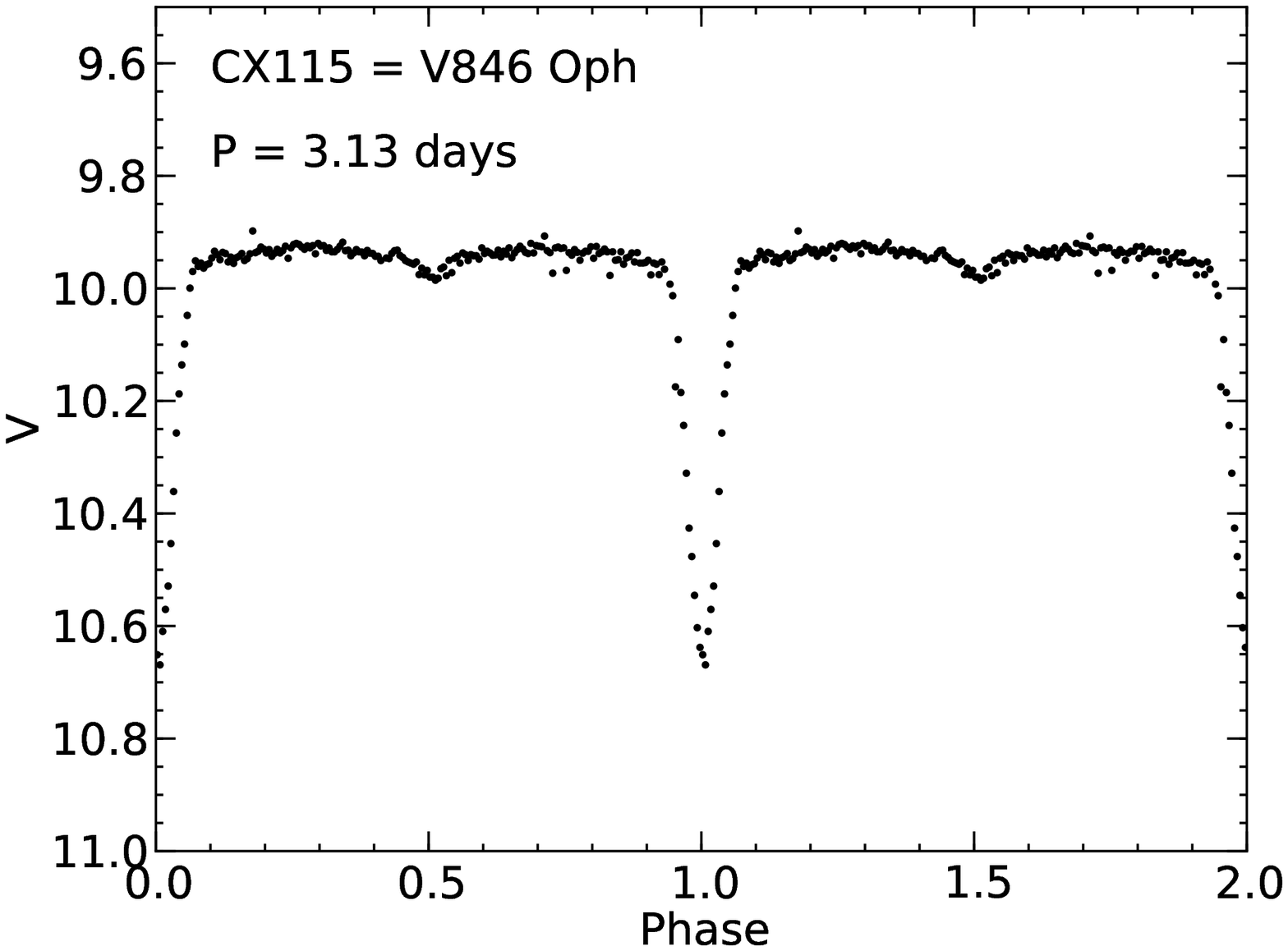}
\includegraphics[width=2.4in]{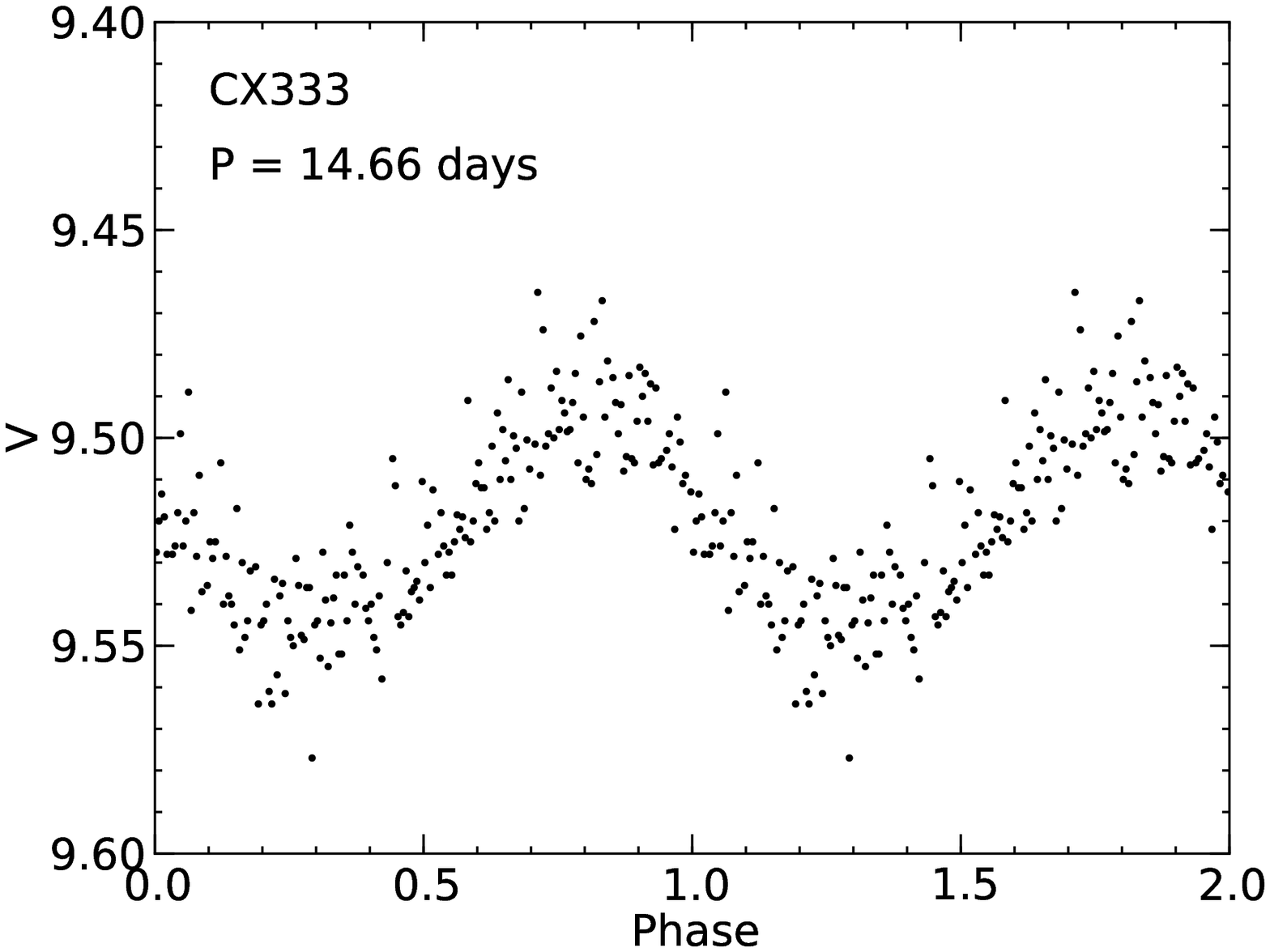}
\includegraphics[width=2.4in]{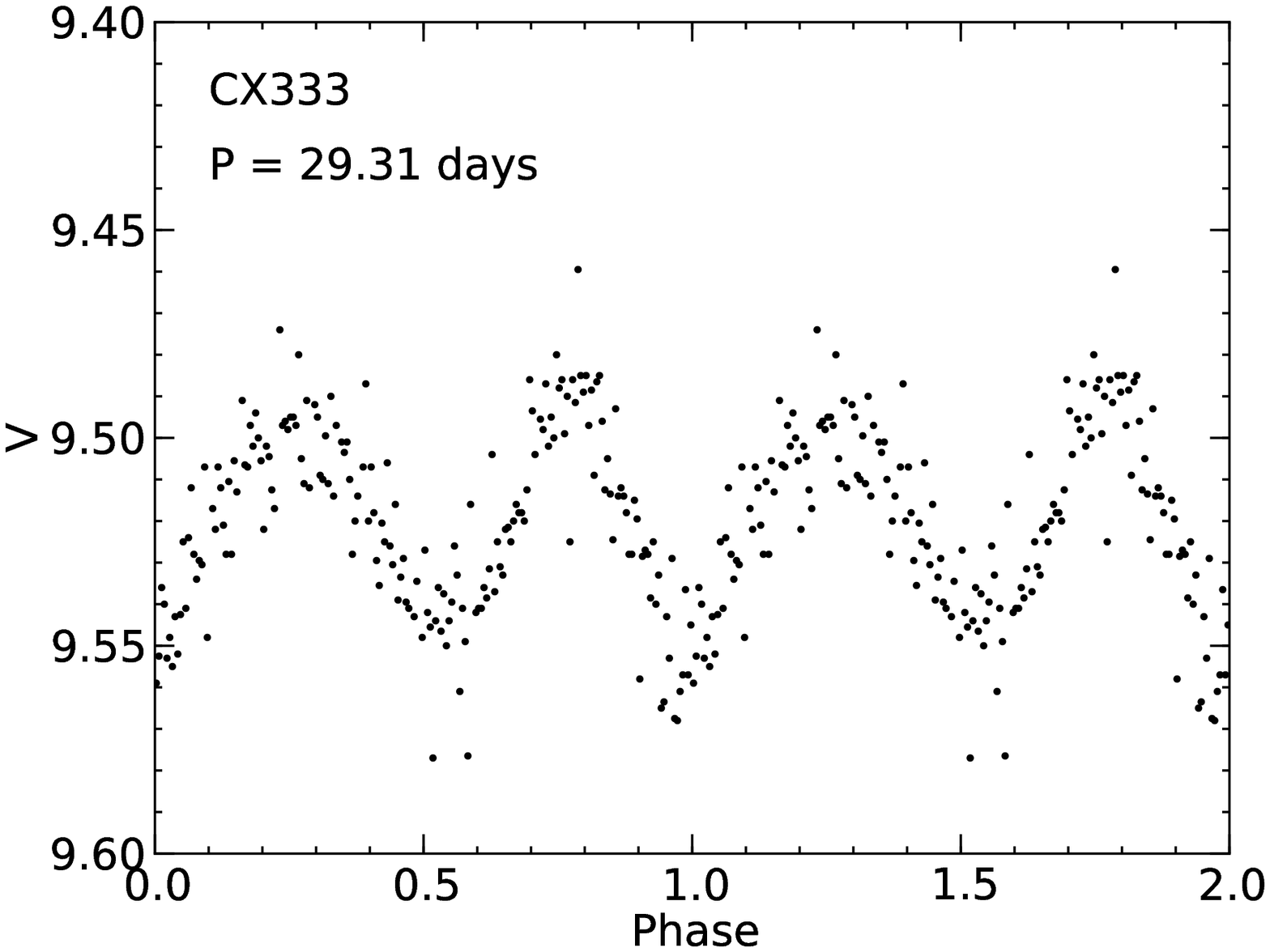}
\includegraphics[width=2.4in]{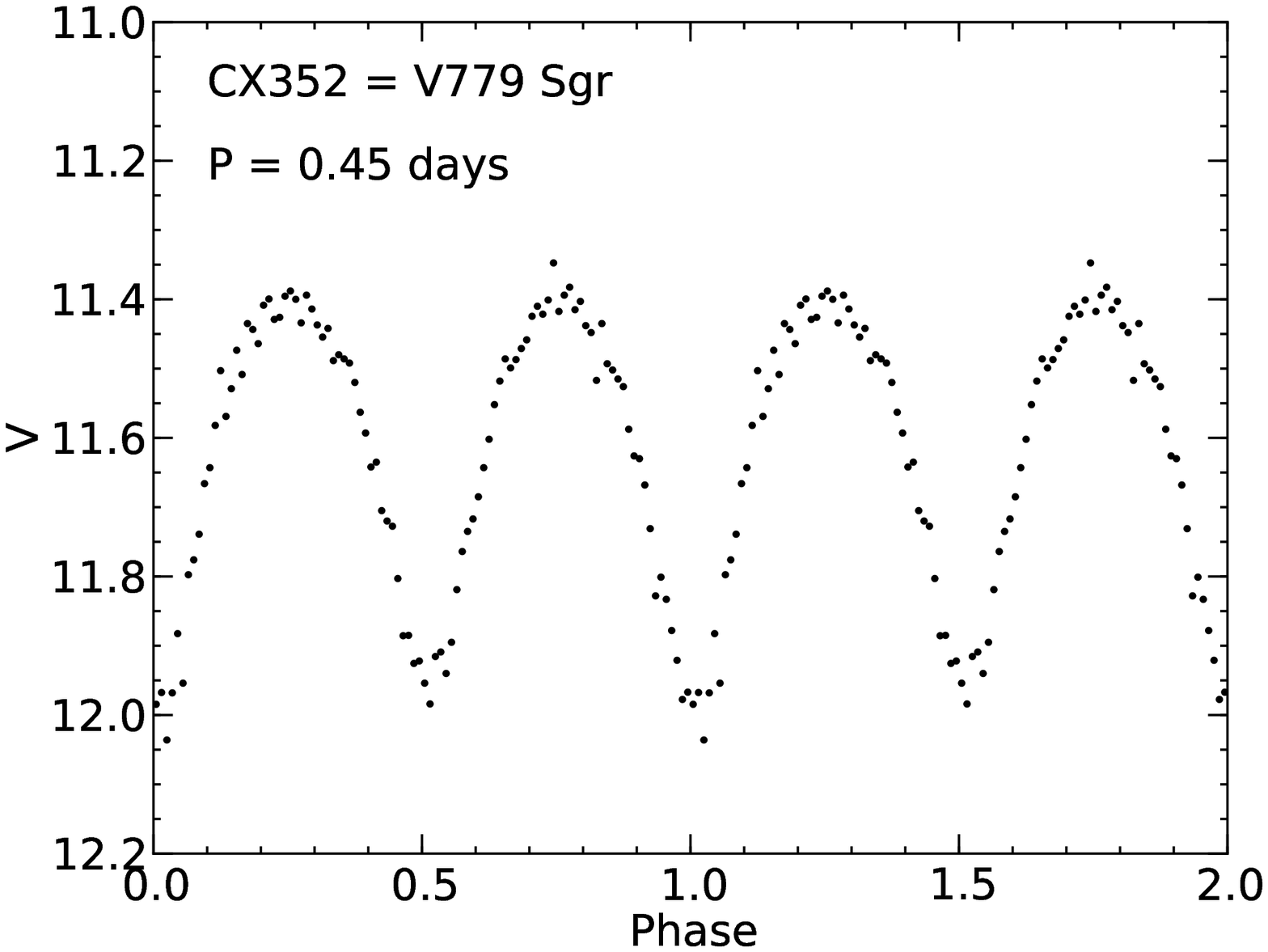}
\includegraphics[width=2.4in]{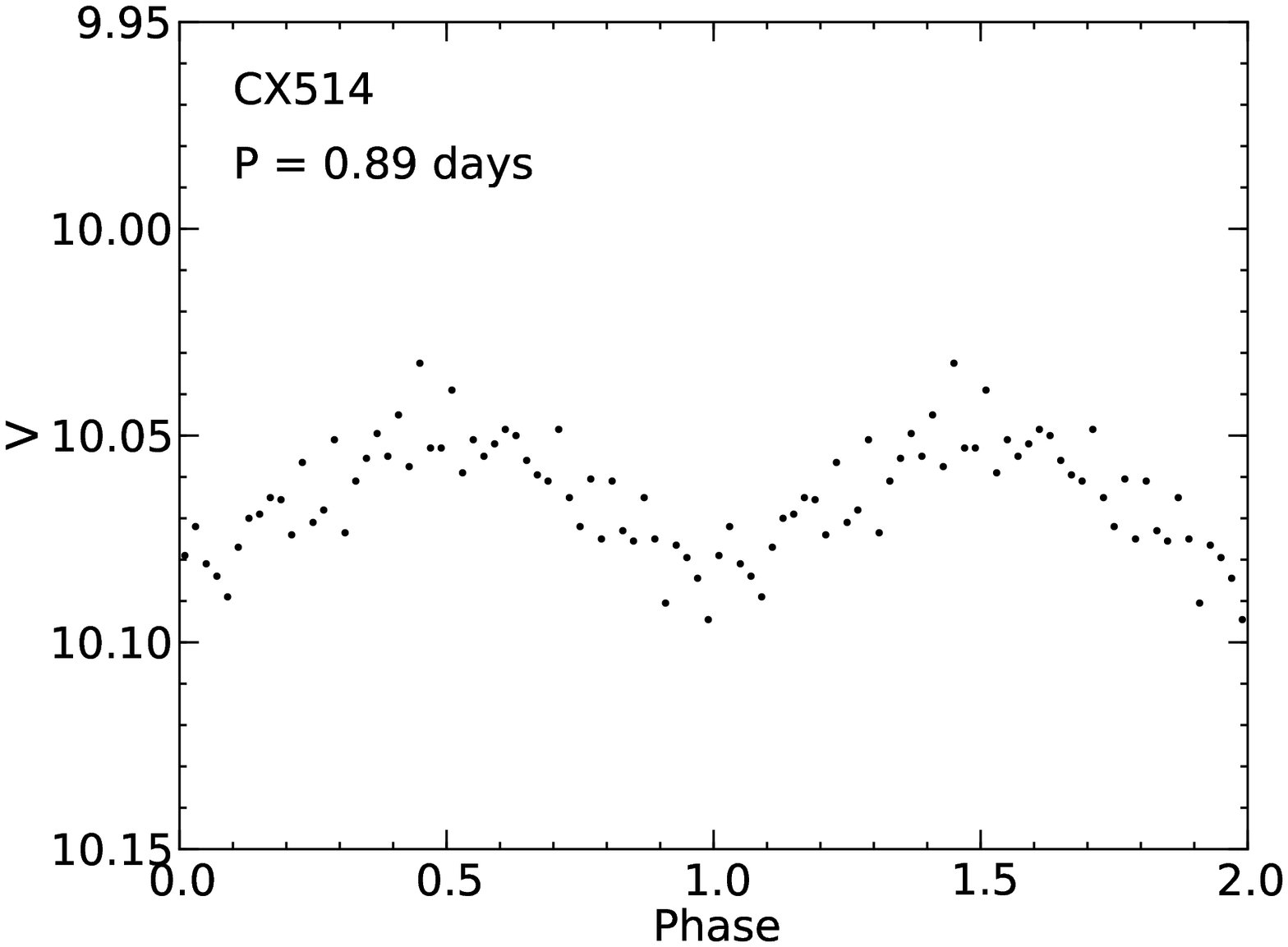}
\caption{ASAS lightcurves of stars with identifiable periods.  Note
  that CX4 also shows aperiodic long-term variability
  (Fig.~\ref{AperiodicFigTwo}), and that CX333 isx shown twice with
  different periods.}
\label{PeriodicFig}
\end{figure*}

\begin{figure*}
\includegraphics[width=2.4in]{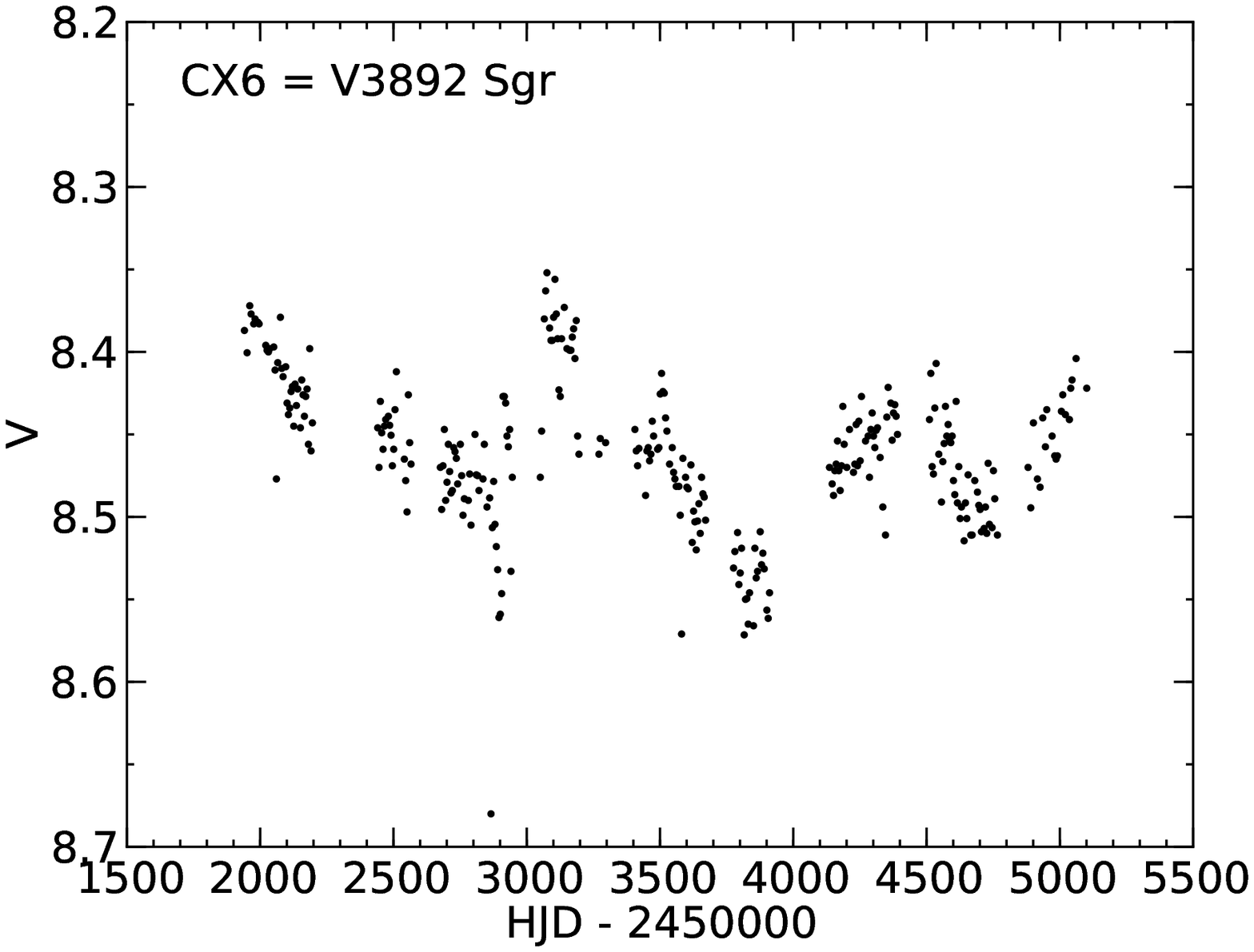}
\includegraphics[width=2.4in]{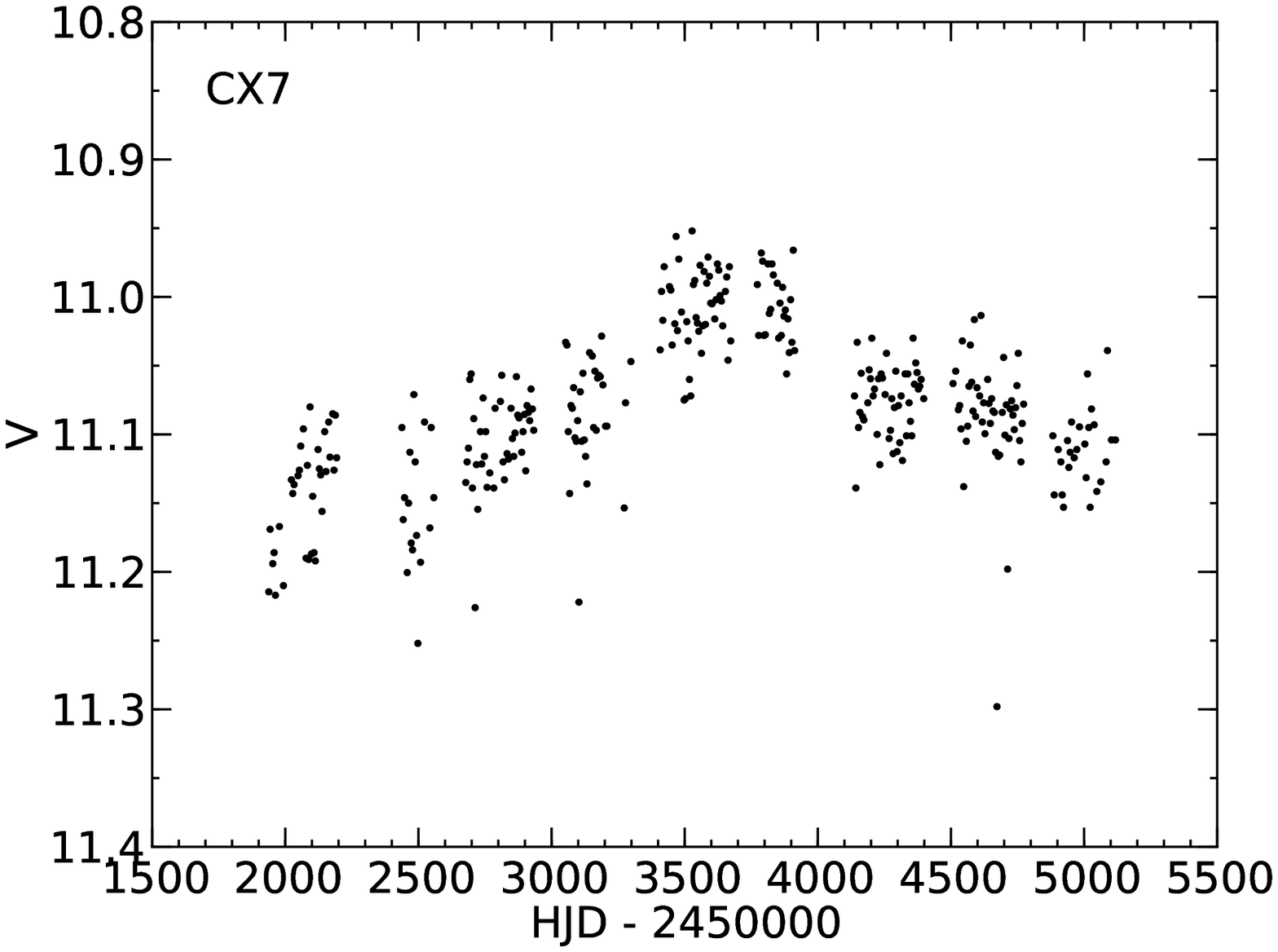}
\includegraphics[width=2.4in]{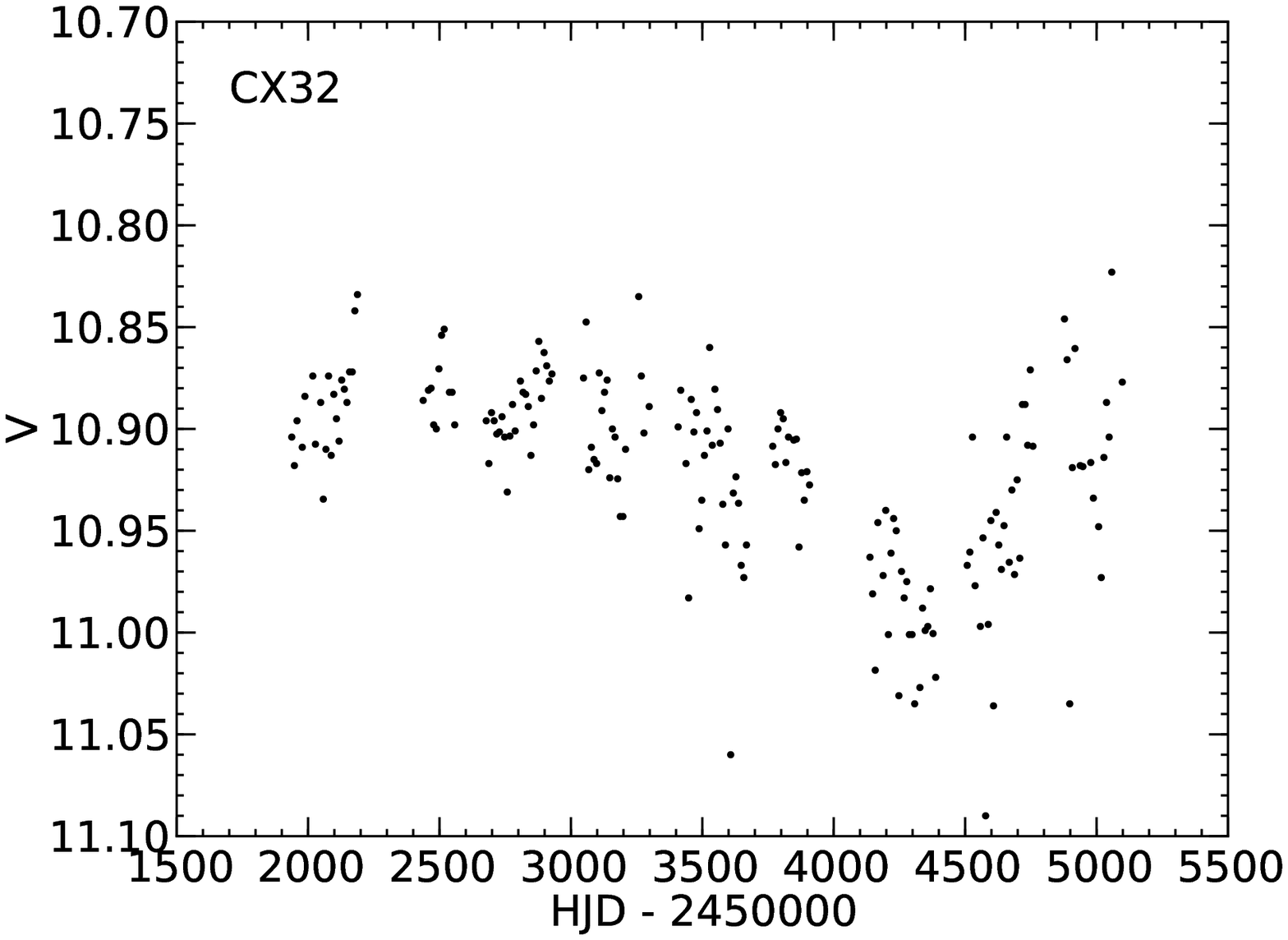}
\includegraphics[width=2.4in]{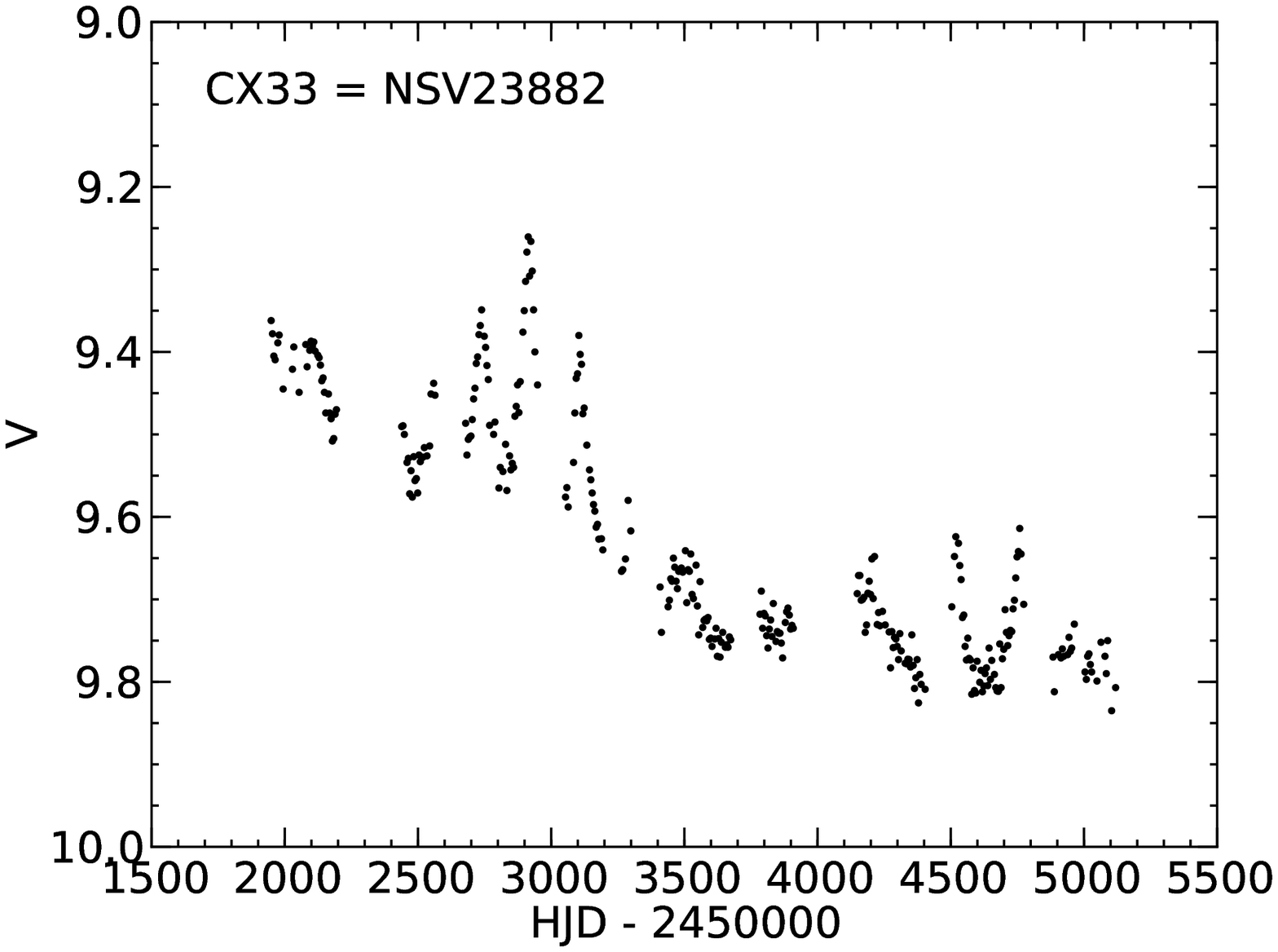}
\includegraphics[width=2.4in]{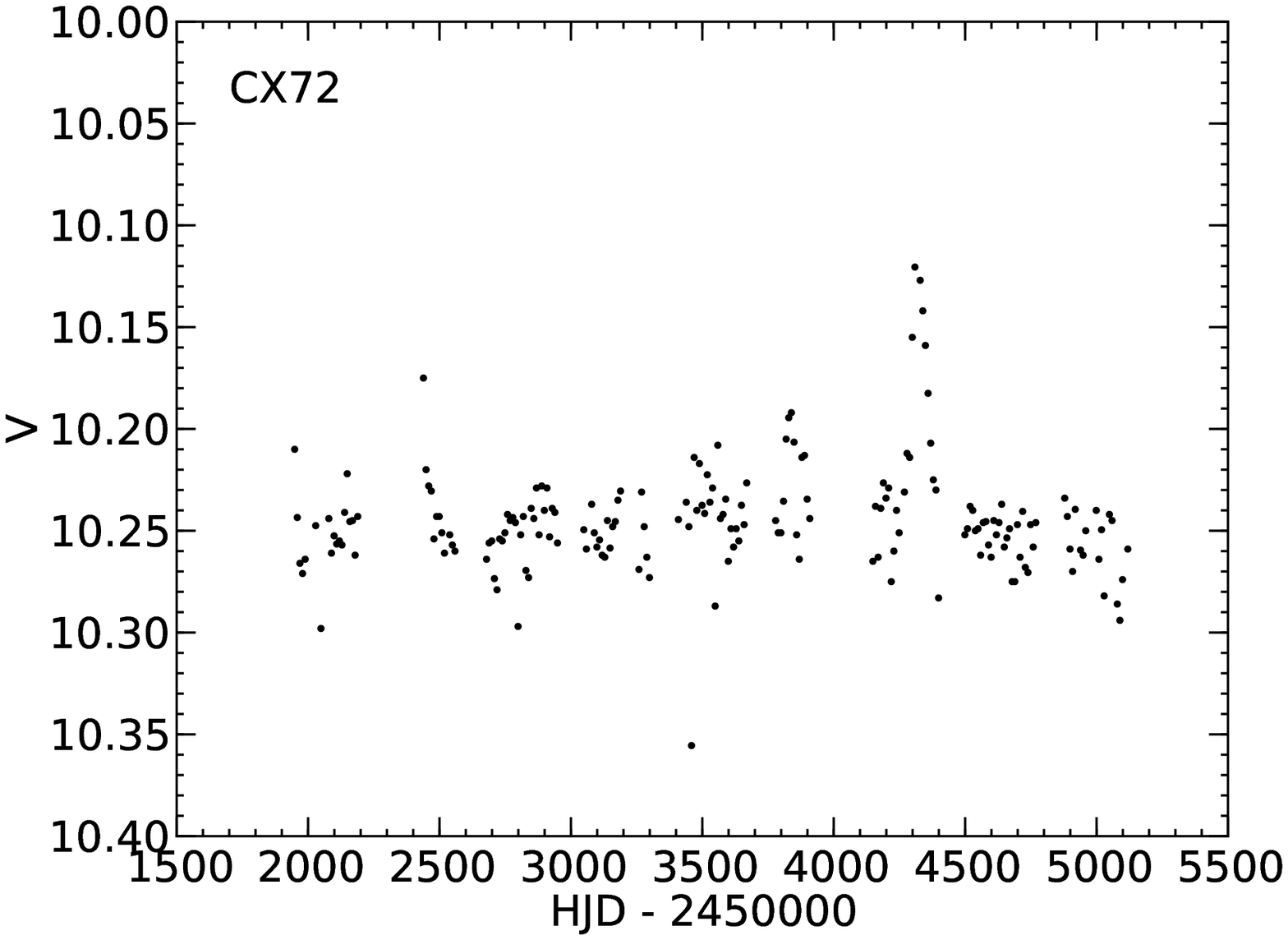}
\includegraphics[width=2.4in]{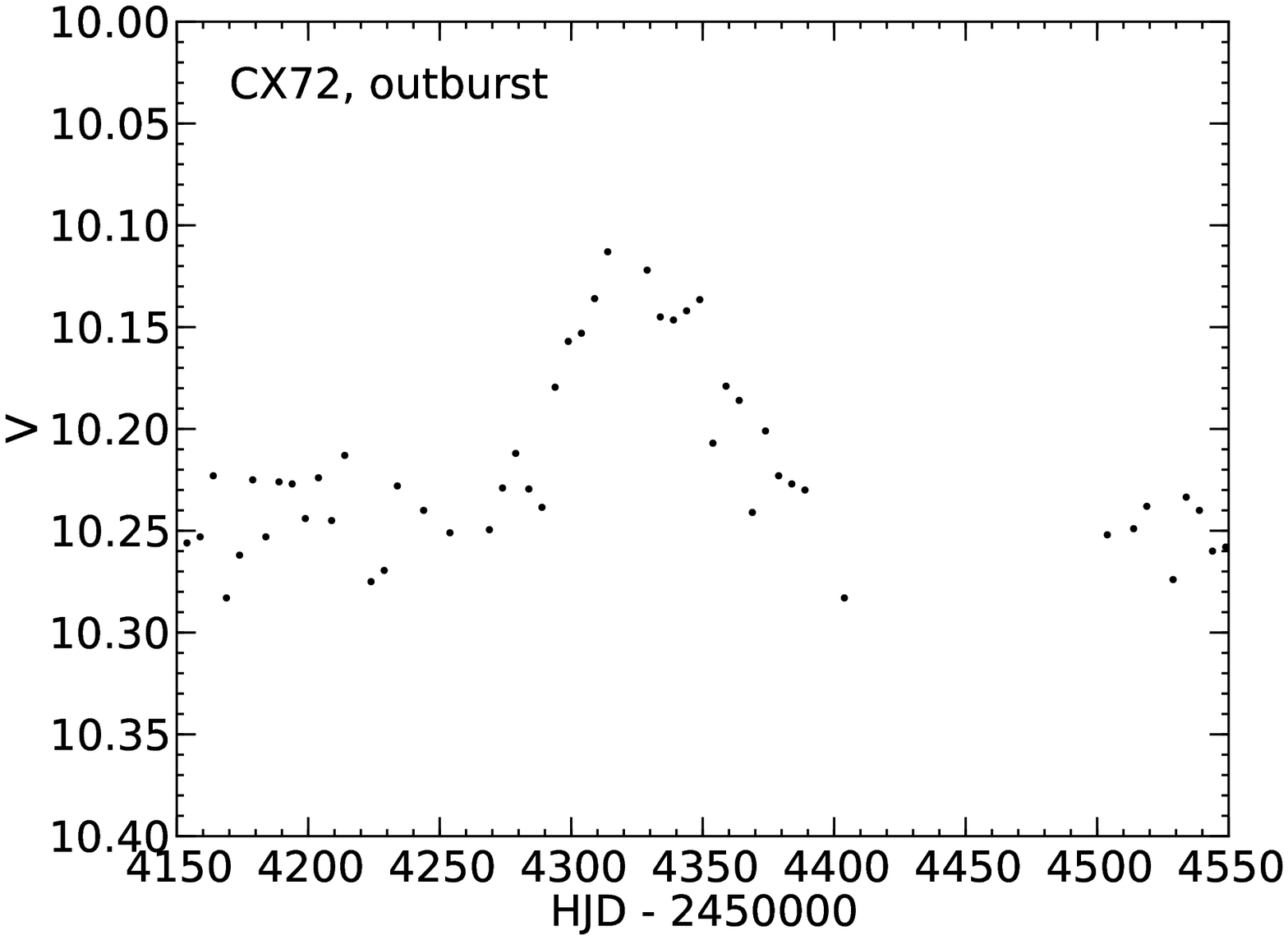}
\caption{ASAS lightcurves of stars showing definite aperiodic
  variability.  Note that CX72 is shown twice, the second as a
  close-up of the flare.}
\label{AperiodicFigOne}
\end{figure*}

\begin{figure*}
\includegraphics[width=2.4in]{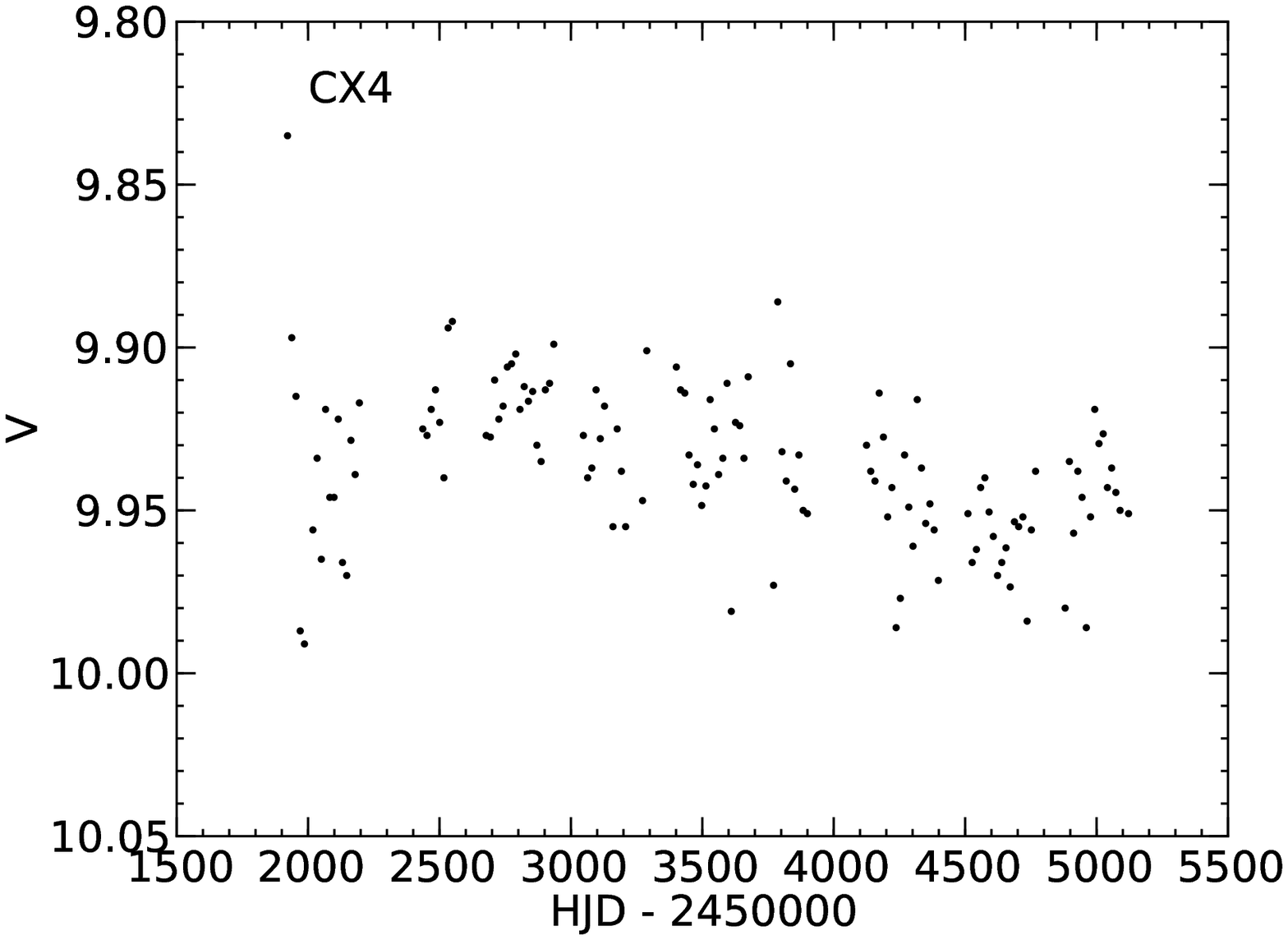}
\includegraphics[width=2.4in]{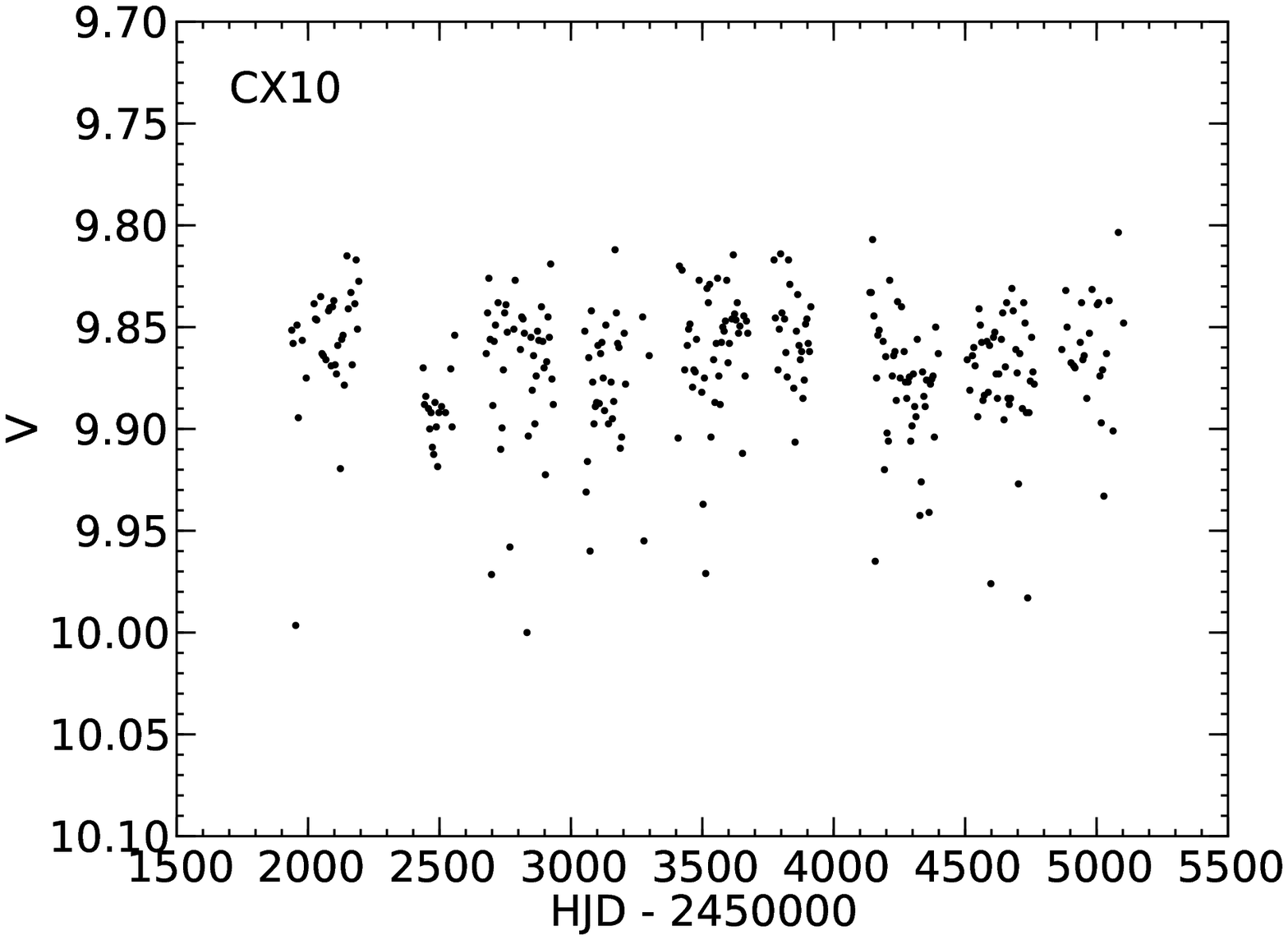}
\includegraphics[width=2.4in]{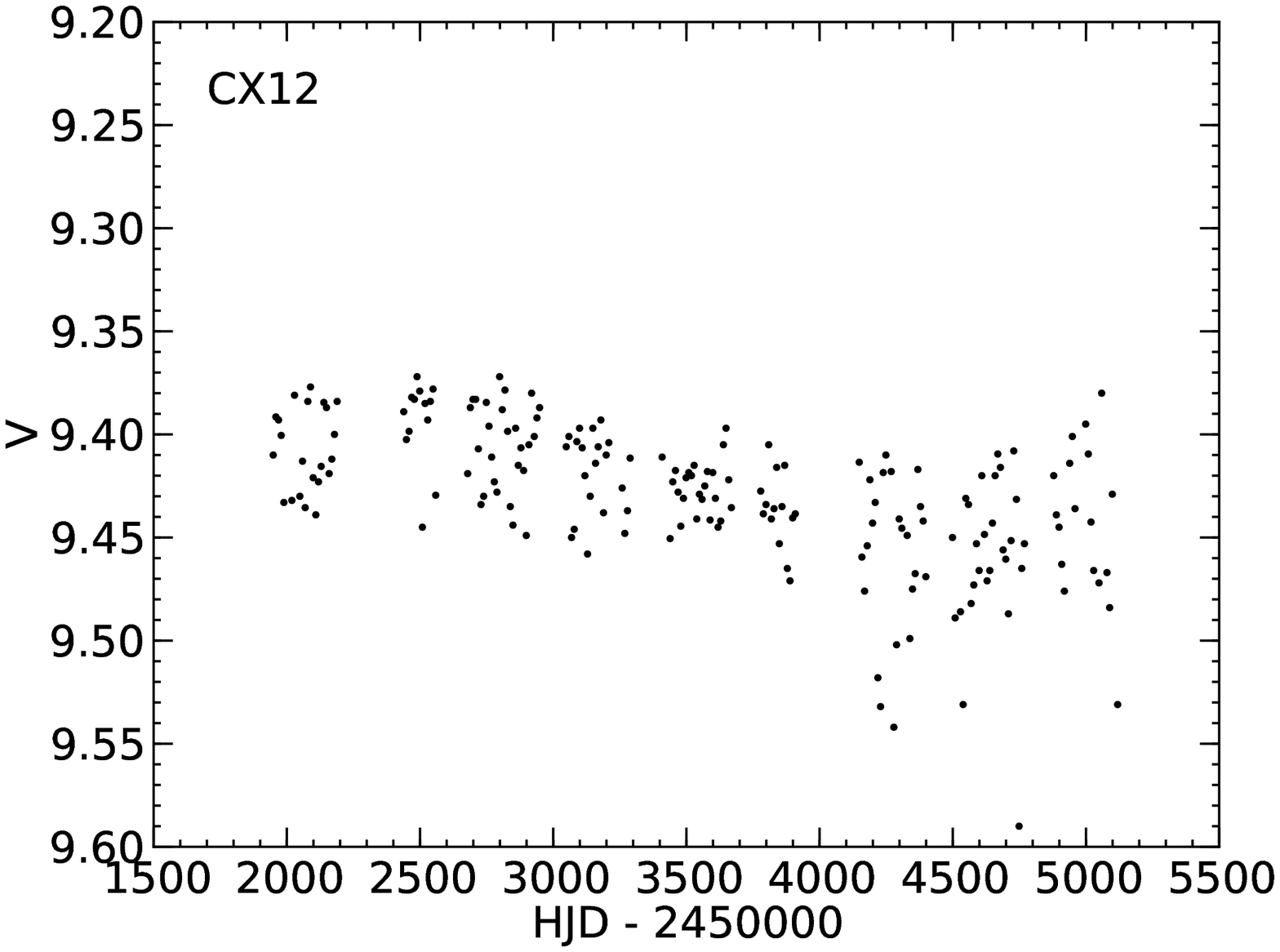}
\includegraphics[width=2.4in]{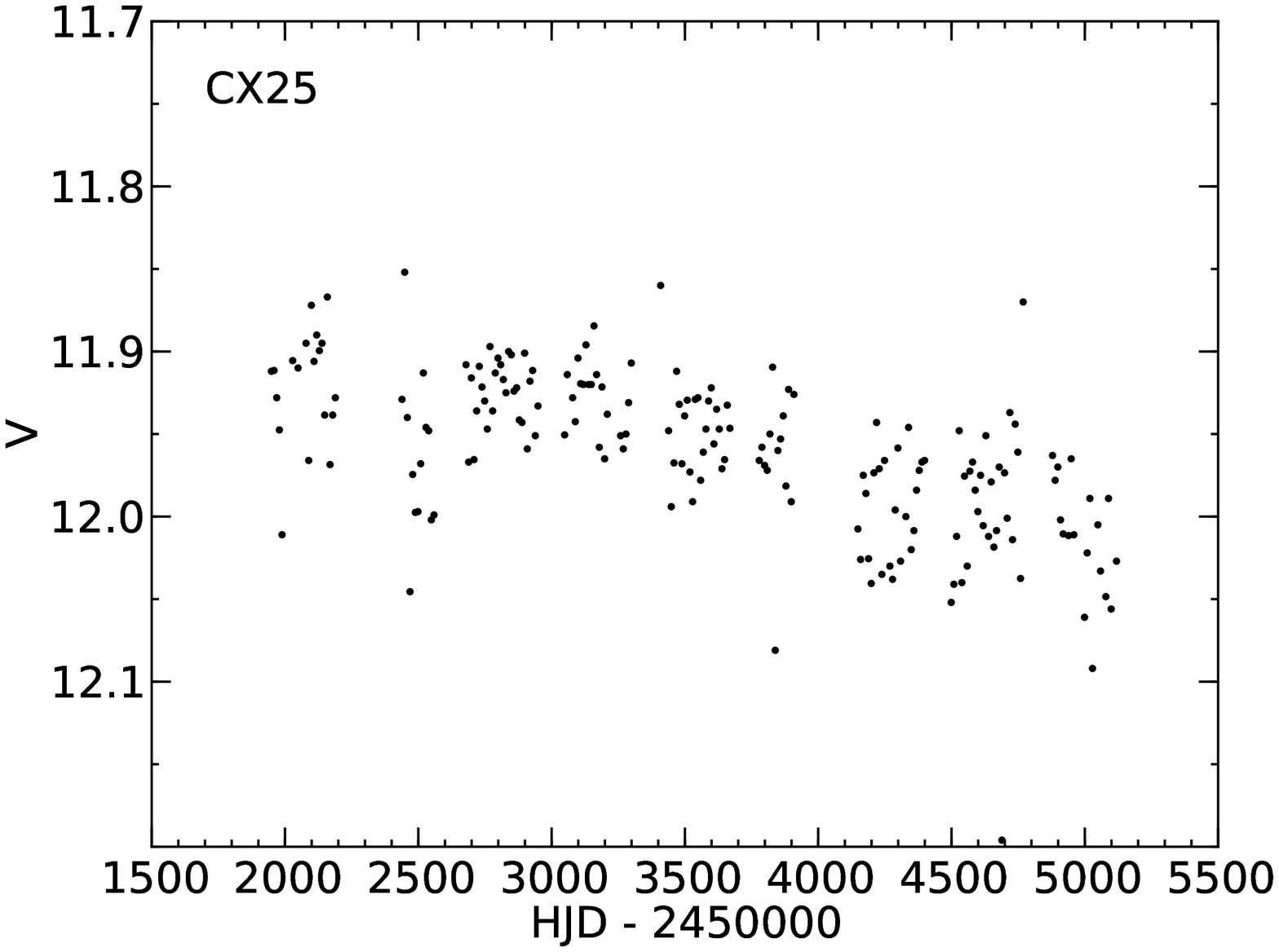}
\includegraphics[width=2.4in]{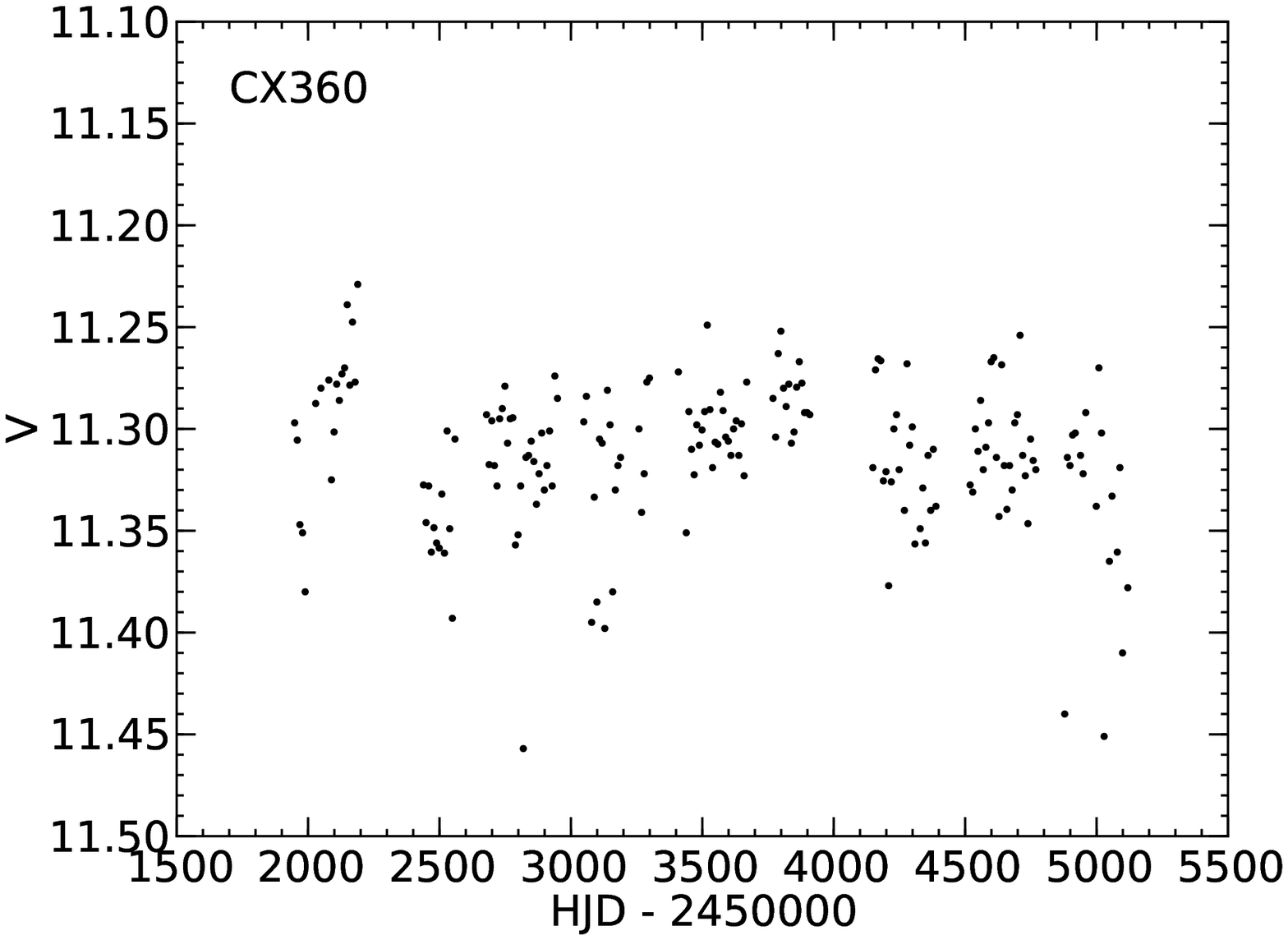}
\includegraphics[width=2.4in]{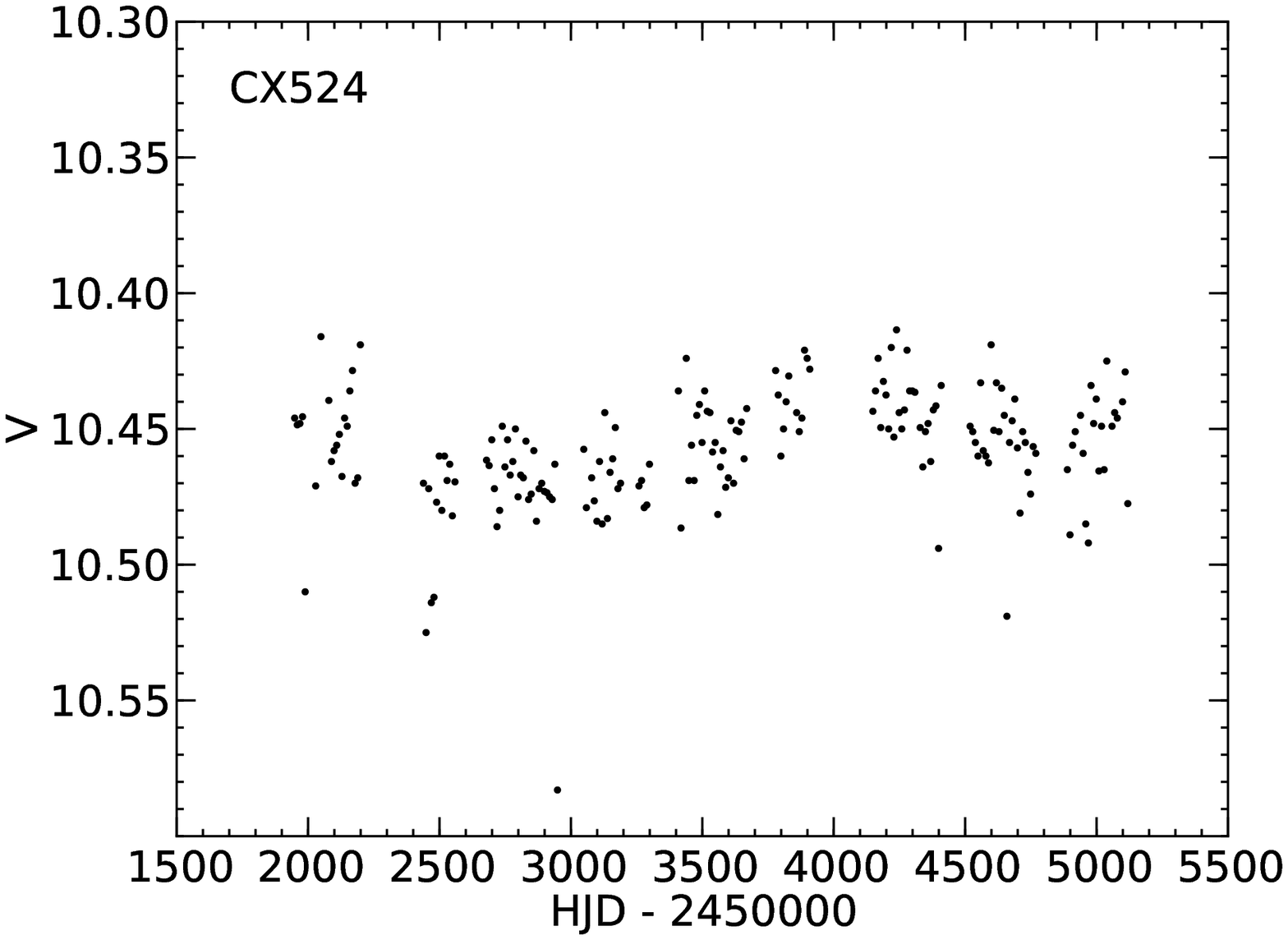}
\includegraphics[width=2.4in]{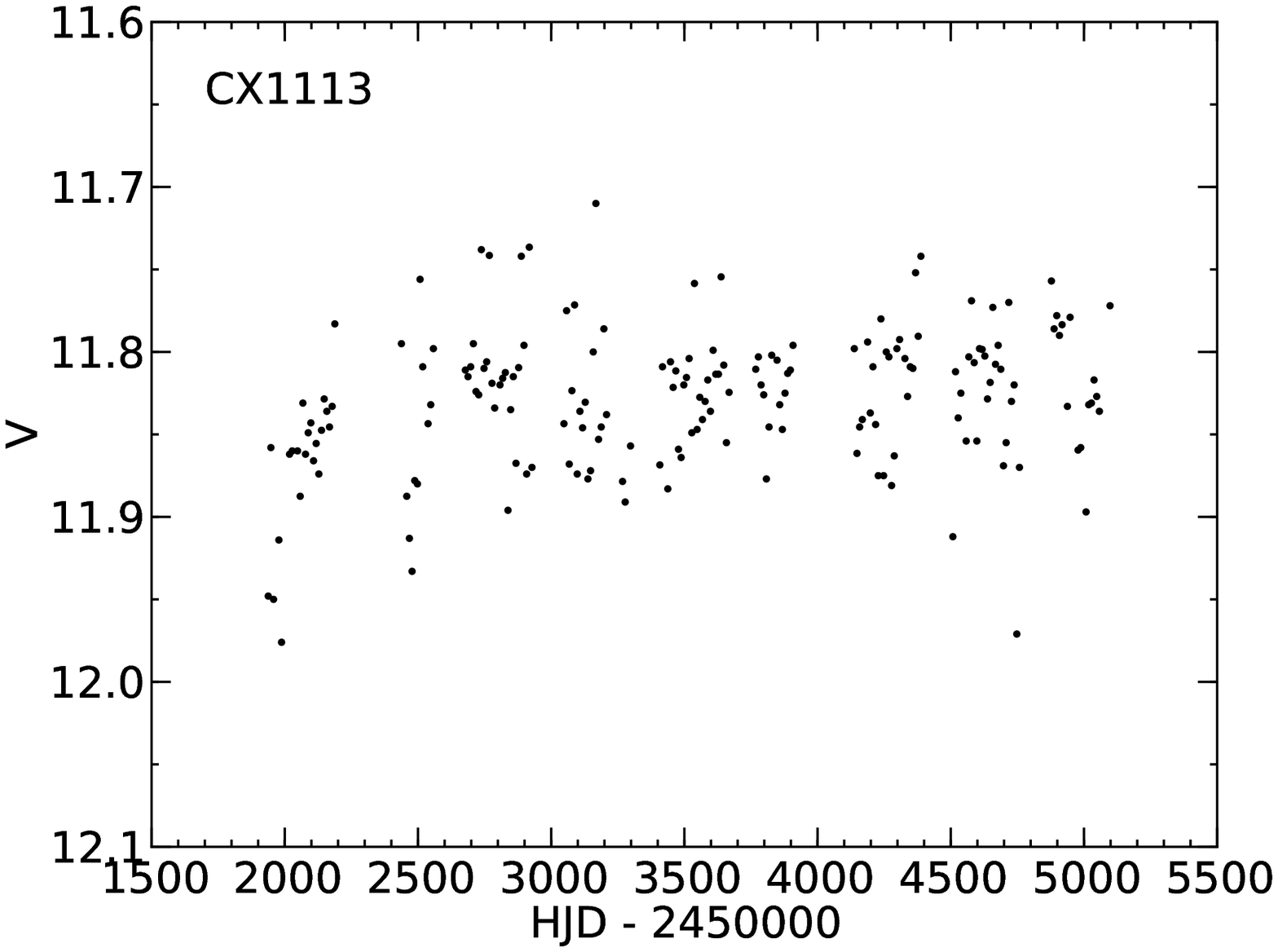}
\includegraphics[width=2.4in]{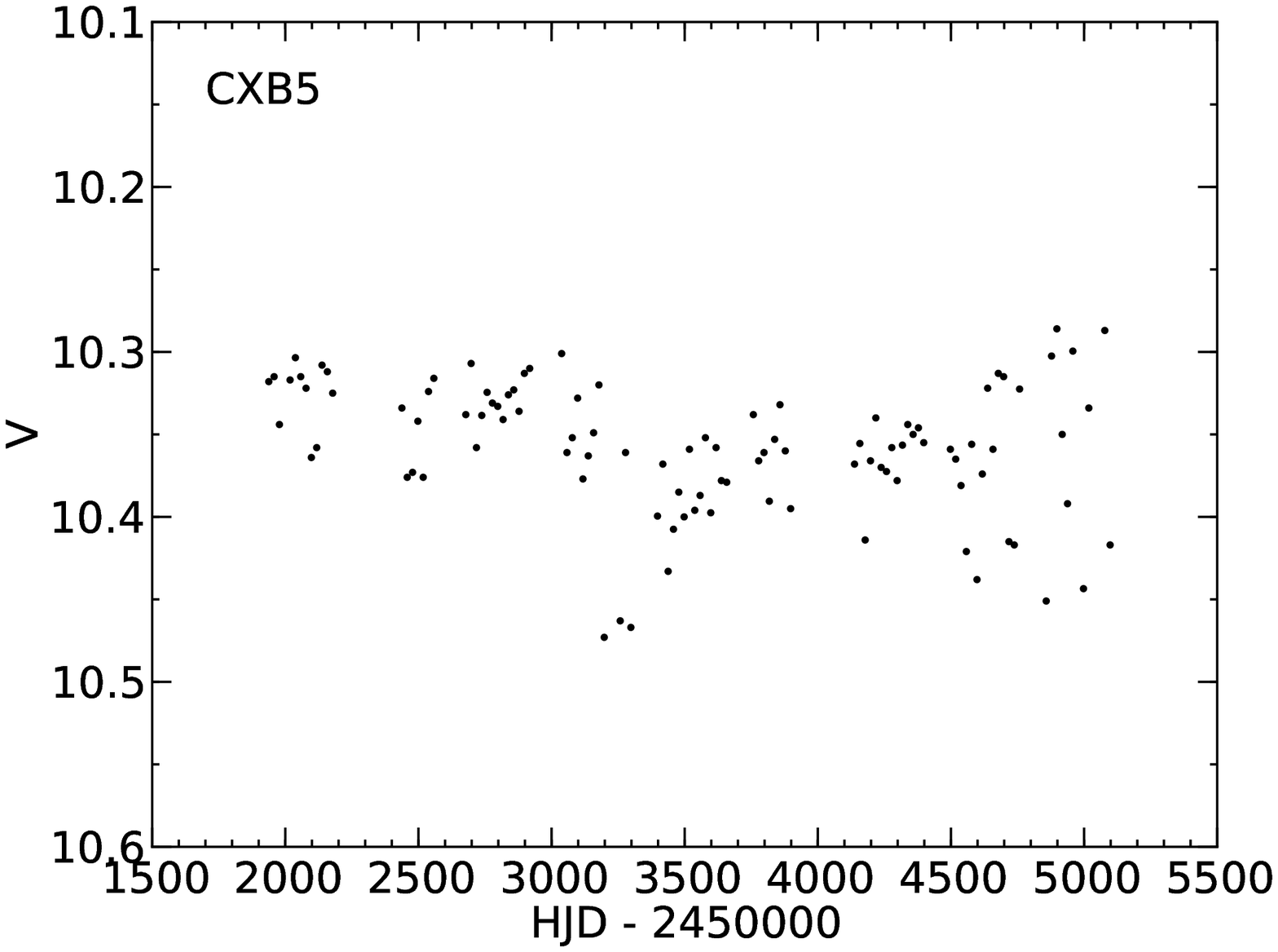}
\includegraphics[width=2.4in]{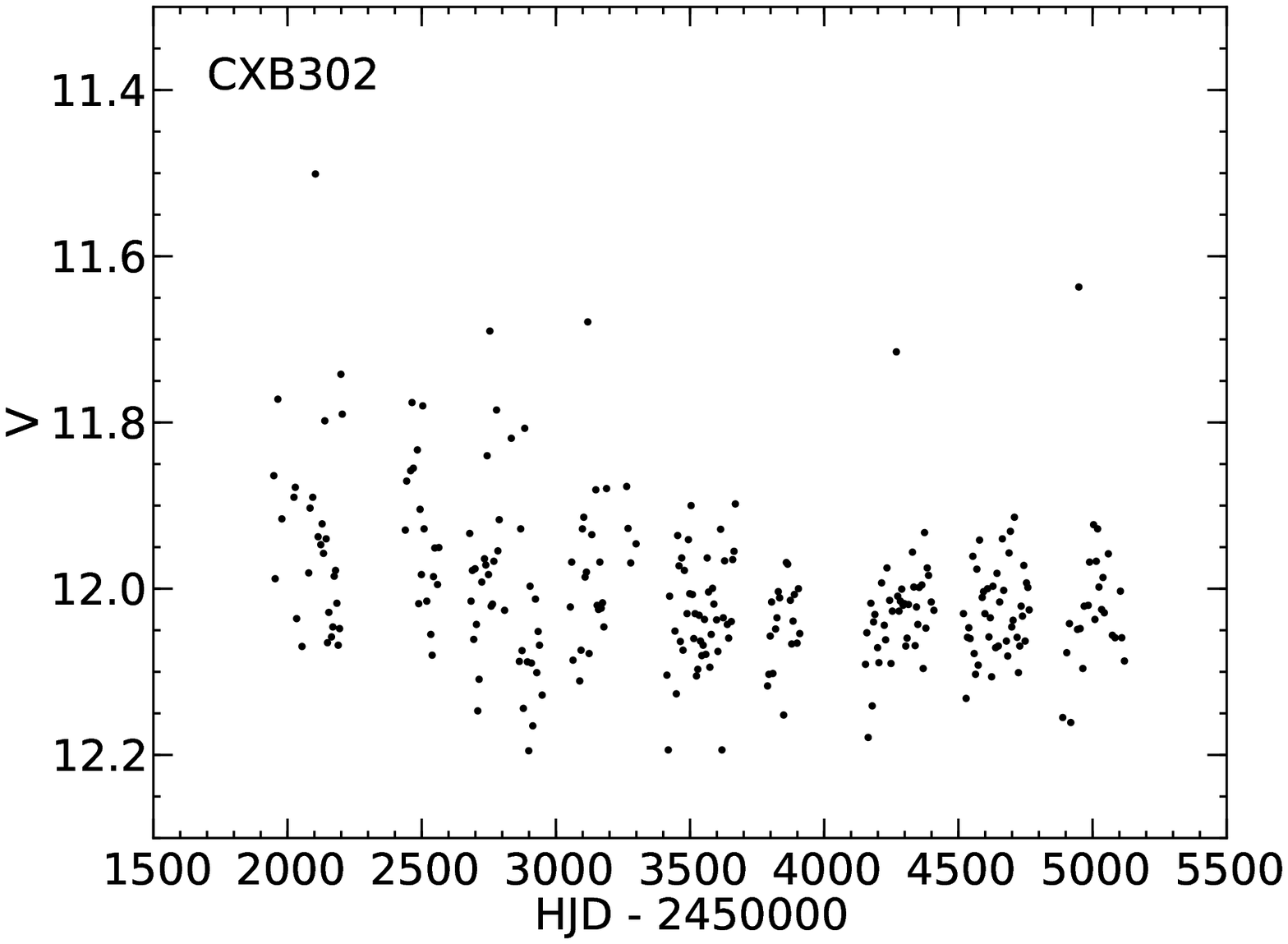}
\caption{ASAS lightcurves of stars showing probable aperiodic
  variability.  Note that CX4 also shows a periodic modulation shown
  in Fig.~\ref{PeriodicFig}.}
\label{AperiodicFigTwo}
\end{figure*}

\subsection{CX4 (CXOGBS~J173931.2--290952)}

This X-ray source was previously detected by {\it ROSAT}/PSPC
\citep{Sidoli:2001a} who noted that the source coincides with the
supernova remnant G359.1+0.9 \citep{Gray:1994a}.  With a better
position, we find that the {\em Chandra} source is inside the remnant,
but off-center, but is a very close match to HD\,316072.  It was also
serendipitously observed by {\it XMM-Newton}, and
\citet{Farrell:2010a} note that the X-ray spectrum is consistent with
thermal plasma.  This, together with the close association with
HD\,316072 supports interpretation of the X-ray source as stellar
coronal emission from this star, rather than being associated with the
supernova remnant.  The X-ray to bolometric flux ratio is unusually
high for a late-type giant ($\log(F_{\rm x} / F_{\rm bol}) = -3.8$),
as is the inferred X-ray luminosity, $L_{\rm X}\simeq
3\times10^{31}$\,erg\,s$^{-1}$, but it is near the top of the range
observed in other late-G giants, rather than outside it
\citep{Gondoin:2005a}.

HD\,316072 is identified as a variable in ASAS data
\citep{Pojmanski:2002a}, strengthening the likelihood of it being the
true counterpart further.  A Lomb-Scargle periodogram yields a single
clear period of $16.0811\pm0.0026$\,days, with the uncertainty
estimated by the bootstrap method.  The periodicity has an amplitude
of about 0.1\,mag, is apparent in individual data points as well as in
phase-binned data, and is present in multiple independent subsets of
the lightcurve.  We show the folded lightcurve in
Fig.~\ref{PeriodicFig}.  The rotation rate is quite rapid for a giant,
so there is no need to look for X-rays from a companion star and we
can comfortably attribute the X-rays to coronal activity in the giant.

\subsection{CX6 (CXOGBS~J174445.7--271344)}

HD\,161103 is one of the prototypical Be stars identified as a class
by \citet{Merrill:1925a}.  Early classifications include B2\,{\sc
  iii--v} \citep{Hiltner:1956a} and B1\,{\sc iii}
\citep{Garrison:1977a}.  Recent works favor a B0.5\,{\sc iii--v}e star
\citep{Steele:1999a,LopesDeOliveira:2006a}.  As expected for a Be
system, it shows a pronounced infrared excess over the colors expected
for an isolated star due to the presence of circumstellar material and
is also a strong outlier in $(J-H)$ vs.\ $(H-K)$.

This object is a long-known variable star, V3892~Sgr, and confirmed as
a variable by ASAS \citep{Pojmanski:2002a}.  We show its lightcurve in
Fig.~\ref{AperiodicFigOne}.  It exhibits relatively low amplitude
irregular variations, at least by comparison with the other Be star in
our sample, CX33.

HD\,161103 emerged as a possibly more interesting object when
\citet{Motch:1997a} matched it with a {\it ROSAT} source and inferred
an unusually high luminosity for its spectral type (unabsorbed $L_{\rm
  x}\simeq 10^{32}$\,erg\,s$^{-1}$), and a quite hard color.  Both
characteristics stand out in our {\it Chandra} data as well.  Based on
these characteristics, \citet{Motch:1997a} identified this as a
candidate X-ray binary.  Since the X-ray luminosity is lower than
known Be + neutron star systems, they suggested a white dwarf
companion.  \citet{LopesDeOliveira:2006a} studied this object further,
identifying it as one of a number of objects analogous to
$\gamma$~Cas.  They reported a 3200\,s pulsation, although it was
unclear if this was truly a coherent pulsation from a rotating compact
object, or a quasi-periodic oscillation as seen in $\gamma$~Cas.  The
nature of this object, and of the $\gamma$~Cas analogs in general,
remains unknown \citep{Motch:2007a}.

\subsection{CX7 (CXOGBS~J173826.1--290149)}
\label{CXSevenSection}

TYC 6839-513-1 was spectroscopically classified as a K0\,{\sc v} star
by \citet{Torres:2006a}.  These authors measured a Li equivalent width
of 80\,m\AA, comparable to the lower end of the distribution seen in
Pleiades stars of the same color, suggesting that this is a young
object.

\begin{figure}
\includegraphics[width=3.6in]{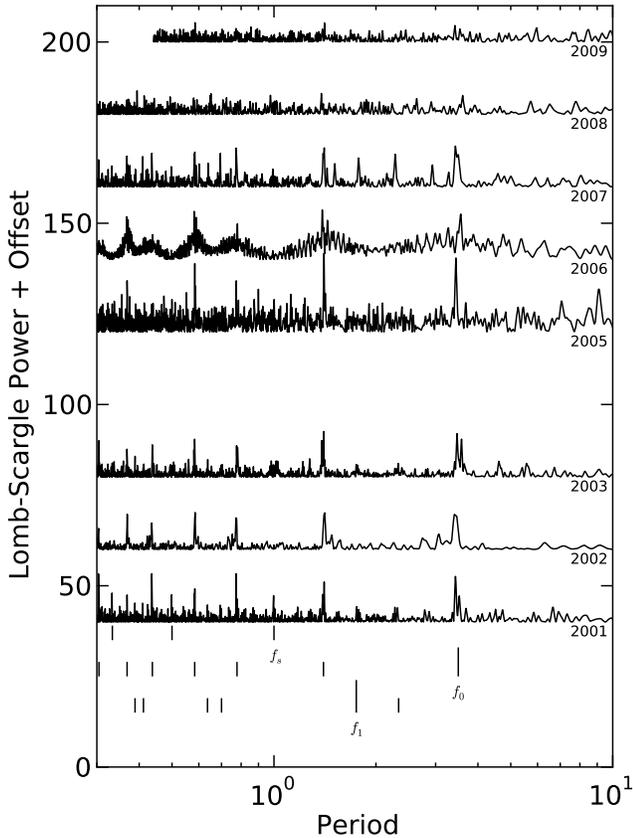}
\caption{Yearly ASAS Lomb-Scargle periodograms for CX7.  2004 has been
  omitted as the sampling within this year compromised the
  periodogram.  We label three sets of frequencies: $f_s$, the one-day
  sampling frequency and its harmonics; $f_0$, the 3.5\,day
  fundamental period and its aliases, and $f_1$, the 1.75\,day first
  harmonic and its aliases.}
\label{CXSevenFig}
\end{figure}

This star is not identified as a variable in the ASAS variability
catalog, but on examination it does show quite large amplitude, very
slow variability (Fig.~\ref{AperiodicFigOne}).  It also does appear to
show periodic variability with a period around 3.5\,days, although
this is only apparent after examining yearly periodograms
(Figure~\ref{CXSevenFig}).  A primary period around $3.5\pm0.1$\,days
is apparent on most years, although the frequency is not stable, and
sometimes two periods are seen.  The quoted uncertainty primarily
reflects the variations seen in the dominant period.  In addition on a
few years there is significant power at the first harmonic.  Together,
these frequencies and their aliases account for all the features seen
in the periodograms.  We interpret this behavior as variations caused
by starspots that form at different latitudes subject to different
rotation rates.  Sometimes multiple spots may result in significant
power at harmonics.

The X-ray to bolometric flux ratio of CX7 is very high, $\log ( F_{\rm
  X} / F_{\rm bol})=-3.33$ (Fig.~\ref{SpTRatioFig}), close to the
limit of coronal saturation, and together with CX9, it also stands out
in X-ray to optical flux ratio above all of the other 67 stars
considered (Fig.~\ref{ColorVsRatioFig}).  We can use the spectral type
to derive a convective turnover time \citep{Wright:2011a} and combine
this with the rotation period to obtain the Rossby number, $R_0 =
P_{\rm rot} / \tau = 0.23$.  Comparing this with Figure~2 of
\citet{Wright:2011a} this is close to the saturated regime, consistent
with the very high value of $\log ( F_{\rm X} / F_{\rm bol}$).  The
strong aperiodic variability, quite rapid rotation, and the high level
of X-ray activity all support the classification of this object as a
young K star.

\subsection{CX9 (CXOGBS~J173508.3--292328)}

Superficially, this object presents a challenge for interpretation as
an A star with the highest X-ray to bolometric flux ratio in our
sample.  The spectrum is among the harder objects in the sample, so if
it is a coronal spectrum it must be rather hot, above 10\,MK, atypical
of later A stars which can show low-luminosity cool coronal emission.
We then expect this system to be a binary of some kind, with the
X-rays attributed to a companion star.  This is indeed the case.  This
is an ASAS variable, showing shallow eclipses with period 2.8723\,days
\citep{Otero:2006a}, see Fig.~\ref{PeriodicFig}.  No secondary eclipse
is seen, and the IR colors are clearly redder than expected for a
single A star (Fig.~\ref{ColorColorFig}), so the companion is a
late-type star, and the X-rays can probably be attributed to coronal
activity.  With $F_{\rm X} / F_{\rm Bol}$ so close to the saturation
line at $\log(F_{\rm X} / F_{\rm Bol})\sim-3$, however, the companion
must itself be very close to saturation.  For the companion to have an
acceptable $F_{\rm X} / F_{\rm Bol}$ ratio, it must have a bolometric
luminosity close to the A star, and so must be larger, and hence an
evolved star.  This may then be an Algol system.

We attempt to fit the infrared excess more quantitatively with a
simple model.  We assume that the primary is an A5 main-sequence star
(we will justify this assumption later), with appropriate absolute
magnitude and colors, and then consider a range of late-type giant
companions, with the radius and spectral type of the companion allowed
to vary freely, along with the reddening.  We find acceptable fits to
the photometry for a wide range of companions of spectral type G6 or
later.  All solutions involve a companion under-luminous for a giant,
so we classify the companion as a G6 or later sub-giant.  Formally,
the best fits are found for a K1\,IV companion with radius around
4.5\,R$_{\odot}$, and with $E(B-V)\simeq0.14$ (close to that found
above neglecting the companion).  This is not a unique solution, but
we will examine it further to show that it is consistent with other
characteristics of the system.  The K1\,IV solution corresponds to a
companion which is about 1 bolometric magnitude fainter, and the X-ray
to bolometric ratio of the companion star alone then increases to
$\log_{10}(F_{\rm X} / F_{\rm bol})=-2.7$, placing the companion on or
a little above the saturation line, but not implausibly so.  With an
A5/K1 binary, the ratio of $V$ band surface brightnesses is about
30:1, so the companion star is too dim for a secondary eclipse to be
seen, consistent with the lightcurve (Fig.~\ref{PeriodicFig}).
Finally, we can estimate the binary size given a 2.87\,hr period and
plausible masses.  For a $\sim2$\,M$_{\odot}$ A5\,V primary, the
expected binary separation is then
$a=10.6(1+q)^{\frac{1}{3}}$\,R$_{\odot}$.  This is quite consistent
with the inferred companion radius; for example if we assume
$M_2=1.0$\,M$_{\odot}$, then we expect its Roche lobe to have radius
3.9\,R$_{\odot}$, certainly consistent within the uncertainties on our
crude estimates above.  We note that if the A5 star were larger than a
main-sequence star, then the companion would also have to be larger,
and would not fit within the binary; the assumption of a main-sequence
primary made above is thus justified.

We conclude that the properties of CX9 can be well explained by an
eclipsing binary with an A5\,V primary and a late-type sub-giant
companion (with K1\,IV adopted as a representative example).  The
companion must be either filling its Roche lobe (and actively
transferring mass) or very close to it.  It will therefore be
tidally-locked, explaining the very high level of coronal activity in
this system, which is then either an Algol, possibly in a
mass-transferring phase, or an Algol-like system.  

\subsection{CX10 (CXOGBS~J173629.0--291028)}

HD\,315992 was initially classified as an F8 star of undetermined
luminosity class \citep{Nesterov:1995a}.  More recently, HD\,315992
was included in the RAVE 3rd Data Release \citep{Siebert:2011a} from
which we obtain $T_{\rm eff} = 5369$\,K and $\log g = 4.18$, pointing
instead to a mildly evolved G7 star.  The location in the $(J-H)$
vs.\ $(H-K)$ diagram ($J-H = 0.50\pm0.04$, $H-K = 0.20\pm0.040$) also
favors a later spectral type.

This appears to be a rather active star, with a high X-ray to optical
flux ratio and quite hard X-ray spectrum.  It appears to show
irregular variability in ASAS data, although this is one of the more
marginal variables in our sample (Fig.~\ref{AperiodicFigTwo}).  There
is no evidence for an IR excess, and it lies on the single-star line
in Fig.~\ref{ColorColorFig}.  While we cannot draw a definite
conclusion, CX10 appears consistent with an active single star with no
evidence for binarity.

\subsection{CX12 (CXOGBS~J174347.2--314025)}

HD\,318207 is one of the stars that appears to have an infrared
excess, and correcting for extinction will only make this more
dramatic, so this is likely a binary with a later-type companion.  The
location in the $(J-H)$ vs.\ $(H-K)$ also suggests the presence of a
companion cooler than G5.  Like CX10, this object shows a high X-ray
to optical flux ratio, and quite a hard X-ray spectrum.  It appears to
show slow long-term variability in ASAS data
(Fig.~\ref{AperiodicFigTwo}).  We tentatively classify this as a
binary containing one or two active stars, although the activity may
in this case be unrelated to the binarity as the luminosity is not
outside the range expected for single stars.

\subsection{CX25 (CXOGBS~J174502.7--315934)}

TYC~7381-792-1 shows no detectable period, but does appear to exhibit
irregular variability in ASAS data.  The colors are consistent with a
single star, with the $JHK$ colors suggesting an early K main-sequence
or late G giant type.  It has a high X-ray to optical flux ratio and
quite hard spectrum (like CX10 and 12).  It is likely an active single
star or binary.

\subsection{CX27 (CXOGBS~J173652.8--284841)}

TYC~6839-636-1 has the third highest X-ray to optical flux ratio in
our sample.  The $JHK$ colors suggest a late spectral type, and are
quite similar to CX4, possibly pointing to a giant.  We identify this
as a variable in ASAS data with a periodic modulation about 0.15\,mag
and period 31.13\,days.  The lightcurve is shown in
Fig.~\ref{PeriodicFig}.  Combined with the colors, this period
suggests this is an active late-type giant.

\subsection{CX31 (CXOGBS~J173803.5--290706)}

CX31 is identifed with LS~4306s, the southern object of a pair of O
stars about 25\,arcsec apart.  It has an O9\,{\sc v} spectral type
\citep{Vijapurkar:1993a}.  We estimate $E(B-V)=0.78\pm0.05$.
\citet{Savage:1985a} included this star in a catalog of UV extinction
determinations and estimate a slightly higher but roughly consistent
$E(B-V)=0.99$.  Our distance estimate for $E(B-V)=0.78$ is 2.6\,kpc,
consistent with \citet{Savage:1985a}, with a height above the plane of
about 80\,pc.

The relatively high luminosity, coupled with an unusually hard X-ray
spectrum for a massive star flag this as an unusual object.  The
inferred interstellar reddening is higher than for the other OB stars
in the sample, but not outside the range considered in
Section~\ref{HardnessSection}, so if there is no additional
absorption, the X-ray colors favor temperatures above 40\,MK.  Local
absorption is not generally seen in O stars, where the X-rays are
believed to originate sufficiently far from the stellar surface to not
be seen through a large column.  If there is substantial local
absorption, then the unabsorbed X-ray luminosity, and hence the
$F_{\rm X} / F_{\rm bol}$ ratio, become even larger.

This object does appear in the Washington Double Star Catalog
\citep{Mason:2001a}.  The companion star is about 2\,arcsec away,
corresponding to about 5000\,AU at 2.6\,kpc, making this an extremely
wide binary if it is indeed a physical association.  The companion is
3.5\,mag fainter than the O star, so if it is a main-sequence star,
would be have a spectral type around B5.  The X-rays clearly originate
from the O star, however, and not from the companion or a point in
between, and it is unlikely that the presence of the companion is
pertinent to the origin of the X-rays.

It remains possible that the O star itself is a much closer inner
binary.  In the optical/IR color-color plot it lies on the single star
track, but that could only indicate two stars of similar spectral
type.  In this case, the unusual X-ray emission could indicate this is
a colliding wind system.  A compact companion is also a possibility,
which would make this a quite low luminosity X-ray binary.

\subsection{CX32 (CXOGBS~J174104.9--281503)}

We find TYC~6839-218-1 to be variable in ASAS data with slow, and
probably irregular variations of about 0.1\,mag.  Its X-ray to optical
flux ratio, and X-ray color are both unremarkable for stellar X-ray
emission.  The $JHK$ colors are consistent with a G or K star.

\subsection{CX33 (CXOGBS~J174835.5--295728)}

HD\,316341 is a known Be star, although rather less well studied than
HD\,161103 (CX6) and has never been proposed to be a Be X-ray binary.
Like CX6, and as expected for a Be system, it shows a pronounced
infrared excess over the colors expected for an isolated star due to
the presence of circumstellar material, and is an outlier in $(J-H)$
vs.\ $(H-K)$.  It is also a known variable, NSV~23882, and was clearly
identified as variable by ASAS.  Its lightcurve is shown in
Fig.~\ref{AperiodicFigOne}.  There is substantial variability, both an
overall decline and several outbursts, with an overall range of around
0.5\,mag.  The outbursts appear to have a characteristic recurrence
time of about 200\,days, but are not strictly periodic and outburst
separations range from 180--230\,days.  CX230 appears to be an
off-axis detection of this object, indicating that substantial X-ray
variability is also present, since this corresponds to a $4\times$
fainter X-ray flux than its detection as CX33.

Beyond being a Be star, it is unclear how to classify this.  The X-ray
luminosity is not remarkable for single Be stars, especially not B0
stars, and is lower than the typical range for $\gamma$~Cas stars
\citep{Motch:2007a}.  The spectrum is harder than is typical, however,
with a hardness ratio comparable to the $\gamma$~Cas analog, CX6.  As
a B0 star, it also lies very close to the narrow range of spectral
types around B0.5 seen in the other $\gamma$~Cas analogs.  It may be a
lower luminosity member of the class, or one temporarily caught in a
low-luminosity state.

\subsection{CX53 (CXOGBS~J174448.8--263523)}

CX53 was matched against HD\,161117, although with a quite large
offset of 3.33\,arcsec.  Initially we flagged this object as a
possible chance alignment.  HD\,161117 is a resolved double star,
however, with a 12th magnitude companion listed in the Tycho Double
Star Catalog \citep{Fabricius:2002a}.  Using the coordinates and
proper motions of this star we find an offset of 0.52\,arcsec from the
{\it Chandra} position at the epoch of the {\it Chandra} observations,
so we believe this is likely to be the true counterpart to the X-ray
source.  If it is a true binary, then the reddening, distance, and
X-ray luminosity will be the same as calculated in
Table~\ref{HipparcosDistanceTable}.  This double is also listed in the
Washington Double Star Catalog \citep{Mason:2001a}.  The two stars
have very similar proper motions, with a common proper motion of
46\,mas\,yr$^{-1}$, and a difference between them of just
7\,mas\,yr$^{-1}$.  The latter is approximately consistent with the
orbital proper motion expected for a binary star separated by about
3\,arcsec at 100\,pc, so we believe this is indeed a true binary.

The companion is 3.4\,mag fainter than HD\,161117, but otherwise no
information is known about it.  If it is a main-sequence star, then we
expect a spectral type around K4\,V.  Associating this star with the
X-ray source implies a true $\log(F_{\rm X} / F_{\rm opt})=-2.52$,
placing it among the X-ray brighter objects in our sample.  The X-ray
hardness, however, is unremarkable for coronal emission from late-type
stars.

\subsection{CX59 (CXOGBS~J174500.5--261228)}

CX59 was matched with TYC 6832-663-1 with a quite large offset of
3.53\,arcsec.  This source was detected only 3\,arcmin off-axis by
{\it Chandra} with 27 photons, so there is no reason to expect such a
large offset.  This star has not been reported as a double, and the
X-ray source is also the hardest of the sources considered here for
which we can calculate a hardness ratio.  We conclude that this is
likely to be a chance alignment.

\subsection{CX72 (CXOGBS~J174820.3--302836)}

The X-ray spectrum of CX72 is quite soft, consistent with coronal
emission.  The object lies above the line in the IR/optical
color-color diagram (Fig.~\ref{ColorColorFig}), with a significant IR
excess relative to its Tycho-2 color.  In general this indicates a
cooler companion, but in this case, the $JHK$ colors are consistent
with its F8 spectral type, with moderate reddening, so the Tycho-2
color may be at fault, leaving no convincing evidence for a cool
companion.

We identified an outburst in the ASAS lightcurve of HD\,316356
(Fig.~\ref{AperiodicFigTwo}).  This is quite significant with
amplitude 0.15\,mag and lasting for three months.  The outburst
behavior is quite different to the flares of coronally active stars
and requires a different explanation further arguing against
associating the X-rays with a cool, active companion.  The outburst
itself is more reminiscent of dwarf nova outbursts in morphology,
although the duration is substantially longer than typical dwarf nova
outbursts.  Aside from the outburst, HD\,316356 shows no significant
variability in the ASAS data, and in particular no evidence for
ellipsoidal variations as would be expected if it were the donor star
in a cataclysmic variable.  A chance alignment with an unrelated dwarf
nova, or a hierarchical triple cannot be ruled out, but it would then
be a quite improbable coincidence with a quite atypical dwarf nova, so
this seems unlikely.  At this point, the nature of the outburst
remains unclear.

\subsection{CX77 (CXOGBS~J173638.3--285945)}

CX77 lies close to TYC~6839-487-1, although with a 6.32\,arcsec offset
this is likely to be a chance alignment.  This star is also HD\,159571
\citep{Fabricius:2002b}, although we note that the SIMBAD coordinates
listed for HD\,159571 lie rather far from TYC~6839-487-1 and do not
appear to coincide with a bright star at all.

As for CX53, this star is listed as a double \citep{Fabricius:2002a}.
Both stars have astrometry included in the UCAC-3 catalog
\citep{Zacharias:2010a}, although the proper motions for the companion
star are flagged as suspect.  The offset of the companion star from
the X-ray source is 1.17\,arcsec.  The companion star has been
detected by both 2MASS and the {\it Spitzer} GLIMPSE Point Source
Catalog\footnote{http://irsa.ipac.caltech.edu/data/SPITZER/GLIMPSE}.
The colors from $J$ to 8\,$\mu$m are flat, and difficult to reconcile
with a cool star.  The colors and apparent magnitude are as would be
expected for an A star at around the distance of HD\,159571.  The
X-ray source itself is strongly variable, being not detected in an
observation closer to on-axis, and is rather hard, and neither these
characteristics, nor the inferred luminosity, would be typical for an
A star.  It seems quite likely it is indeed an unrelated object and
that this object should be classified as a chance coincidence.

\subsection{CX82 (CXOGBS~J175709.5--272532)}

CX82 is matched with TYC~6849-1294-1.  There is very little
information about this star and no published spectral type.  The
Tycho-2 colors are consistent with a G type star or earlier, but have
large uncertainties so a cooler star cannot be ruled out. The colors
lie above the single-star line in Fig.~\ref{ColorColorFig}, but the
$(B-V)$ color is too uncertain to say this with confidence.  The $JHK$
colors favor a K3--5 classification, with a $V-K$ color also
consistent with this.

We find this object to be variable in ASAS data with a 6.3\,day period
and 0.1\,mag amplitude (Fig.~\ref{PeriodicFig}).  The X-ray to optical
flux ratio is among the highest objects in the sample.  Both
characteristics would point to a rapidly rotating active star or
binary.  Combining the 6.3\,day period with a possible spectral type
around K4, we would expect a Rossby number of $R_0\sim0.29$ (see
Section~\ref{CXSevenSection}), consistent with a star near coronal
saturation.

\subsection{CX91 (CXOGBS~J175610.9--271426)}

HD\,314883 shows no peculiarities in our data, with no variability and an
insignificant deviation from the single-star line in
Fig.~\ref{ColorColorFig}.  It is however a noted (close) resolved
binary (see e.g.\ \citealt{Fabricius:2002a}).  The two components are
separated by 0.87\,arcsec with both consistent with the X-ray position to
within an arcsecond.  They appear to share common proper motion, and have
very similar $V_{\rm T}$ and $B_{\rm T}$ magnitudes.  This appears to
be a real binary with two very similar stars.  The X-rays could
originate from either component, or from a combination of the two.
The X-ray to optical flux ratio is unremarkable for one, or two, late
F stars, and so most probably reflects normal coronal emission.

\subsection{CX115 (CXOGBS~J173940.8--285111}

CX115 is clearly the X-ray counterpart to V846~Oph (HD\,316070), a
known eclipsing Algol system.  \citet{Budding:2004a} list this object
as having an A2 primary and G7\,{\sc iv} secondary star.  Their fits
to the lightcurve give mass ratio $q=0.15$, $i=78^{\circ}$, and a
reasonable likelihood that this is a semi-detached system.

CX115 is one of the objects showing a pronouned infrared excess
(Fig.~\ref{ColorColorFig}) relative to single stars.  If we assuming
an A2 primary and G7\,IV secondary as above, we estimate that the
companion contributes about 15--20\,\%\ of the light in the $V$ band
and has a negligible effect on the reddening estimate, so the distance
estimate above should be increased by only about 10\,\%.  For
comparison, \citet{Singh:1996a} cite a distance of 580\,pc, and a {\it
  ROSAT}/PSPC X-ray luminosity of $6\times10^{30}$\,erg\,s$^{-1}$.

The eclipses are very clearly seen in the ASAS data, although it is
surprisingly omitted from the catalog of variables.  We show the
folded lightcurve in Fig.~\ref{PeriodicFig}.  The period we derive
from this photometry is $3.12678\pm0.00014$\,days, in perfect
agreement with the period of 3.1268\,days cited by
\citet{Budding:2004a}.  The secondary eclipse is small, but
detectable, and the lightcurve shows additional variability out of
eclipse.

\subsection{CX156 (CXOGBS~J174928.3--291859)}

CX156 appears to be a chance alignment with HD\,161907, with an offset
of over 6\,arcsec.  The most likely match to the X-ray source is
2MASS~J17492831--2918593, located just 0.09\,arcsec from the X-ray
position.  The colors ($J-K=0.65\pm0.11$) would suggest an early-K
spectral type.  At $K=9.30$, this star would then be about the right
brightness for an early K main-sequence star at a distance of 100\,pc.
It is therefore possible there is a physical association of this star
with CX156, although the only evidence in favor of this is that they
plausibly lie at the same distance.

\subsection{CX183 (CXOGBS~J175041.1--291644)}

HD\,162120 lies off the single-star track in Fig.~\ref{ColorColorFig},
although it is not one of the most significant IR excess cases.  It
shows no apparent variability or other peculiarities.  It probably has
a late-type companion star contributing the X-ray emission, although
there is no evidence for or against identifying this as a close
Algol-like system.

\subsection{CX205 (CXOGBS~J174917.2--303551)}

HD\,161852 is the brightest star in the sample, and saturated in ASAS
data.  This is classified as an F2\,{\sc iv/v} star
\citep{Houk:1982a}.  The optical brightness would suggest a
main-sequence rather than sub-giant classification at the {\it
  Hipparcos} distance of 50\,pc.

HD\,161852 was included in the X-ray survey of {\it Hipparcos} F stars
using {\it ROSAT} by \citet{Suchkov:2003a}.  The {\it ROSAT} estimate
of the X-ray luminosity of $7.6\times10^{28}$\,erg\,s$^{-1}$ is in a
softer band than our {\it Chandra} observation.  It was found to be a
little below average for F stars of approximately Solar metallicity as
this is ($[{\rm Fe}/{\rm H}]=-0.1$), and indeed this is among the
lowest $F_{\rm X} / F_{\rm bol}$ objects in our sample.

\subsection{CX256 (CXOGBS~J175348.1--284118)}

With $\log (F_{\rm X} / F_{\rm bol}) = -5.9$, HD\,162761 is quite
typical of single K0\,III stars ($\log (F_{\rm X} / F_{\rm bol}) =
-5.6\pm0.5$; \citealt{Hunsch:1998a}).  \citet{Richichi:2008a} observed
a Lunar occulation of this object, and identified a companion
0.11\,arcsec away, with a brightness ratio of approximately 1:120
relative to the giant.  If this is a physical binary, the separation
at 240\,pc would be about 25\,AU.  A K0 main sequence companion star
would be consistent with the brightness ratio, but would then have an
X-ray to bolometric flux ratio of around $10^{-4}$, possible, but
amongst the most active late-type stars in our sample.  Identifying
the X-rays with the visible giant star is more plausible.

\subsection{CX275 (CXOGBS~J174205.3--265046)}

CX275 lies about 2\,arcsec away from HD\,160627.  While the offset is
quite large, this was observed at quite a large off-axis angle, and
the offset is consistent with the X-ray positional uncertainty
\citep{Jonker:2011a}.  The X-ray luminosity is quite high for an F0\,V
star.  HD\,160627 is quite rapidly rotating, however ($v\sin i\simeq
30$\,km\,s$^{-1}$; \citealt{Nordstrom:1997a}), and the X-ray
luminosity is consistent with other F dwarfs with similar rotation
\citep{Maggio:1987a}.

\subsection{CX296 (CXOGBS~J174951.1--295611)}

HD\,316432 is a very good positional match to CX296.  It lies above
the single-star line in Fig.~\ref{ColorColorFig}, however the
uncertainty in its Tycho-2 color is large, and the Tycho-2 color is
bluer than an F0 star should be.  If we instead assume a typical F0
color (consistent within uncertainties), then there is no significant
IR excess.  The only real peculiarity about this object is the high
implied X-ray luminosity, and the high $F_{\rm X} / F_{bol}$.  This
suggests an unusually active F star, possibly a binary, although there
is no detected variability to support this classification.

\subsection{CX333 (CXOGBS~J173617.5--283417)}

HD\,159509 is a resolved double \citep{Fabricius:2002a}, with the
companion lying closer to the X-ray position, so it is possible that
HD\,159509 is not the true counterpart.  This was observed far enough
off-axis that neither star can be ruled out with confidence.  The
brighter star is a suspected variable (NSV~22873;
\citealt{Samus:2009a}).  The ASAS data confirm a clear modulation with
period of 14.6565 days, or more likely twice that with ellipsoidal
variations (Fig.~\ref{PeriodicFig}).  The optical/IR colors, too,
indicate that this is not a single A-type giant as there is a clear
infrared excess, corresponding to IR magnitudes about 1\,mag brighter
than expected for the A star.  Together, these facts would suggest
that HD\,159509 has a late-type companion.  Since an A3/5\,III star
should be about ten times too small to fill its Roche lobe in a
30\,day binary, it may be that the (low-amplitude) ellipsoidal
variations actually arise from the small contribution to the visible
light from a larger late-type giant companion.  This would make this
another Algol-like system with an evolved late-type companion to an A
star, although in this case a giant.  Fixing the primary to A4\,III,
we find a best fit to the $BVJK$ photometry for a K1\,IV/III
companion.  Given that it appears to be an Algol-like system, it seems
natural to associate the X-rays with HD\,159509 rather than the
slightly closer alternate star.  There are comparable Algol systems.
For example, RZ~Cas has a 32.3\,day period, an A5\,IIIe primary, and
K1\,III secondary \citep{Malkov:2006a}.

\subsection{CX337 (CXOGBS~J173527.2--293046)}

CX337 also matches an A star, HD\,315995.  Like many of the A stars in
our sample, this shows colors inconsistent with a single star
(Fig.~\ref{ColorColorFig}) indicating the presence of a cooler
companion.  If we assume a main-sequence primary, we can fit the
$BVJK$ photometry well for a range of late-G to early K sub-giant
companions, much as for the other A stars in the sample.

\subsection{CX352 (CXOGBS~J175537.1--281759)}

CX352 clearly coincides with the contact binary, V779~Sgr
(HD\,316675).  Its spectral type in the Henry Draper Charts is listed
as F8 \citep{Nesterov:1995a}, with the companion inferred to be F9
\citep{Svechnikov:2004a}.  Its Tycho-2 color is anomalous, presumably
due to the large amplitude variability, so we can make no reddening
estimate.

V779~Sgr is included in the ASAS Catalog of Variable Stars.  A
Lomb-Scargle periodogram gives a very sharp peak at 0.222515\,days.
This is the first harmonic as dominates in all ellipsoidal variables,
and the orbital period is twice this, $0.44503040\pm0.00000018$\,days,
with a bootstrap uncertainty.  The folded lightcurve is shown in
Fig.~\ref{PeriodicFig}, and shows a modulation amplitude of about
0.6\,mag.

Estimates of system parameters are presented by
\citet{Brancewicz:1980a}, who attribute 56\,\%\ of the light to the
brighter component, which has a radius 1.48\,R$_{\odot}$.
\citet{Svechnikov:2004a} alternatively attribute the brighter
component 70\,\%\ of the light, and radius 1.44\,R$_{\odot}$.  We then
expect an absolute magnitude about 1\,mag brighter than a single F8
star if the light from both stars is included.  At maximum light, when
we seen both stars well, it is observed by ASAS to be around
$V\simeq11.4$, so allowing for the light from both stars, we estimate
a distance around 500\,pc, neglecting the unknown extinction. This
would correspond to $L_{\rm X}\simeq7\times10^{29}$\,erg\,s$^{-1}$.

\subsection{CX360 (CXOGBS~J175114.4--291912)}

CX360 lies 4.07\,arcsec from TYC~6840-525-1.  CX191 also appears to be
an off-axis detection of this source.  CX360 was observed at only
2.8\,arcmin off-axis, so there is no reason to expect such a large
positional error and this star is almost certainly not the true
counterpart.  It is not noted as a double, so this is likely a chance
alignment with an unrelated object.  We note that TYC~6840-525-1 does
appear to show irregular variability in the ASAS lightcurve
(Fig.~\ref{AperiodicFigTwo}), although it is one of the weaker
detections in our sample.

\subsection{CX388 (CXOGBS~J173659.3--290603)}

CX388 lies 4.79\,arcsec from TYC~68390615-1.  This was observed at
only 0.9\,arcmin off-axis, so the positional offset is highly
significant and this star is almost certainly not the true
counterpart.  It is not noted as a double so, again, this is likely a
chance alignment with an unrelated object.

\subsection{CX402 (CXOGBS~J173533.2--302336)}

The optical counterpart to CX402 appears to be the G5 star HD\,316033.
The activity level is not atypical for G dwarfs, but the optical/IR
colors would suggest the presence of a cooler companion as well,
although this is not one of the more significant IR excesses in our
sample.

\subsection{CX485 (CXOGBS~J173718.7--282925)}

The optical counterpart to CX485 appears to be the F8 star HD\,316059.
This object shows pronouncedly redder IR than optical colors,
indicating the presence of a cooler companion.  The X-rays can most
likely be attributed to coronal activity from one or both of the
stars.

\subsection{CX506 (CXOGBS~J174045.9--280148)}

CX506 is well aligned with HD\,160390.  This star actually lies rather
close to the single-star line in Fig.~\ref{ColorColorFig}, unlike most
of the A stars in our sample.  A small IR excess may be present, but
we find by fitting the $BVJK$ colors that this is only sufficient to
accomodate a late-type main-sequence or slightly evolved star, no more
than a magnitude above the main-sequence.  Late-type giant, and
probably sub-giant companions can probably be ruled out.  The allowed
companions (later than F0) all have bolometric magnitudes at least
3.2\,mags fainter than the A giant, and so if we attribute the X-rays
to the (putative) companion, they would have $F_{\rm X} / F_{\rm Bol}
\ga -3.5$, so would be close to coronal saturation and very active.
Such a case cannot be ruled out, but we should stress that there is no
evidence (other than the presence of X-ray emission) for such a
companion.

\subsection{CX514 (CXOGBS~J175637.0--271145)}

CX514 lies near the B9 star HD\,314884.  With an X-ray luminosity of
at least $10^{30}$\,erg\,s$^{-1}$, and $F_{\rm X} / F_{\rm bol} =
-5.4$, this falls well outside the norms for late B stars, so either
this is a chance alignment, or a companion is present.

We do find HD\,314884 to be variable in ASAS data, showing a 0.89\,day
period (Fig.~\ref{PeriodicFig}).  The periodicity is stable in
amplitude and phase through the period covered by ASAS.  Besides
variability and X-ray emission, there is no other evidence for
multiplicity in this system.  The optical/IR colors are consistent
with a single B9 star, ruling out late-type sub-giant or giant
companions (if the primary is main-sequence), but a main-sequence
companion could still be present, or an evolved companion to a B
giant.  If the putative companion is an early K or earlier type then
the X-ray emission can be accounted for with $\log (F_{\rm X} / F_{\rm
  bol}) < -3$.  Such a companion would have to be quite active, but
this is expected since it would also have to be quite young if in a
binary with a B9 star.

\subsection{CX524 (CXOGBS~J175419.3--283654)}

TYC~6853-80-1 appears to be an irregular variable in ASAS data
(Fig.~\ref{AperiodicFigTwo}).  Its $JHK$ colors favor a K star, or
possibly a G giant like CX4.  This then appears to be a late-type,
somewhat active star.

\subsection{CX622 (CXOGBS~J173654.1--295231)}

CX622 is offset 2.37\,arcsec from TYC 6839-19-1.  This was observed at
3\,arcmin off-axis, so we would not expect such a large error in the
X-ray position.  There is very little information available about TYC
6839-19-1 but it shows no peculiarities that would support identifying
it as the true counterpart to the X-ray source.  This then, is likely
to be a chance alignment.

\subsection{CX632 (CXOGBS~J173353.9--292355)}

CX632 lies 1.28\,arcsec from HD\,315998.  This was originally
classified as a K5 star \citep{Nesterov:1995a}.  It is one of our
stars that is included in the RAVE 3rd Data Release
\citep{Siebert:2011a} from which we obtain $T_{\rm eff} = 4533$\,K and
$\log g = 2.08$.  From this, it is clearly a giant and the temperature
then indicates a spectral class around K1.  Hence we classify this as
K1\,{\sc iii}.  With $\log(F_{\rm X} / F_{\rm bol})=-5.6$, the X-ray
emission is quite typical of single K1\,III stars for which
\citet{Hunsch:1998a} find a distribution of ratios $-5.5\pm0.5$.  The
optical/IR colors are consistent with a single giant star.

\subsection{CX698 (CXOGBS~J175220.8--290427)}

TYC~6853-1354-1 matches well with the position of CX698.  The optical
color would suggest a mid-F star or earlier, whereas the IR color is
redder, possibly indicating the likely presence of a cooler companion,
although the IR colors are flagged by 2MASS as suspect.

\subsection{CX785 (CXOGBS~J174249.9--275038)}%

CX785 coincides with HD\,160769.  This object falls above the
single-star line in Fig.~\ref{ColorColorFig}, but the 2MASS $J$ and
$H$ photometry is flagged as suspect.  The DENIS
survey\footnote{VizieR On-line Data Catalog: II/263} is in reasonable
agreement, with $J-K_{\rm S}=0.55\pm0.09$, so the IR excess in this
object appears to be real indicating a cooler companion.  We can find
a reasonable fit to the $BVJK$ photometry with a late-K companion star
a little above the main-sequence.  It is quite possible that this is
the origin of the X-rays instead of the F3 star.

\subsection{CX863 (CXOGBS~J173303.4--293902)}

The optical counterpart to CX863 is clearly HD158902.  This is a mid-B
supergiant, originally classified as B8 by \citet{Cannon:1924a}.  More
recent two-dimensional classifications from objective prism surveys
have given spectral types of B5\,{\sc i}a \citep{MacConnell:1976a},
B5\,{\sc i}b \citep{Garrison:1977a}, and B3\,{\sc ii}
\citep{Houk:1982a}.  The implied distance would then range from
1.3\,kpc (B3\,{\sc ii}) to 4.5\,kpc (B5\,{\sc i}a).  Given that the
GBS is observing out of the plane, and this source is at
$b=+1.94^{\circ}$, less luminous stars seem most likely, since they
imply shorter distances from the plane.  For a distance of 1.3\,kpc,
the distance above the plane is about 70\,pc, whereas for 4.5\,kpc it
is 160\,pc.  We therefore suggest this is a B3\,{\sc ii} star at a
distance of 1.3\,kpc, although a more distant and more luminous object
remains possible.  The properties of this star are quite normal for
its spectral type.  Its $F_{\rm X} / F_{\rm bol}$ ratio is quite
reasonable, it shows no variability, and its optical/IR colors are
close to the single star line.  The IR colors are a little redder than
expected but the deviation is not highly significant.

\subsection{CX904 (CXOGBS~J175625.3--271043)}

CX904 is a good match to TYC~6849-1144-1.  It does fall above the
single-star line in Fig.~\ref{ColorColorFig} suggesting it may be a
binary, but the IR excess is not significant without either a spectral
type or a more reliable optical color.  The $JHK$ colors alone would
suggest a K spectral type.

\subsection{CX916 (CXOGBS~J175543.3--280443) and CX917 (CXOGBS~J175543.0--280446)}

CX916 matches TYC 6849-227-1 (HD\,316666A) within 0.69\,arcsec.  CX917
matches HD\,316666B within 1.5\,arcsec.  HD\,316666A does show
something of an IR excess in Fig.~\ref{ColorColorFig}, although it is
not highly significant.  It may itself be an unresolved binary.  This
was identified as a candidate variable by \citet{Piquard:2001a} with a
period of 0.173946\,days.  We could not reproduce this, or any other
periodicity in the ASAS data.

We have much less information on the companion.  It is independently
detected by 2MASS with $K=9.878\pm0.033$, 1.17\,mags fainter than
HD\,316666A.  The difference in magnitudes would be appropriate for a
G0\,V star, and the $JHK$ colors are consistent with such a
classification.

\subsection{CX1113 (CXOGBS~J174011.4--280212)}

CX1113 matches well with TYC~6835-312-1.  
This star shows marginal evidence for irregular variability and its
$JHK$ colors would suggest a G star, but we have no other constraining
information.

\subsection{CXB5}

CXB5 coincides very closely to HD\,315961, originally classified as a
K5 star \citep{Nesterov:1995a}.  \citet{Neese:1988a} instead
classified this star (listed as PSD~157--562;
\citealt{BarbierBrossat:2000a}) as K0\,III, with $E(B-V)=0.3$, and a
distance of 504\,pc, based on DDO photometry. \citet{Neese:1988a}
compare their photometric classifications with those from objective
prism spectroscopy, and find scatters of up to around five spectral
subclasses, and two luminosity classes, so this should be considered
rather uncertain.  The quite high X-ray luminosity, $L_{\rm
  X}\simeq1.0\times10^{31}$\,erg\,s$^{-1}$, together with a high $\log
( F_{\rm X} / F_{\rm bol})=-4.3$ (compared to $5.6\pm0.5$;
\citealt{Hunsch:1998a}) points to a quite active giant.

We see no evidence for a periodicity in the ASAS data, but the star
does seem to be an irregular variable (Fig.~\ref{AperiodicFigTwo}).
The X-ray color is rather hard, comparable to other stars at high
$\log ( F_{\rm X} / F_{\rm bol})$ such as CX4, 10, and 12.  It appears
to be quite comparable to CX4 in some ways, and that also shows
irregular variability in addition to its periodicity.

This is noted as a double with a fainter, yet bluer companion
1.6\,arcsec away \citep{Fabricius:2002a}.  The X-rays are clearly
associated with the K star, however, and it is unclear if the nearby
star is physically associated.

\subsection{CXB9}

CXB9 is associated with HD\,161853.  This is clearly an OB star, but
there is uncertainty about its classification.  It was originally
listed as B3 by \citet{Cannon:1924a}, and then revised to B0/1\,{\sc
  ii} in the Michigan HD classification \citep{Houk:1982a}.  Other
estimates have included O7.5 \citep{Crampton:1971a}, O8\,{\sc v}\,{\it
  ((n))} \citep{Walborn:1973a}, O7.5\,{\sc v}\,{\it n}
\citep{Garrison:1977a}, O9\,{\sc i} \citep{Hu:1993a}, and finally an
O8\,{\sc iii} post-AGB star \citep{Parthasarathy:2000a} based on its
association with IRAS~17460--3114. \citet{Suarez:2006a} instead argue
that it is not a post-AGB star, but instead a young object.

Based on UV data, \citet{Savage:1985a} estimated $E(B-V)=0.54$, in
fair agreement with our estimate of $E(B-V)=0.44\pm0.02$.  The implied
distance is then 2.7\,kpc and the distance below the plane is about
70\,pc \citep{Straizys:1981a}.  This would be consistent with the
Scutum-Centaurus Arm.  Main sequence types would situate it closer to
2.0\,kpc, between Scutum-Centaurus and Sagittarius, while a supergiant
classification would put it at 4--6\,kpc, consistent with either the
Norma Arm or the 3\,kpc Arm.  In the latter cases, the distance below
the plane is, however, larger, 110--190\,pc.  Overall, an O8\,{\sc
  iii} star at around 2.7\,kpc in the Scutum-Centaurus Arm appears to
fit best.  For this case, the X-ray luminosity is about
$3\times10^{32}$\,erg\,s$^{-1}$.  This is quite high, but the $F_{\rm
  X} / F_{\rm bol}$ ratio is not excessive, and with quite soft colors
this appears consistent with normal emission for O star winds.  There
is no variability seen in this object, and no evidence of binarity is
reported.  While the optical/near-IR colors we examine here are normal
for a single star, it does have an mid-IR excess \citep{Clarke:2005a}.
As noted above, \citet{Suarez:2006a} suggest that this is evidence
that it is a young object.

\subsection{CXB17}

CXB17 lies near the F8 star HD\,316565 \citep{Nesterov:1995a}, but
with an offst of 4.98\,arcsec, this is likely to be a chance
alignment.  We note that a 2MASS~J17543011--2923572 is coincident with
the X-ray source and is a much more likely counterpart.

\subsection{CXB93 and CXB36}

CXB93 coincides with a nearby M dwarf, LTT\,7072 = GJ\,2130A =
HD~318327A. It is one component of a 20\,arcsec separation binary,
with the companion GJ\,2130B = HD~318327B also detected by the GBS as
CXB36.  The spectral type of GJ\,2130A has been variously estimated as
M0 \citep{Nesterov:1995a}, M1.5 \citep{Bidelman:1985a}, M2
\citep{Stephenson:1977a}, and M3 \citep{Raharto:1984a}, with the order
reflecting that the determinations, not of the publication.  The same
works cite spectral types of M2--5 for GJ\,2130B.  GJ\,2130A appears
above the single-star line in Fig.~\ref{ColorColorFig}.  In this case,
it appears to reflect the large uncertainty in the Tycho-2 $B-V$
color, or inaccuracy of the transformations from $B_{\rm T}-V_{\rm T}$
for very red objects.  An independent estimate gives $B-V=1.465$
\citep{Koen:2010a}, in better agreement with its spectral type.

The X-ray luminosity of GJ\,2130A is actually rather low among M
dwarfs, for which \citet{Schmitt:1995a} find a range from
$10^{26}-10^{29}$\,erg\,s$^{-1}$.  While there is uncertainty in the
{\it Hipparcos} parallax distance, spectroscopic parallax distances
favor short distances, and low luminosities as well, ranging from
16\,pc for M0 to 12\,pc for M3.  Partly this discrepancy may reflect
the softer bandpass used by {\it ROSAT}, with our {\it Chandra}
observations possibly missing much of the X-ray flux.  The companion,
is X-ray brighter, with $L_{\rm X}\simeq
1.5\times10^{27}$\,erg\,s$^{-1}$.

Optically, GJ\,2130A was studied by \citet{Rauscher:2006a} as part of
a survey of chromospheric emission in late K and M dwarfs.  It was
found to show the strongest Ca\,{\sc ii} emission among stars of
comparable absolute magnitude.

\subsection{CXB211}

CXB211 lies 2.61\,arcsec from HD\,160826, but this was observed at a
large off-axis angle (8.6\,arcmin), so this is still a plausible
match.  HD\,160826 is classified as a B9\,V star \citep{Houk:1982a},
so should not be an X-ray source in its own right.  This object shows
no noticeable IR excess, but we do note this is a resolved double with
a separation of 0.6\,arcsec \citep{Fabricius:2002a}.  The nearby star
has $V=10.15\pm0.04$ and $B-V=0.43\pm0.05$, and lies at 2.90\,arcsec
from CXB211.  If this is a physical binary with the same reddening and
distance then this would be a late-F or early G giant or sub-giant.
It is likely that this is the true X-ray counterpart rather than
HD\,160826 itself as both stars are consistent with the poorly
constrained X-ray position.

\subsection{CXB287}

CXB287 lies 1.39\,arcsec from HD\,158982, but the offset is not
significant as this was observed 5.9\,arcmin off-axis.  HD\,158982 is
classified as a A2\,IV/V star by \citet{Houk:1982a}, so should not be
an X-ray source.  This is listed as a resolved double by
\citet{Mason:2001a}, with a separation 0.3\,arcsec.  The companion may
be the true counterpart.  It is only 0.4\,mag fainter than the A2
star, so is also unlikely to be a significant X-ray source in its own
right unless it is a cooler sub-giant or giant, rather than a
main-sequence star of type A4 or so, or the companion is closer than
the A star.

\subsection{CXB296}

CXB296 lies 5.95\,arcsec from HD\,161839, making this star itself
inconsistent with the X-ray source given a modest off-axis angle
(2.1\,arcmin).  This star was originally classified as B5, but the
Michigan Catalog updates this to B5/7\,{\sc ii/iii}
\citep{Houk:1982a}.  An alternative classification of B2\,{\sc v} was
proposed by \citet{Garrison:1977a}.  All of these classifications
would correspond to $E(B-V)$ between 0.25 and 0.35, and a range of
distances of 1--5\,kpc.  HD\,161839 has a pronounced infrared excess,
one of the largest in our sample (Fig.~\ref{ColorColorFig}).  It was
previously identified by \citet{Clarke:2005a} as an IR excess object
based on matching {\it MSX} and Tycho-2 detections.  Combining 2MASS
and GLIMPSE IR photometry, with colors of $H-K=0.23$ and $K-[8]=1.40$,
it appears consistent with a Be star (see Fig.~4 of
\citealt{Clarke:2005a}).  While a Be star might be expected to be an
X-ray source, it is not consistent with the X-ray position, and is not
noted as being a double, so this is probably a chance alignment.

\subsection{CXB302}

TYC~6849-01627-1 appears to be the optical counterpart to CXB302.
There is very little information about this star.  The optical colors
would be consistent with a range of spectral types earlier than early
G, depending on reddening, and the $JHK$ colors suggest early F.  It
appears to show marginal irregular variability in ASAS.

\subsection{CXB306}

CXB306 coincides with HD\,163613.  This appears to be a normal early B
supergiant.  Its optical/IR colors are consistent with a single normal
star, it shows no variability and there is no evidence for binarity.
It was originally classifed as B5 \citep{Cannon:1924a}, and updated to
B1\,{\sc i--ii} by \citet{Hoffleit:1956a}.  Its $F_{\rm X} / F_{\rm
  bol}$ ratio is quite reasonable.  The expected distance ranges from
2.8\,kpc (B1\,{\sc ii}) to 7.0\,kpc (B1\,{\sc i}a).  Given that the
GBS is observing out of the plane, and this source is at
$b=-1.96^{\circ}$, less luminous stars seem most likely, since they
imply shorter distances from the plane.  For a distance of 2.8\,kpc,
the distance below the plane is about 80\,pc, whereas for 7.0\,kpc it
is 240\,pc.  We then suggest this is a B1\,{\sc ii} star at about
2.8\,kpc in the Scutum-Centaurus arm.

\subsection{CXB422}

CXB422 is well matched with HD\,315956.  This is classified as an F2
star \citep{Nesterov:1995a}.  Its parallax distance and brightness
suggest a dwarf to giant luminosity class, and hence spectral class
F2\,III--V.

\section{Discussion}
\label{DiscussionSection}

\subsection{OB and Be stars}

Four apparently normal OB stars, and two Be stars were found in our
sample, all with very close matches between X-ray and optical
positions.  In addition, a third X-ray source (CXB296) was found near
a Be star, but too far away to be a credible X-ray counterpart to it.
There is every reason to believe the other six early type stars are
all true optical counterparts to the X-ray sources.

Two of the OB stars CX31 and CXB8 have late-O spectral classes.  They
have quite high $F_{\rm X} / F_{\rm bol}$ ratios, but still plausible
for single, normal OB stars.  Both have optical/IR colors consistent
with normal stars, although CXB8 does have a mid-IR excess and appears
to be a young object.  CX31 has a rather hard X-ray spectrum (by the
standards of our sample), but this is at least in part due to a higher
$N_{\rm H}$ than the other OB stars in the sample, judging by its
$(B-V)$ color.  It remains possible that it is a more unusual object
such as a colliding wind binary or low-luminosity X-ray binary.  CX863
and CXB306 are slightly later in spectral class, early-B.  Their
$F_{\rm X} / F_{\rm bol}$ ratios are lower, and quite typical of OB
stars.  CXB306 has normal stellar colors.  CX863 may have slightly
redder colors than a single star, but this only marginally significant

One of the GBS predictions was that $\sim9$ Be X-ray binaries would be
identified \citep{Jonker:2011a}.  None have been found in the (nearby)
Tycho-2 sample, but the sources predicted were expected to be further
away and subject to more reddening, and so would not be expected to
have Tycho-2 counterparts.  Instead, the two Be stars we have found
are fainter in X-rays.  One, CX6, is an intermediate luminosity
$\gamma$~Cas system, possibly harboring a white dwarf.  The other,
CX33, is most likely a normal, single Be star.

\subsection{Algol Candidates and other A stars}

\begin{figure}
\includegraphics[width=3.6in]{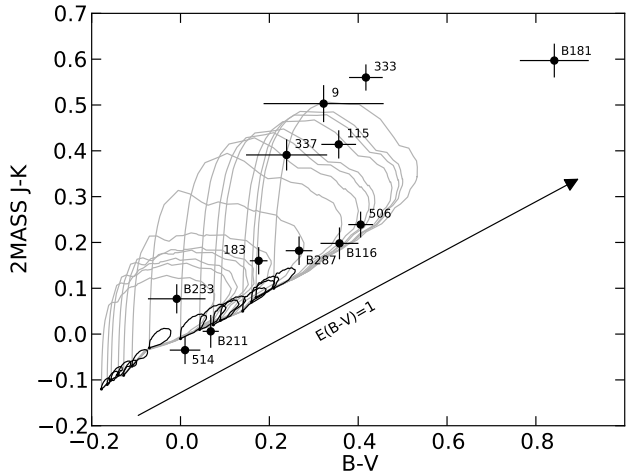}
\caption{Optical/IR color-color diagram for spectroscopically
  classified late B and A stars in the sample.  Black loops trace the
  effect of adding later-type main-sequence companions on the colors,
  for B3--A7 primaries.  Grey loops correspond to companions
  4\,magnitudes brighter than a main-sequence star.  In both cases,
  the loop is traced counter-clockwise from early-type companions to
  later types.  This is adapted from Figure~\ref{ColorColorFig}; see
  that figure for further explanation.  }
\label{AStarColorColorFig}
\end{figure}

There are a surprising number of A stars in our sample.  To some
extent, this is a selection effect since our Tycho-2 sample is
effectively an optical flux-limited sample and so has a larger sample
volume for earlier spectral types.  In most cases, as expected, there
is evidence from IR colors and/or variability for a companion star,
usually inferred to be larger in radius than the A star.
Figure~\ref{AStarColorColorFig} shows just the spectroscopically
confirmed late B and A stars from Figure~\ref{ColorColorFig}.  We show
the B3--A7 main-sequence, together with two sets of loops, one
corresonding to main-sequence companions of a range of spectral types,
the other to more luminous companions.  We see that the sources appear
to divide into three groups.  Neither of the two B9 stars, CX514 and
CXB211 show evidence for an IR excess, although a main-sequence
companion cannot be ruled out.  CX183, CX506, CXB116, CXB233, and
CXB287 all show weak excesses, suggesting main-sequence or mildly
evolved companions.  Finally CX9, CX115, CX333, CX337, and CXB181 all
require luminous companions $\ga4$\,mags above the main sequence.

The preponderance of evolved companions is not expected, since later
type companions should not have had time to evolve off of the
main-sequence. It could arise in two ways.  First, where a large
difference in spectral types is seen, it is possible that the
later-type star is still a very active pre-main-sequence star, and so
is larger than it will eventually be.  Secondly, an A star plus a
late-type sub-giant or giant could be a typical Algol system.

CX115 is, indeed, a known eclipsing Algol.  CX9 is also eclipsing and
we propose that it is a new Algol candidate, and possibly a
mass-transferring system.  CX333 appears to show ellipsoidal
variations suggesting a close, late-type companion in a 30\,day
period.  It may be a long-period Algol.  For the remainder of the
sources, although a late-type companion is implied, there is no
evidence that it is close.

\subsection{Active late type stars}

We attempt to distinguish active late-type stars (i.e.\ FGKM stars
with very strong coronal emission) from those showing more normal
coronal emission.  Criteria for `activity' include a high X-ray to
optical flux ratio, hard X-ray spectra, or irregular variability.  

The presence of radio emission would be a very powerful diagnostic as
magnetically active stars usually follow the G\"{u}del-Benz relation,
whereas inactive stars are much weaker in the radio, while retaining
significant X-ray emission \citep[][and references
  therein]{Forbrich:2011a}.  \citet{Maccarone:2012a} looked for radio
counterparts to GBS X-ray sources using the NVSS catalog
\citep{Condon:1998a}.  12 matches were found with the original CX
sources of \citet{Jonker:2011a}, but none correspond to objects in
this paper.  Neither are any of the CXB sources in this paper detected
by the NVSS.  Unfortunately, while the NVSS is the deepest radio
survey covering the whole GBS area, it is not sensitive enough to
detect radio emission from these objects, as the G\"{u}del-Benz
relation would predict that all of the sources considered in this
paper would fall well below the 2.5\,mJy sensitivity of the NVSS.

Among the late-type stars, CX7 stands out.  This is at or near coronal
saturation and appears to be a very young K dwarf.  CX352 is also
distinguished by being the only W~UMa contact binary in the sample.  A
few other stars show signs of strong activity.  CX4 is a very active
giant.  CX10 and 12, and CXB5 are all active G or K stars with no
luminosity class, showing high $F_{\rm X} / F_{\rm bol}$, quite hard
X-ray spectra, and irregular variability.  CX12, at least, appears to
have a later type companion as well.

Finally we note two somewhat unusual F stars, both with moderately
high $F_{\rm X} / F_{\rm bol}$.  CX72 exhibited an unclassified
outburst, with no other peculiarities.  CX296 is unusual in its lack
of peculiarities since F0 stars typically have quite weak X-ray
emission; indeed, in Figure~\ref{SpTRatioFig} it appears somewhat
isolated, and lies an order of magnitude above other stars of the same
spectral type.

\subsection{Normal late type stars}

Many of the late-type stars show no signs of strong activity; modest
$F_{\rm X}/ F_{\rm bol}$, and no variability.  These include two
giants, CX256 and CX632, and several dwarfs, CX26, CX205, CX275, and
CX785.  The remainder, CX91, CX402, CX485, CX916, and CXB422, have no
luminosity classifications.

Our sample only included one Tycho-2 M dwarf, CXB93, together with its
binary companion CXB36 which is not in the Tycho-2 catalog.  Since the
sample is selected by optical brightness, this is not surprising.  The
X-ray luminosities of these two stars are unremarkable, and if
anything unusually low.

\subsection{Chance alignments}

After checking for known doubles, a few apparently single Tycho-2
stars are significantly offset from the nearby {\it Chandra} source,
and these remain as probable chance alignments.  These are CX59,
CX156, CX360, CX388, CX622, CXB17, and CXB296.  CXB296 is a
particularly interesting case since it appears to lie near a Be star,
yet that star rather securely is not the detected X-ray source.  In
addition, CX77 seems likely to not be associated with either
HD\,157571 or the star close to it, bringing us to a total of eight
likely coincidences.  There are additionally a few objects for which a
real match could not be ruled out at the 95\,\%\ confidence level, but
which may nonetheless be close coincidences.

We estimated in Section~\ref{CoincidenceSection} that there should be
about 5 true coincidences in our sample.  While finding 8 is not a
substantial excess, with the high incidence of binarity among field
stars, it is likely that some of these could be undetected resolved
binary companions.

\section{Conclusions}
\label{Conclusion}

We have examined the properties of 69 stars from the Galactic Bulge
Survey (\citealt{Jonker:2011a}, Jonker et al. in preparation) which
match with Tycho-2 stars.  All have quite low $F_{\rm X} / F_{\rm
  opt}$ ratios in the range $10^{-6} - 10^{-2}$, and can be adequately
explained without invoking compact companions.  The sample includes
one case where we appear to detect a resolved companion rather than
the Tycho-2 star, and a further eight which appear to be chance
alignments.  This is comparable to the expectation of five
coincidences and may include a few more undetected resolved
companions.

The hot stars include two Be stars, one of which is a $\gamma$~Cas
analog. The remaining four spectroscopically identified OB stars show
$F_{\rm X} / F_{\rm bol} \sim 10^{-7}$ as is commonly found for X-rays
from the winds of these stars.

A substantial fraction of our sources, 12 objects out of 38 that have
spectroscopic classifications, are A stars.  One of these is a known
Algol, V846~Oph, and the others include a disproportionate number of
objects showing infrared excesses, and we expect that the X-rays in
most or all of these objects come from cooler companions.  These
systems appear divided between some with very pronounced IR excesses
that appear to have sub-giant or giant companions, and others with
weaker or no IR excess consistent with main-sequence or mildly evolved
companions.

Among the late-type stars a few stand out as having high $F_{\rm X} /
F_{\rm bol}$ ratios, and hard X-ray spectra implying high temperature
coronae.  These include the active giant, CX4, and a young K dwarf,
CX7.  Several other stars show quite high levels of activity.

The remainder of the late-type stars have lower $F_{\rm X} / F_{\rm
  bol}$ ratios, and typically no evidence for variability.  Some of
these sources are rather close, in particular the one M dwarf in our
sample, CXB93, and the modest X-ray luminosities implied can be
accounted for by normal coronal emission.

\acknowledgments

This work was supported by the National Science Foundation under Grant
No. AST-0908789 and by grant GO2-13044B from the Smithsonian
Astrophysical Observatory.  P.G.J. and G.N. acknowledge support from
the Netherlands Organisation for Scientific
Research. D.S. acknowledges an STFC Advanced Fellowship. T.J.M. thanks
STFC for support via a rolling grant to the University of Southampton.
R.I.H. would like to thank the Universities of Warwick and
Southampton, and SRON, the Netherlands Institute for Space Research in
Utrecht for their hospitality while much of this work was being done.

This publication makes use of data products from the Two Micron All
Sky Survey, which is a joint project of the University of
Massachusetts and the Infrared Processing and Analysis
Center/California Institute of Technology, funded by the National
Aeronautics and Space Administration and the National Science
Foundation.  This research has made use of the SIMBAD database,
operated at CDS, Strasbourg, France, and NASA's Astrophysics Data
System.  Finally we are very grateful to Dr. Eric Mamajek for making
available his compilation of stellar colors and temperatures for dwarf
stars.

{\it Facilities:} \facility{ASAS}, \facility{CTIO:2MASS}, \facility{CXO (ACIS)}, \facility{HIPPARCOS}.

\end{document}